\documentclass[twocolumn]{aastex61}
\pdfoutput=1

\received{May 30, 2017}
\revised{August 28, 2017}
\accepted{August 31, 2017}
\submitjournal{\apj}

\shorttitle{Masses and Stellar Content of Nuclei}
\shortauthors{Spengler et al.}

\begin{document}


\title{Virgo Redux: The Masses and Stellar Content of Nuclei in Early-Type Galaxies
    from Multi-Band Photometry and Spectroscopy}


\begin{abstract}
We present an analysis of 39 nuclei and their early-type hosts in the Virgo Cluster using ten broadband filters: F300W, F475W, F850LP, F160W, $u^*griz$, and $K_s$. We describe the {\it Virgo Redux} program, which provides high-resolution UV and NIR imaging. Combining this data with optical and NIR imaging from the {\it ACS Virgo Cluster Survey} and the {\it Next Generation Virgo Cluster Survey}, we estimate masses, metallicities and ages using simple stellar population (SSP) models. For 19 nuclei, we compare to SSP parameters derived from Keck and Gemini spectra and find reasonable agreement between the photometric and spectroscopic metallicity: the RMS scatter is 0.3~dex. We reproduce the nucleus-galaxy mass fraction of $0.33^{+0.09}_{-0.07}$ percent for galaxy stellar masses $10^{8.4}-10^{10.3} M_\odot$ with a typical precision of $\sim$35\% for the nuclei masses. Based on available model predictions, there is no single preferred formation scenario for nuclei, suggesting that nuclei are formed stochastically through a mix of processes. Nuclei metallicities are statistically identical to those of their hosts, appearing $0.07 \pm 0.3$~dex more metal-rich on average --- although, omitting galaxies with unusual origins, nuclei are $0.20\pm0.28$~dex more metal-rich. Nuclei appear to be $0.56 \pm 0.12$~dex more metal rich than ultra-compact dwarf galaxies (UCDs) at fixed mass. We find no clear age difference between nuclei and their galaxies, with nuclei displaying a broad range of ages. Interestingly, we find that the most massive nuclei may be flatter and more closely aligned with the semi-major axes of their hosts, suggesting that they formed through predominantly dissipative processes.
\end{abstract}

\correspondingauthor{Chelsea Spengler}
\email{spengler@uvic.ca}

\author{Chelsea Spengler}
\affil{Department of Physics and Astronomy, University of Victoria, Victoria, BC V8P 5C2, Canada}
%\email{spengler@uvic.ca}

\author{Patrick C\^ot\'e}
\affil{Herzberg Institute of Astrophysics, National Research Council of Canada, Victoria, BC V9E 2E7, Canada}

\author{Joel Roediger}
\affil{Herzberg Institute of Astrophysics, National Research Council of Canada, Victoria, BC V9E 2E7, Canada}

\author{Laura Ferrarese}
\affil{Herzberg Institute of Astrophysics, National Research Council of Canada, Victoria, BC V9E 2E7, Canada}

\author{Rub\'en S\'anchez-Janssen}
\affil{Herzberg Institute of Astrophysics, National Research Council of Canada, Victoria, BC V9E 2E7, Canada}
\affil{STFC UK Astronomy Technology Centre, The Royal Observatory Edinburgh, Blackford Hill, Edinburgh, EH9 3HJ, UK}

\author{Elisa Toloba}
\affil{Department of Physics, University of the Pacific, 3601 Pacific Avenue, Stockton, CA 95211, USA}

\author{Yiqing Liu}
\affil{Sub-department of Astrophysics, Department of Physics, University of Oxford, Denys Wilkinson Building, Keble Road, Oxford OX1 3RH, UK }

\author{Puragra Guhathakurta}
\affil{UCO/Lick Observatory, Department of Astronomy and Astrophysics, University of California Santa Cruz, 1156 High Street, Santa Cruz, CA 95064, USA}

\author{Jean-Charles Cuillandre}
\affil{CEA/IRFU/SAp, Laboratoire AIM Paris-Saclay, CNRS/INSU, Universit\'e Paris Diderot, Observatoire de Paris, PSL Research University, F-91191 Gif-sur-Yvette Cedex, France}

\author{Stephen Gwyn}
\affil{Herzberg Institute of Astrophysics, National Research Council of Canada, Victoria, BC V9E 2E7, Canada}

\author{Andrew Zirm}
\affil{Greenhouse Software, 3rd Floor, 110 5th Avenue, New York, NY 10011, USA}

\author{Roberto Mu\~noz}
\affil{Institute of Astrophysics, Pontificia Universidad Cat\'olica de Chile, Av. Vicu\~na Mackenna 4860, 7820436 Macul, Santiago, Chile}

\author{Thomas Puzia}
\affil{Institute of Astrophysics, Pontificia Universidad Cat\'olica de Chile, Av. Vicu\~na Mackenna 4860, 7820436 Macul, Santiago, Chile}

\author{Ariane Lan\c con}
\affil{Observatoire Astronomique de Strasbourg, Universit\'e de Strasbourg, CNRS, UMR 7550, 11 rue de l'Universit\'e, F-67000 Strasbourg, France}

\author{Eric W. Peng}
\affil{Department of Astronomy, Peking University, Beijing 100871, China}
\affil{Kavli Institute for Astronomy and Astrophysics, Peking University, Beijing 100871, China}

\author{Simona Mei}
\affil{LERMA, Observatoire de Paris,  PSL Research University, CNRS, Sorbonne Universit\'es, UPMC Univ. Paris 06, F-75014 Paris, France}
\affil{University of Paris Denis Diderot, University of Paris Sorbonne Cit\'e (PSC), 75205 Paris Cedex 13, France}
\affil{Jet Propulsion Laboratory, Cahill Center for Astronomy \& Astrophysics, California Institute of Technology, 4800 Oak Grove Drive, Pasadena, California, USA}

\author{Mathieu Powalka}
\affil{Observatoire Astronomique de Strasbourg, Universit\'e de Strasbourg, CNRS, UMR 7550, 11 rue de l'Universit\'e, F-67000 Strasbourg, France}


%% subject header.  Objects should be in the appropriate "individual"
%% headers (e.g. quasars: individual, stars: individual, etc.) with the
%% additional provision that the total number of headers, including each
%% individual object, not exceed six.  The \objectname{} macro, and its
%% alias \object{}, is used to mark each object.  The macro takes the object
%% name as its primary argument.  This name will appear in the paper
%% and serve as the link's anchor in the electronic edition if the name
%% is recognized by the data centers.  The macro also takes an optional
%% argument in parentheses in cases where the data center identification
%% differs from what is to be printed in the paper.


\section{Introduction} \label{sec:intro}

\begin{figure*}[htbp]
%%\epsscale{1.1}
\plotone{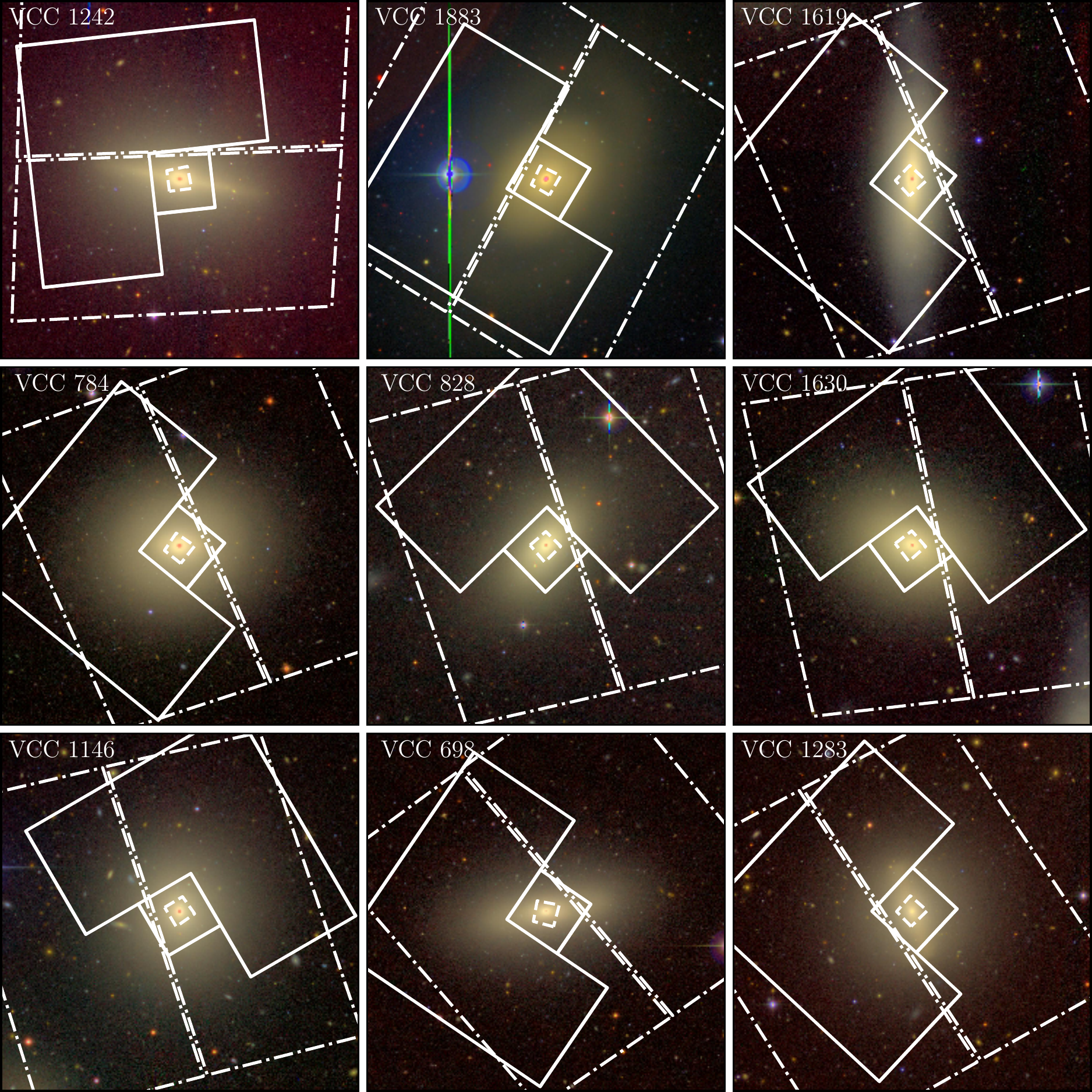}
\caption{CFHT/MegaCam $giz$ color images with HST instrument footprints overlaid. Galaxies are shown in order of decreasing luminosity in the F475W filter (from left to right and top to bottom). Note that the colormap scaling is not absolute across all panels. Each image measures $3\farcm75 \times 3\farcm75$ ($18 \times 18$\,kpc) and thus covers only a small fraction of the MegaCam 1\,$\deg^2$ field. ACS/WFC footprints are shown as dashed-dotted lines, NICMOS footprints are show as dashed lines, and WFPC2 footprints are shown as solid lines. In all cases, north is up and east is to the left.}
\label{fig:color}
\end{figure*}

\begin{figure*}[htbp]
\setcounter{figure}{0}
%%\epsscale{1.1}
\plotone{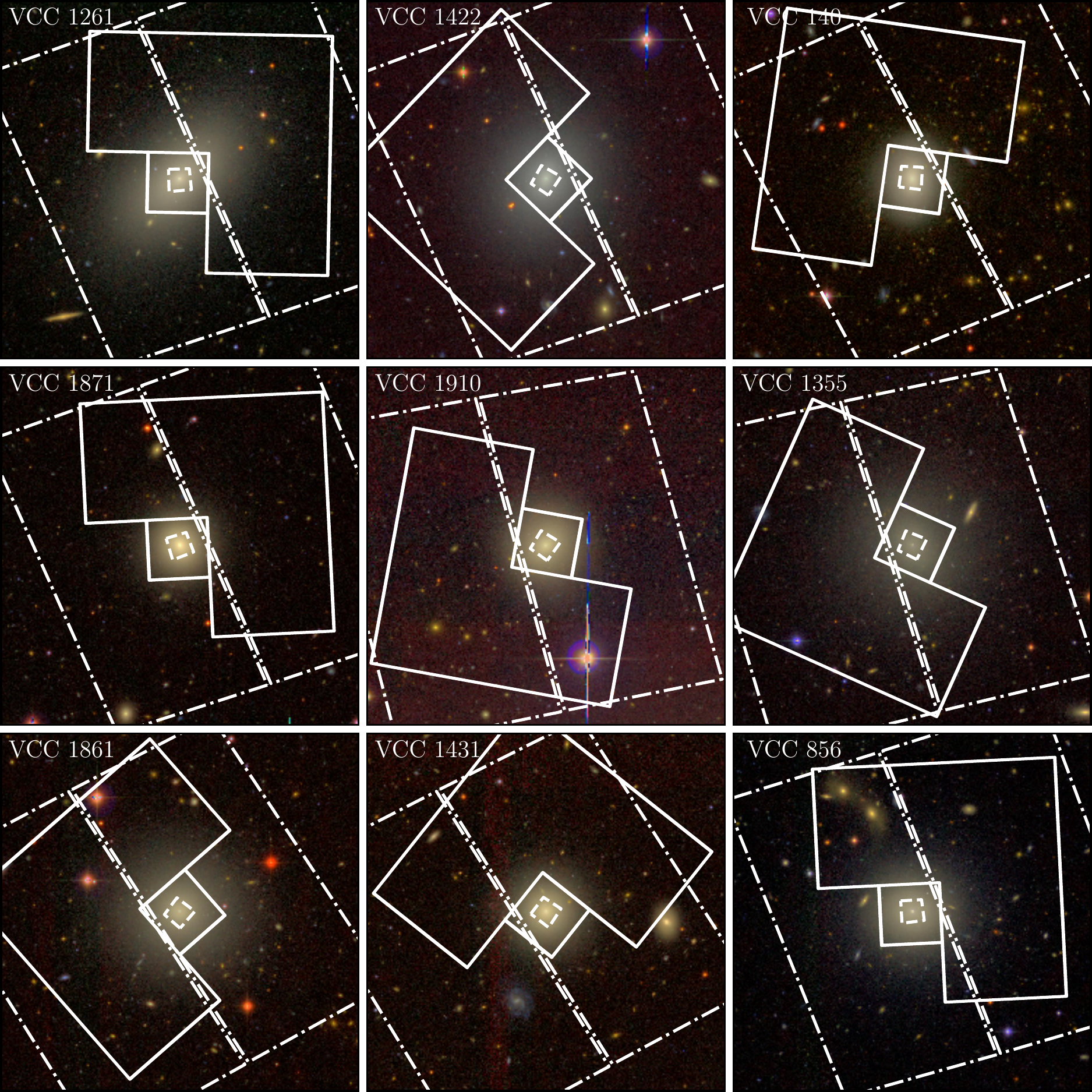}
\caption{\textit{Continued.}}
\end{figure*}

\begin{figure*}[htbp]
\setcounter{figure}{0}
%%\epsscale{1.1}
\plotone{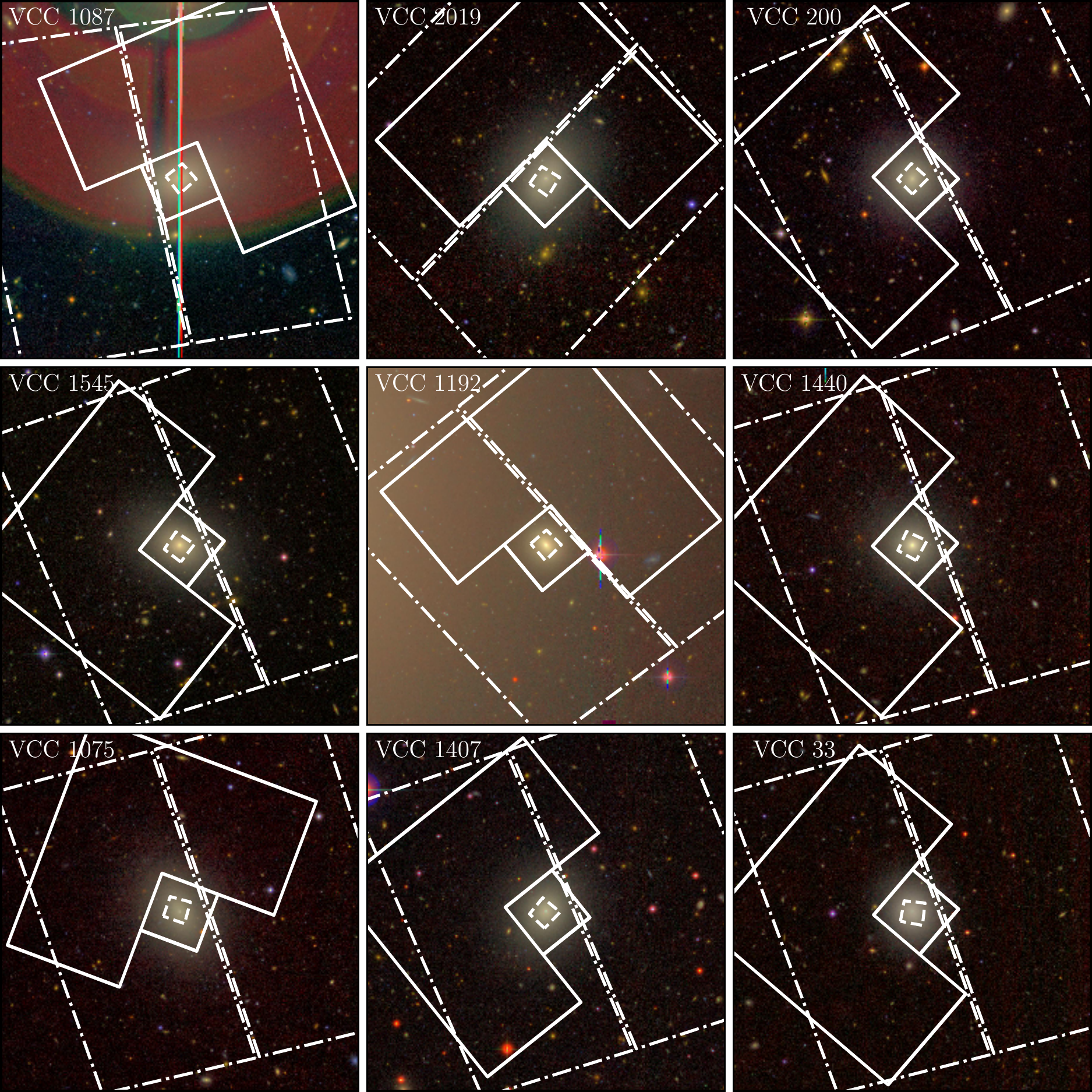}
\caption{\textit{Continued.}}
\end{figure*}

\begin{figure*}[htbp]
\setcounter{figure}{0}
%%\epsscale{1.1}
\plotone{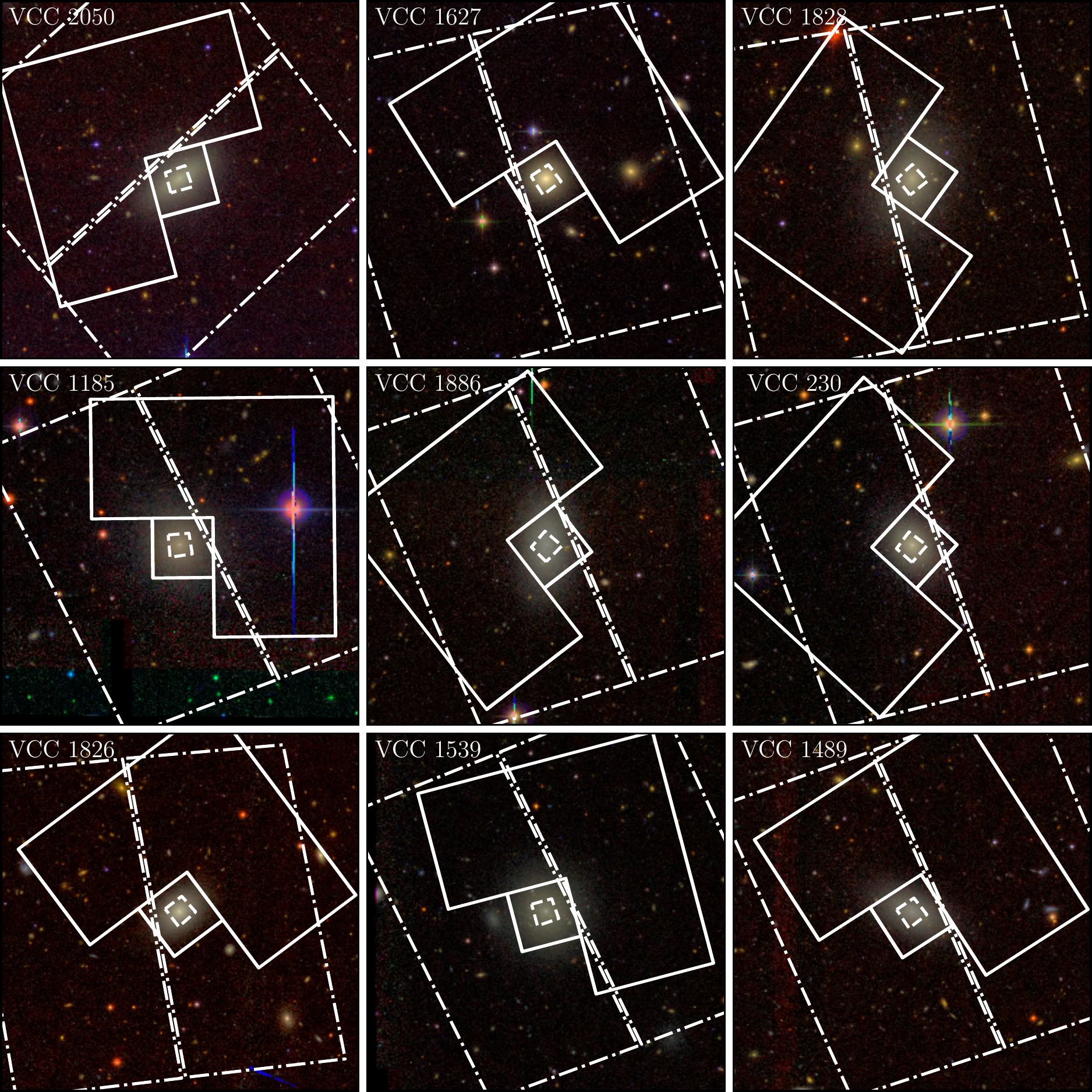}
\caption{\textit{Continued.}}
\end{figure*}

\begin{figure*}[t]%[htbp]
\setcounter{figure}{0}
%%\epsscale{1.1}
\plotone{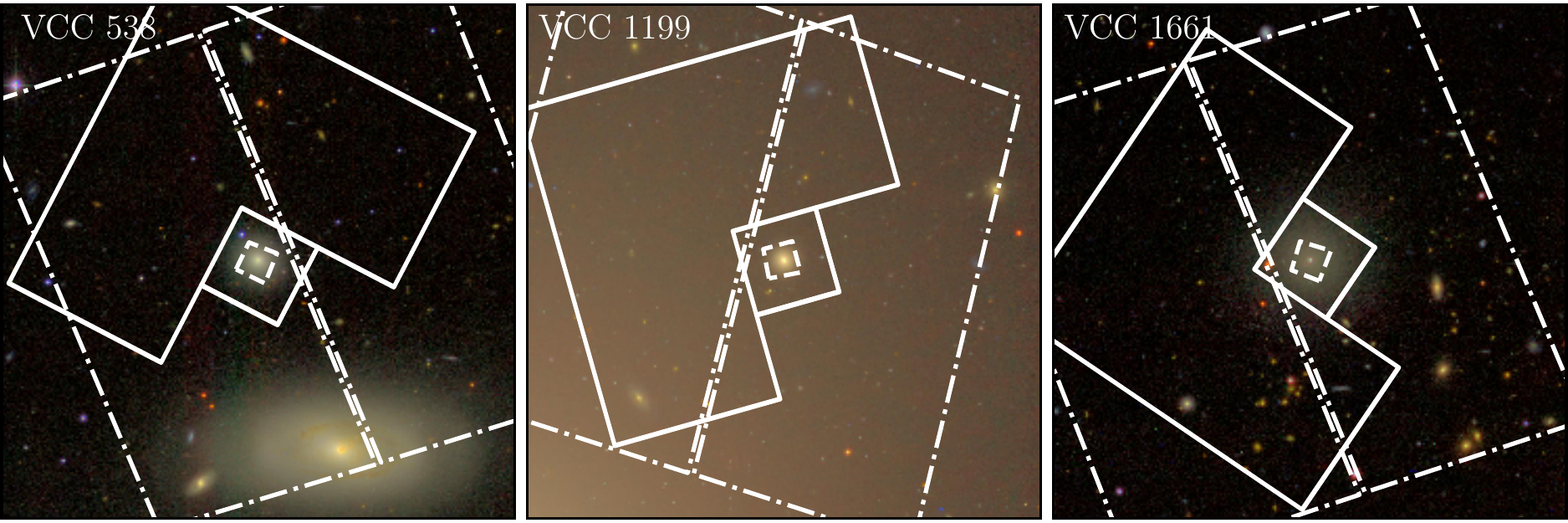}
\caption{\textit{Continued.}}
\end{figure*}

The spectral energy distribution (SED) of a stellar population is dictated by a host of properties, including its initial mass function (IMF), chemical composition, dust content, and detailed star formation history. The method of SED fitting aims to recover these properties by comparing observed SEDs to theoretical spectra. While a detailed knowledge of the full spectrum is necessary for a complete understanding of an object and its evolutionary history, even a rough sampling of the SED with broadband photometry can provide useful constraints on important properties such as stellar mass \citep[e.g.,][]{Taylor:2011aa, Mendel:2014aa}, age, and metallicity \citep[e.g.,][]{Li:2007aa, Salim:2007aa, Crockett:2011aa, Kaviraj:2012aa, Fan:2014aa}. The inclusion of ultraviolet (UV) or infrared (IR) wavelengths are especially useful for improved age and metallicity measurements \citep[e.g.,][]{Anders:2004aa, Kaviraj:2007aa, Georgiev:2012aa, de-Meulenaer:2014aa}, or estimates of the star formation history \citep[e.g.,][]{Yi:2005aa, Kaviraj:2007ab}. No matter what data are used to sample the SED, the precise choice of comparison model --- and some assumptions applied during the SED fitting procedure --- may introduce ambiguities in the derived parameters \citep{Conroy:2010aa, Fan:2012aa, Powalka:2016aa}. Nevertheless, SED fitting using broadband photometry can be a powerful method of characterizing the stellar populations of a stellar system, particularly in situations where spectroscopic measurements are challenging or impractical.

One such application of SED fitting using broadband photometry is the study of compact stellar nuclei --- objects whose origins and properties have been the focus of numerous studies during the last decade. These nuclei, which are sometimes referred to as nuclear star clusters, are found nestled in the cores of galaxies spanning wide ranges in morphology, mass, and size. Unlike supermassive black holes (SMBHs) --- which occupy similar locations at the bottoms of their host galaxy gravitational potential wells --- nuclei can be observed {\it directly}, providing insight into the formation and evolution of galactic cores. 

Early imaging surveys with the Hubble Space Telescope (HST) found nuclei in $\sim$50--60\% of late-type galaxies, with slightly higher nucleation fractions among the later morphologies \citep{Phillips:1996aa,Carollo:1997aa,Carollo:1998aa}. More recent surveys have increased that fraction to 65--80\% \citep{Boker:2002aa,Seth:2006aa,Georgiev:2014aa}. For early-type galaxies, the nucleation fraction is similar, at 70--80\% in the luminosity range $-19.5\lesssim M_B\lesssim-11$ \citep{Cote:2006aa,Turner:2012aa, den-Brok:2014aa}.

As the name implies, compact nuclei are small, dense objects. Typical half-light radii are 2--5 pc \citep{Boker:2004aa, Cote:2004aa, Cote:2006aa} with some as large as tens of parsecs \citep{Geha:2002aa, Georgiev:2014aa}. Estimated masses fall in the range ${\sim}10^{5}-10^{8} M_{\odot}$ \citep{Boker:2004aa, Walcher:2005aa}, and appear to be related to their host galaxy masses, following roughly the same relation that exists for SMBHs \citep{Cote:2006aa, Wehner:2006aa, Rossa:2006aa, Turner:2012aa}. The existence of similar mass relationships involving nuclei and SMBHs implies that these central massive objects (CMOs) may share similar formation processes, with a gradual transition from SMBH- to nucleus-dominated CMOs as galaxy profiles smoothly transition from central light deficits to excesses \citep{Glass:2011aa}. However, recent work suggests that the nucleus mass relation can vary with galaxy morphology, with late-type galaxies having a shallower mass relation than early-types \citep{Georgiev:2016aa} and more concentrated galaxies having brighter --- and presumably more massive --- nuclei \citep{den-Brok:2014aa}. Other studies have found that SMBHs and nuclei follow relations with different slopes \citep{Balcells:2007aa,Scott:2013aa, Leigh:2012aa, Graham:2012aa}, so the exact nature of CMOs remains unclear.

Nuclei and galaxy colors seem to be loosely connected as well. While nuclei display a broad range of colors, they are usually somewhat bluer than their hosts \citep{Lotz:2004aa, Cote:2006aa}, suggesting that their stellar populations are younger than the underlying galaxy or that they may have a steeper IMF \citep{Goudfrooij:2014aa}. Detailed investigations of nuclei ages, however, have yielded mixed results. Some nuclei show evidence of multiple stellar populations \citep{Rossa:2006aa, Walcher:2006aa, Carson:2015aa}, although this can only be determined for resolved objects. Spectroscopic studies have measured ages ranging from 10 Myr to 12 Gyr, although, with a few exceptions, the nuclei ages are usually found to be younger than their host galaxies \citep{Butler:2005aa, Seth:2006aa, Chilingarian:2007aa, Chilingarian:2009aa, Paudel:2011aa, Guerou:2015aa}.

The relationship between nuclei and other compact stellar systems (such as globular clusters and ultra compact dwarf galaxies; GCs and UCDs) is also a matter of  interest. Nuclei are quite similar in size to most GCs, but tend to be brighter by ${\sim}4$ magnitudes \citep{Boker:2004aa, Georgiev:2014aa}. In contrast, UCDs are somewhat larger than nuclei, with half-light radii of 10--100 pc \citep{Drinkwater:2003aa, Mieske:2008aa}, and yet have similar masses ($2\times10^{6} \leq M_{\star} \leq 10^{8} M_{\odot}$). The optical colors of nuclei, GCs and UCDs in the central region of the Virgo cluster are remarkably similar \citep{Roediger:2017aa}. A number of groups have proposed that GCs could be the progenitors of nuclei (see below), and at least some UCDs are thought to be the stripped nuclei of disrupted nucleated dwarf galaxies \citep{Goerdt:2008aa, Pfeffer:2013aa}.

How nuclei form is still not well understood. Generally speaking, there are two broad scenarios for their formation: star cluster infall or \textit{in situ} formation. In the cluster infall scenario, GCs spiral into the galaxy's core via dynamical friction and then merge to form a massive central star cluster \citep[e.g.,][]{Tremaine:1975aa, Oh:2000aa, Lotz:2001aa, Capuzzo-Dolcetta:2008aa, Antonini:2012aa, Gnedin:2014aa}. The alternative scenario is that the nuclei develop from gas funneled into the galactic center, possibly as the result of a merger \citep[e.g.,][]{Mihos:1994aa, Milosavljevic:2004aa, Schinnerer:2008aa, Bekki:2015aa}. In this picture, stellar feedback can regulate the growth of the nucleus, potentially producing multiple stellar populations and leading to the $M-\sigma$ relation, involving the galaxy stellar mass $M$ and velocity dispersion $\sigma$, via the same mechanisms proposed for the growth of SMBHs \citep{McLaughlin:2006aa, Bourne:2016aa}. Recently, \citet{Guillard:2016aa} proposed a {\it wet migration} model in which massive clusters form outside the galaxy center, but retain gas reservoirs to continue forming stars as they fall to the center, merging with other clusters in the process. In reality, nucleus formation is likely more complex that idealized models suggest, and some studies have indicated that nuclei form probably through a mixture of  scenarios \citep{den-Brok:2014aa, Antonini:2015aa, Cole:2016aa}. 

While refinements to the simulations are always welcome, a robust test of any formation model will be impossible until we have a large database of compact stellar nuclei with accurately measured parameters based on high-quality, homogenous data. Unfortunately, such studies are observationally challenging. Given their compact sizes, nuclei are only marginally resolved, even with HST, in all but the nearest galaxies. Bright galaxies present an additional challenge, as their nuclei must be separated from the high underlying surface brightness of the host. In addition, large sample sizes are required for a meaningful statistical analysis of nuclei properties. While it is possible to acquire spectra with sufficient signal for age and metallicity measurements, most spectroscopic studies of nuclei have concentrated on small samples of nearby galaxies \citep[e.g.,][]{Seth:2006aa} or limited surveys of more distant systems \citep{Paudel:2011aa}. Multi-band imaging is thus an attractive alternative since it avoids the long observation times needed for spectroscopy, making it possible to efficiently characterize statically meaningful samples.

At a distance of 16.5 Mpc \citep{Mei:2007aa,Blakeslee:2009aa}, the Virgo Cluster is a convenient target for studying nuclei and their parent galaxies. It is near enough for nuclei, with typical sizes of ${\sim}0\farcs05$ \citep[4\,pc][]{Cote:2006aa}, to be marginally resolved by HST. The cluster contains a vast collection of nucleated galaxies that is especially useful for studying nucleation in early-type galaxies. 

Three past or ongoing surveys of Virgo Cluster galaxies can provide both high-resolution, space-based imaging and deep, ground-based imaging in broadband filters that span the UV to near-IR wavelength region. The first of these studies used the Advanced Camera for Surveys (ACS) instrument on HST to carry out the {\it ACS Virgo Cluster Survey} \citep[ACSVCS,][]{Cote:2004aa,Cote:2006aa,Ferrarese:2006aa,Ferrarese:2006ab}. A follow-up HST program, {\it Virgo Redux}, expanded the ACSVCS dataset by adding UV and IR imaging. The latest, and most extensive, program is the {\it Next Generation Virgo Cluster Survey} \citep[NGVS,][]{Ferrarese:2012aa} which used the MegaCam instrument on the 3.6m Canada France Hawaii Telescope (CFHT) to acquire deep, wide-field $u^*giz$ imaging over 104 $\mbox{deg}^{2}$ of the Virgo Cluster. Using the NGVS, it is possible to identify and study nuclei belonging to galaxies of unprecedented faintness (S\'anchez-Janssen et al. 2017, in preparation). The NGVS also makes it possible to study the structural and photometric properties of not just nuclei, but also GCs and UCDs \citep{Durrell:2014aa, Liu:2015aa, Zhang:2015aa}. The NGVS also includes deep $r$-band and infrared ($K_s$) imaging for a subset of the NGVS fields \citep{Munoz:2014aa}.

In this study, we combine all available data from the ACSVCS, Virgo Redux, and NGVS (including NGVS-IR) for 39 nucleated galaxies observed in the various surveys. The combined dataset consists of observations in up to 10 filters spanning the UV, optical, and near-IR regions. With high-resolution imaging from HST, and deep, wide-field imaging from CFHT, we are able to estimate masses, ages and metallicities for the nuclei and their host galaxies in a systematic and homogeneous way. Additionally, for a subset of our targets, we use high-quality optical spectra acquired with the 10m Keck and 8m Gemini telescopes to validate our photometrically derived parameters.

This article is organized as follows. \S\ref{sec:data} summarizes our sample and observations, while \S\ref{sec:photometry} describes the isophotal and 2D decomposition methods for measuring structural and photometric parameters. In \S\ref{sec:spectrometry}, we describe the reduction and analysis of various ground-based spectroscopic observations available for a subset of the nuclei. In \S\ref{sec:results}, we describe our SED-fitting process and present results on nuclei properties measured from photometry and spectroscopy. We summarize our findings in \S6 and conclude with some directions for future work.

\section{Data and Observations} \label{sec:data}

%% In a manner similar to \objectname authors can provide links to dataset
%% hosted at participating data centers via the \dataset{} command.  The
%% second curly bracket argument is printed in the text while the first
%% parentheses argument serves as the valid data set identifier.  Large
%% lists of data set are best provided in a table (see Table 3 for an example).
%% Valid data set identifiers should be obtained from the data center that
%% is currently hosting the data.

\subsection{Sample Selection and Properties} \label{subsec:sample} 

\begin{figure*}[htbp]
   \centering
   \epsscale{1.1}
   \plotone{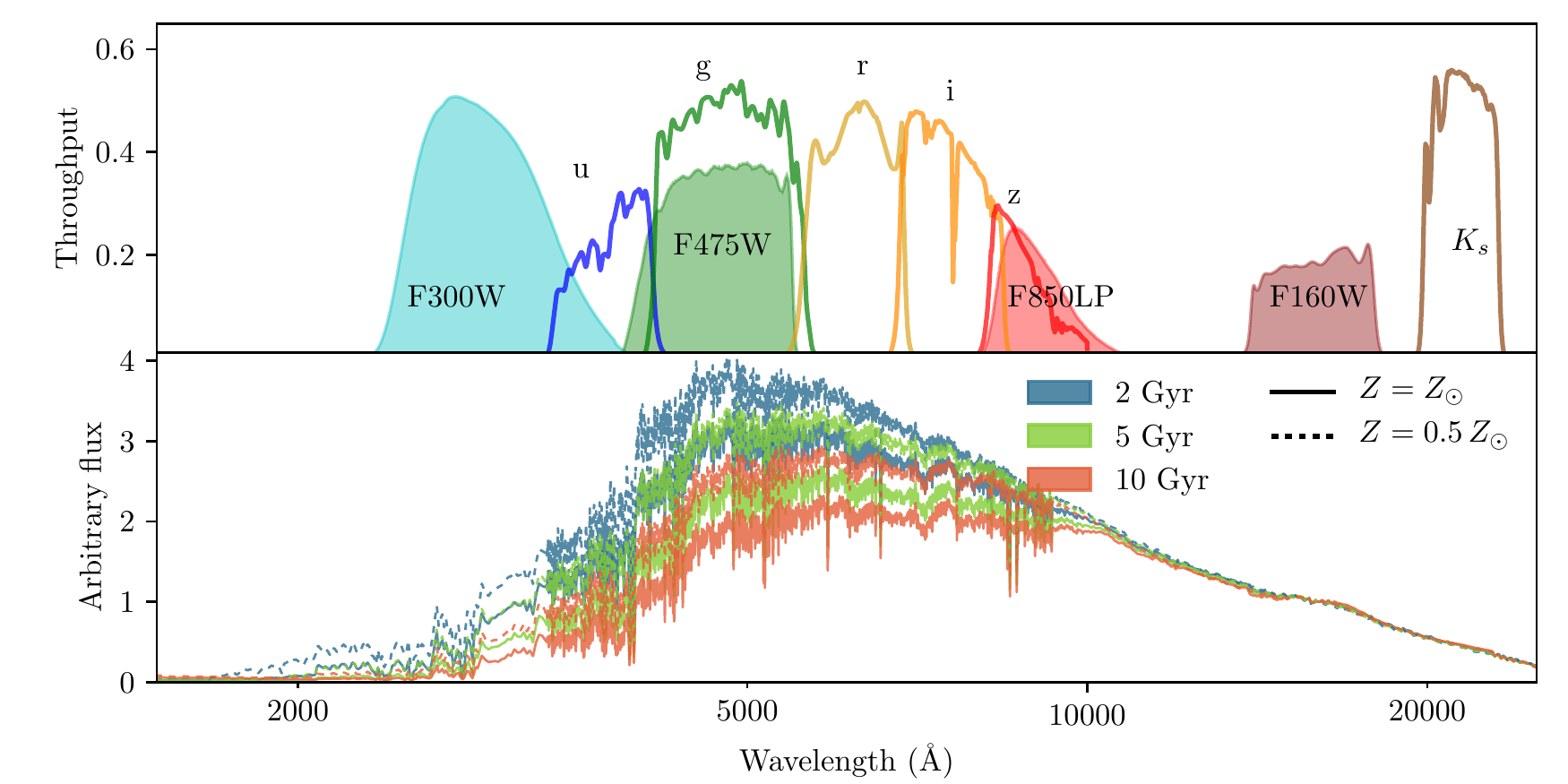} 
   \caption{\textit{(Top panel).} Passbands for the different filters used in this study. Filled curves show the HST filters while open curves show the CFHT filters. Note that the $K_s$ filter is only available for the six galaxies that fall inside the 2 $\deg \times\,2 \deg$ region around M87. \textit{(Bottom panel).} Model spectra for selected SSPs using the BC03 models with a Chabrier IMF. Three different ages are shown: 2, 5, and 10 Gyr (as the blue, green, and red lines, respectively). Solid lines denote SSPs with solar metallicity, while dotted lines correspond to populations with half solar metallicity. The spectra have been normalized at 1.6 $\mu$m in the F160W filter.}
   \label{fig:filters}
\end{figure*}

Our 39 program galaxies were selected from three imaging surveys of the Virgo cluster that together span the UV, optical and IR regions (i.e., wavelength in the range 0.3--2.2~$\mu$m). Figure~\ref{fig:color} shows $giz$ color images created from NGVS data with the different HST instrument footprints overlaid. The wide spectral coverage of the data enables more precise determination of stellar population properties, particularly ages and metallicities, which have a well-known degeneracy for old or intermediate-age populations, such as those expected for many compact stellar nuclei. Figure~\ref{fig:filters} demonstrates the sensitivity of our filter set to differences in theoretical spectra for simple stellar populations (SSPs) of various ages and metallicities. The observational details of each program are explained in the following subsections, with some general information summarized in Table~\ref{tab:obs}.

\begin{deluxetable*}{llclccc}
\tablecaption{Summary of Imaging \label{tab:obs}}
\tablehead{
\colhead{Telescope} & \colhead{Instrument} & \colhead{Field of View} & \colhead{Filters} & \colhead{Scale} & \colhead{FWHM} & \colhead{N$_{\textrm{gal}}$} \\
\colhead{} & \colhead{} & \colhead{} & \colhead{} & \colhead{(arcsec~px$^{-1}$)} & \colhead{(\arcsec)} & \colhead{}
}
\colnumbers
\startdata
HST & ACS-WFC & $202\arcsec \times 202\arcsec $ & F475W, F850LP & 0.05 & 0.1 & 39 \\
HST & WFPC2-PC & $35\arcsec \times 35\arcsec $ & F255W, F300W & 0.05 & 0.08 & 37 \\
HST & NICMOS-NIC1 & $11\arcsec \times 11\arcsec $ & F160W & 0.03 & 0.095 & 38 \\
CFHT & MegaCam & $0\fdg96 \times 0\fdg94$ & $u^*griz$ & 0.187 & $\leq1$ & 39 \\
CFHT & WIRCam & $21\arcmin \times 21\arcmin $ & $K_s$ & 0.186 & $\leq$~0.7 & 6 \\
\enddata
\tablecomments{Summary of telescopes and instruments used to collect the images analyzed in this paper. All MegaCam images have seeing better than 1\arcsec~ but FWHM varies with filter; the median seeing ranges from 0\farcs54 in $i$ to 0\farcs88 in $u^*$. Note that two galaxies (VCC~1185 and VCC~1627) are missing WFPC2 observations due to a loss of guiding during the observation; similarly, VCC~1627 is missing NICMOS data due to a guiding failure. Only six objects have $K_s$-band imaging because WIRCam observations are available for only the central 4~$\deg^2$ of the Virgo cluster \citep{Munoz:2014aa}.}
\end{deluxetable*}

Here we focus on our target selection, which is largely determined by the ACSVCS sample and classifications. The ACSVCS imaged 100 early-type galaxies in the Virgo Cluster in the F475W (${\sim}g$) and F850LP (${\sim}z$) filters \citep{Cote:2004aa}, covering a range of early-type morphologies (E, S0, dE, dE,N, dS0, dS0,N) with magnitudes $9.3\lesssim B_T \lesssim 15.7$. The survey is 44\% complete down to its limiting magnitude of $M_B = -15.2$.

\begin{figure}[htbp]
   \centering
   \epsscale{1.1}
   \plotone{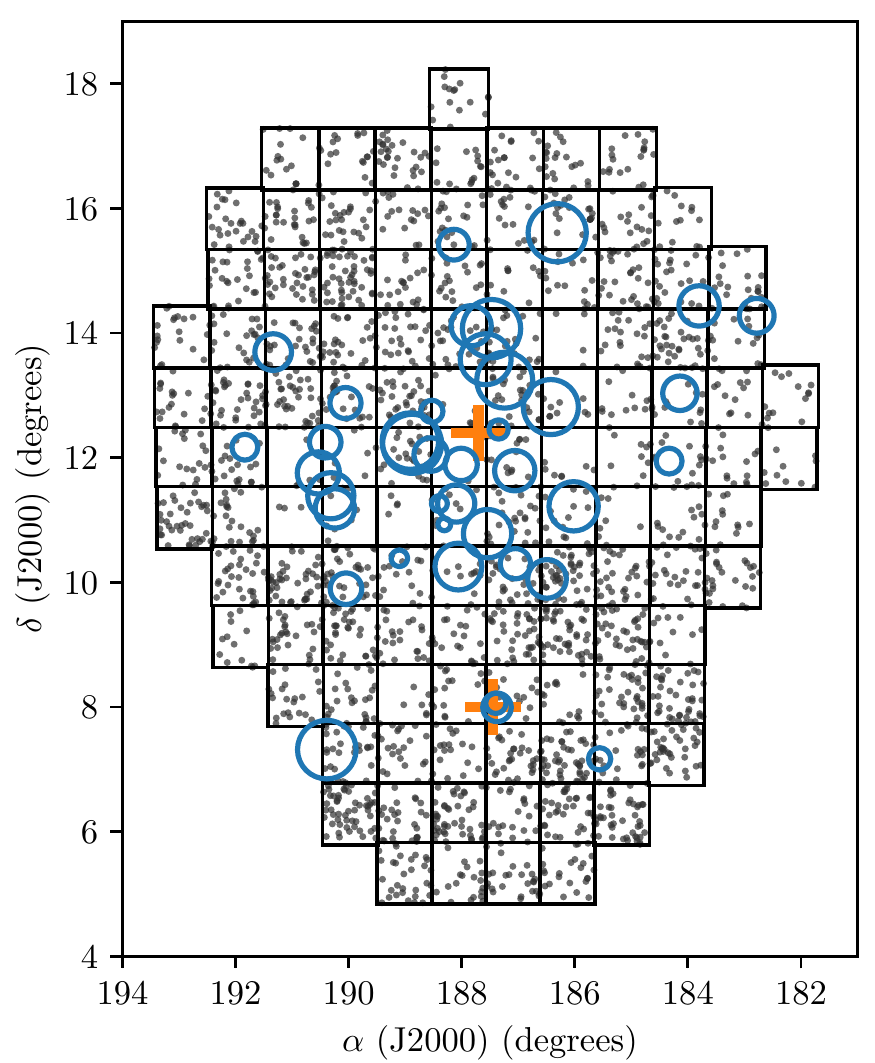} 
   \caption{Distribution of the 39 galaxies selected for this analysis overlaid on the NGVS fields. Open blue circles indicate each sample galaxy. The size of the circles corresponds to galaxy brightness. M87 (VCC~1316) and M49 (VCC~1226) are labeled with orange crosses. Gray points show NGVS galaxies brighter than $M_g \simeq -14.5$.}
   \label{fig:footprint}
\end{figure}

\begin{figure}[htbp]
   \centering
   \epsscale{1.1}
   \plotone{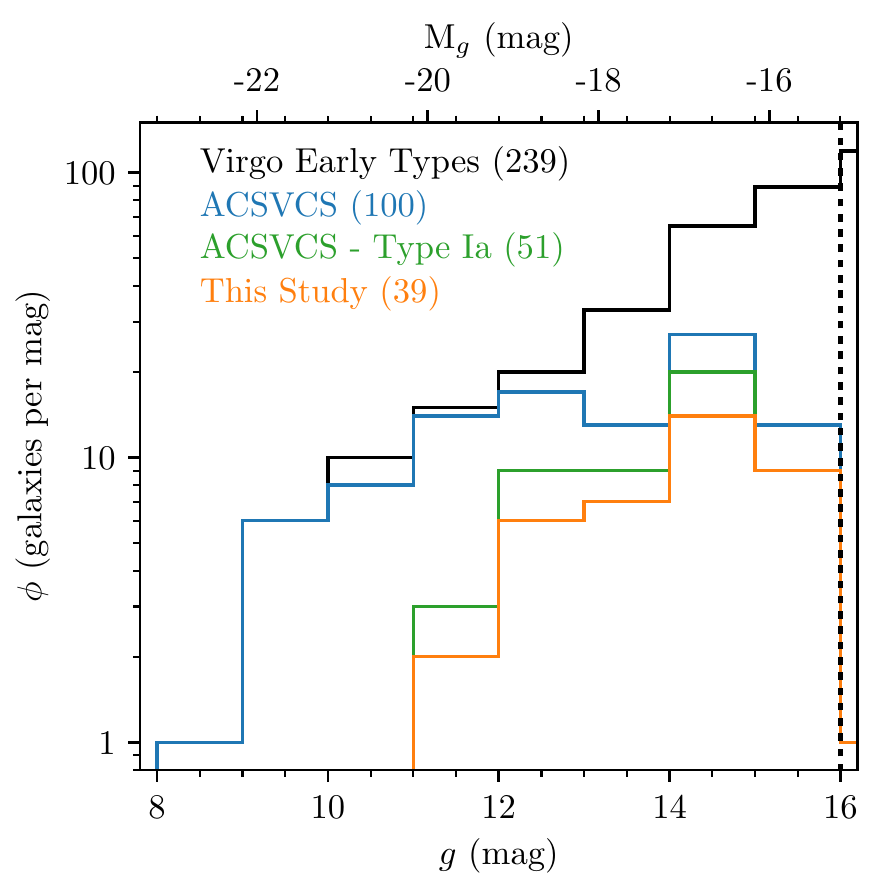} 
   \caption{Magnitude distribution for the full ACSVCS sample, 51 nucleated galaxies ({\tt Type~Ia}) and 39 {\tt Type~Ia} galaxies analyzed in this work. For comparison, we also show the complete sample of Virgo early-type galaxies from \citet{Janz:2008aa, Janz:2009aa}.}
   \label{fig:maghist}
\end{figure}

Our sample originates from the 51 galaxies in ACSVCS that were classified as clearly nucleated ({\tt Type~Ia}) in \citet{Cote:2006aa}, meaning that a King model profile \citep{King:1966aa} was successfully fitted to the galaxy's nuclear component. While other ACSVCS galaxies were classified as likely, or possibly, nucleated, we opted to focus only on the unambiguously nucleated galaxies, as these nuclei can be most easily modeled and separated from their host galaxies. The sample was further reduced by restricting ourselves to galaxies within the $\sim$ 100 deg$^2$ NGVS survey footprint --- a total of 39 galaxies. Some basic information for these galaxies, including coordinates, velocities from the NASA/IPAC Extragalactic Database (NED), and morphologies from \citet[hereafter BST85]{Binggeli:1985aa}, NED and \citet{Kim:2014aa} is given in Table~\ref{tab:coords}. The more recent numerical classifications from \citet{Kim:2014aa}, which are based on SDSS imaging, confirm that these are predominantly early-type systems: 21 are dwarf ellipticals (classifications in the form 4XX), while another eight are considered ellipticals (1XX). The remaining nine galaxies classified by \citet{Kim:2014aa} are disk galaxies (2XX), or lenticulars in the other classifications listed here. The sample galaxies are distributed throughout the cluster, as shown in Figure~\ref{fig:footprint}. Figure~\ref{fig:maghist} shows the magnitude distribution of the galaxies selected for this analysis compared to the full set of \texttt{Type 1a} galaxies, the rest of the ACSVCS, and the general population of early-type galaxies in Virgo. The program galaxies span the full magnitude range of nucleated galaxies detected in the ACSVCS and are well distributed across this range.

\begin{deluxetable*}{ccccccclll}
\tabletypesize{\footnotesize}
\tablecaption{Basic Data for Program Galaxies \label{tab:coords}}
\tablehead{
\colhead{VCC} & \colhead{Other} & \colhead{$\alpha$(2000)} & \colhead{$\delta$(2000)} & \colhead{$B_T$} & \colhead{$E(B-V)$} & \colhead{$V_r$} & \colhead{BST85} & \colhead{NED} & \colhead{EVCC} \\
\colhead{} & \colhead{} & \colhead{(h:m:s)} & \colhead{($^\circ$:$\arcmin$:$\arcsec$)} & \colhead{(mag)} & \colhead{(mag)} & \colhead{(km\,s$^{-1}$)} & \colhead{} & \colhead{} & \colhead{}
}
\colnumbers
\startdata
~~33 & IC3032 & 12:11:07.8 & +14:16:29.3 & 14.67 & 0.037 & 1186 & d:E2,N: & E? & 411\\
~140 & IC3065 & 12:15:12.6 & +14:25:58.3 & 14.30 & 0.037 & 993 & SO1/2(4) & S0? & 200\\
~200 & \ldots & 12:16:33.7 & +13:01:53.7 & 14.69 & 0.030 & 16 & dE2,N & dE2,N & 411\\
~230 & IC3101 & 12:17:19.7 & +11:56:36.5 & 15.20 & 0.028 & 1429 & dE4:,N: & dE4:,N: & 401\\
~538 & NGC4309A & 12:22:14.7 & +07:10:01.7 & 15.40 & 0.020 & 750 & E0 & E0 & 100\\
~698 & NGC4352 & 12:24:05.0 & +11:13:05.1 & 13.60 & 0.026 & 2070 & S01(8) & SA0: sp & 200\\
~784 & NGC4379 & 12:25:14.7 & +15:36:26.7 & 12.67 & 0.024 & 1074 & S01(2) & S0- pec: & 200\\
~828 & NGC4387 & 12:25:41.7 & +12:48:37.9 & 12.84 & 0.033 & 565 & E5 & E5 & 100\\
~856 & IC3328 & 12:25:57.9 & +10:03:13.5 & 14.25 & 0.024 & 1025 & dE1,N & dE,N & 411\\
1075 & IC3383 & 12:28:12.3 & +10:17:51.5 & 15.08 & 0.027 & 1844 & dE4,N & dE4,N & 401\\
1087 & IC3381 & 12:28:14.9 & +11:47:23.3 & 14.31 & 0.027 & 675 & dE3,N & dE,N & 401\\
1146 & NGC4458 & 12:28:57.6 & +13:14:30.9 & 12.93 & 0.023 & 677 & E0-1 & E0-1 & 100\\
1185 & \ldots & 12:29:23.5 & +12:27:02.9 & 15.68 & 0.023 & 500 & dE1,N & dE1 & 401\\
1192 & NGC4467 & 12:29:30.3 & +07:59:34.3 & 15.04 & 0.023 & 1423 & E3\tablenotemark{a} & E2 & 200\\
1199 & \ldots & 12:29:35.0 & +08:03:28.8 & 15.50 & 0.022 & 1401 & E2\tablenotemark{a} & E2 & 100\\
1242 & NGC4474 & 12:29:53.6 & +14:04:06.9 & 12.60 & 0.042 & 1611 & S01(8) & S0 pec: & 200\\
1261 & NGC4482 & 12:30:10.3 & +10:46:46.1 & 13.56 & 0.029 & 1871 & d:E5,N & dE,N & 400\\
1283 & NGC4479 & 12:30:18.4 & +13:34:39.4 & 13.45 & 0.029 & 876 & SB02(2) & SB(s)0!0!:? & 210\\
1355 & IC3442 & 12:31:20.2 & +14:06:54.7 & 14.31 & 0.034 & 6210 & dE2,N & E0: & \ldots\\
1407 & IC3461 & 12:32:02.7 & +11:53:24.3 & 15.49 & 0.032 & 1019 & dE2,N & dE,N & 401\\
1422 & IC3468 & 12:32:14.2 & +10:15:05.2 & 13.64 & 0.031 & 1288 & E1,N: & E1,N: & 210\\
1431 & IC3470 & 12:32:23.4 & +11:15:46.7 & 14.51 & 0.051 & 1505 & E? & E? & 401\\
1440 & IC798 & 12:32:33.4 & +15:24:55.5 & 15.20 & 0.028 & 382 & E0\tablenotemark{a} & E0 & 100\\
1489 & IC3490 & 12:33:13.9 & +10:55:42.5 & 15.89 & 0.034 & 80 & dE5,N? & E? & 401\\
1539 & \ldots & 12:34:06.7 & +12:44:29.7 & 15.68 & 0.032 & 1491 & dE0,N & dE0,N & 401\\
1545 & IC3509 & 12:34:11.5 & +12:02:56.2 & 14.96 & 0.042 & 2000 & E4\tablenotemark{a} & E4 & 401\\
1619 & NGC4550 & 12:35:30.6 & +12:13:15.0 & 12.50 & 0.040 & 459 & E7/S01(7) & SB0!0!:sp LINER & 200\\
1627 & \ldots & 12:35:37.3 & +12:22:55.3 & 15.16 & 0.039 & 236 & E0\tablenotemark{a} & E0 & 100\\
1630 & NGC4551 & 12:35:38.0 & +12:15:50.4 & 12.91 & 0.039 & 1176 & E2 & E: & 100\\
1661 & \ldots & 12:36:24.8 & +10:23:04.8 & 15.97 & 0.020 & 1457 & dE0,N & dE0,N & 401\\
1826 & IC3633 & 12:40:11.3 & +09:53:46.0 & 15.70 & 0.017 & 2033 & dE2,N & dE2,N & 401\\
1828 & IC3635 & 12:40:13.4 & +12:52:29.1 & 15.33 & 0.037 & 1569 & dE2,N & dE,N & 401\\
1861 & IC3652 & 12:40:58.6 & +11:11:04.2 & 14.37 & 0.029 & 629 & dE0,N & E & 401\\
1871 & IC3653 & 12:41:15.7 & +11:23:14.0 & 13.86 & 0.030 & 588 & E3 & E3 & 100\\
1883 & NGC4612 & 12:41:32.8 & +07:18:53.5 & 12.57 & 0.025 & 1775 & S01(6) & (R)SAB0!0! & 200\\
1886 & \ldots & 12:41:39.4 & +12:14:50.6 & 15.49 & 0.033 & 914 & dE5,N & dE5,N & 401\\
1910 & IC809 & 12:42:08.7 & +11:45:15.3 & 14.17 & 0.031 & 206 & dE1,N & E & 401\\
2019 & IC3735 & 12:45:20.4 & +13:41:33.6 & 14.55 & 0.022 & 1895 & dE4,N & E? & 411\\
2050 & IC3779 & 12:47:20.6 & +12:09:59.1 & 15.20 & 0.023 & 1156 & dE5:,N & dE5:,N & 400\\
\enddata
\tablecomments{Key to columns: (1) VCC identification number, (2) Alternate names in the NGC, IC or UGC catalogs, (3) right ascension, (4) declination, (5) total $B$ magnitude from BST85, (6) extinction from \citet{Schlafly:2011aa}, (7) recessional velocity from NED, (8) morphological classification from BST85, (9) morphological classification from NED, and (10) morphological classification from \citet{Kim:2014aa}.}
\tablenotetext{a}{Compact, low-luminosity E (M32-type) galaxy from Table XIII of BST85.}
\end{deluxetable*}

\subsection{HST/ACS Imaging} \label{subsec:acs}
\begin{figure*}[htbp]
\epsscale{1.1}
\plotone{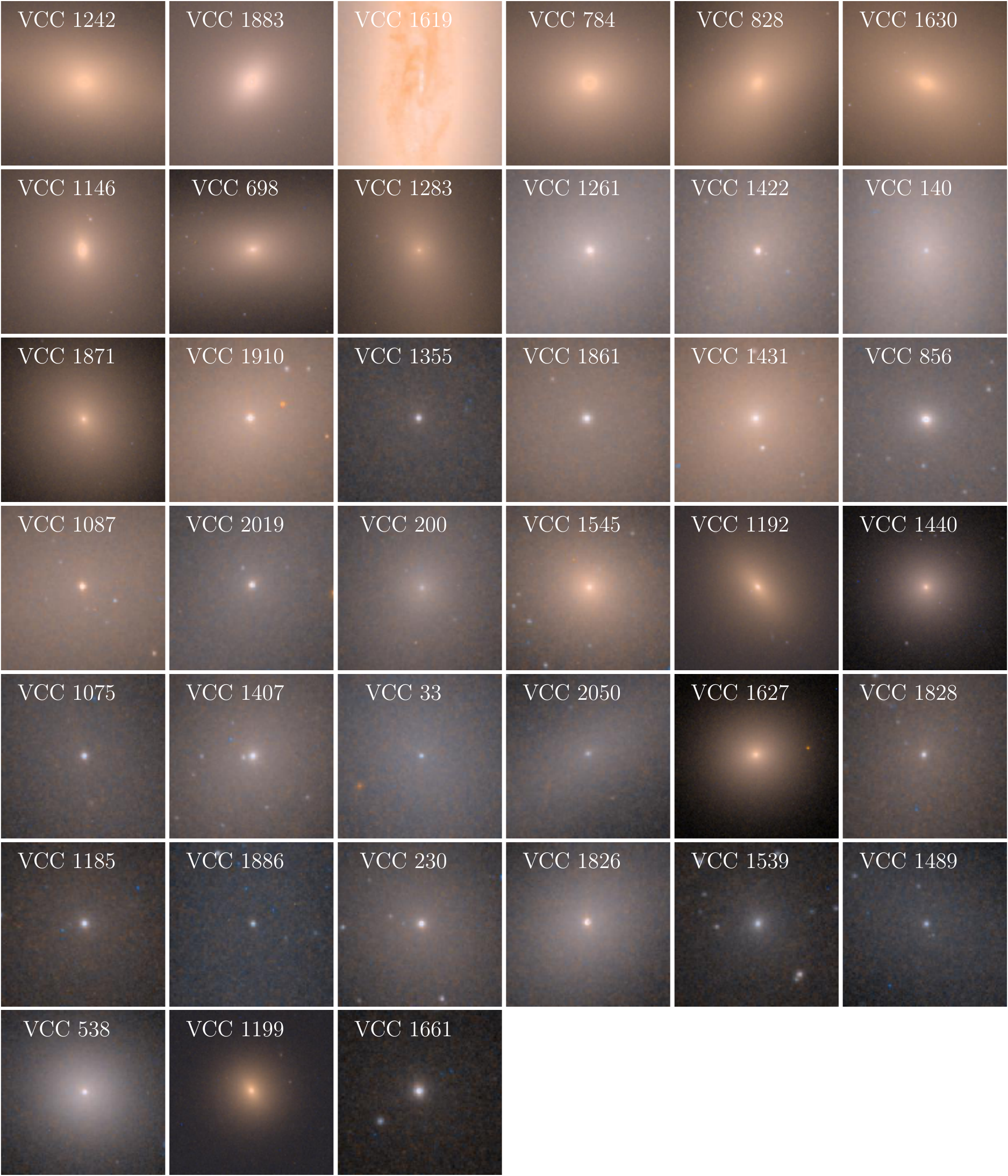}
\caption{HST color images focusing on the central 20\arcsec~$\times$~20\arcsec{} ($1.6 \times 1.6$\,kpc) region of each program galaxy, sorted by decreasing F475W luminosity. In all images, north is up and east is to the left.}
\label{fig:subcolor}
\end{figure*}

The ACSVCS carried out imaging with the ACS instrument \citep{Ford:1998aa} in its Wide Field Channel (WFC) mode (Program ID = 9401). ACS/WFC provides high resolution (FWHM $\approx$ 0\farcs1) imaging across a 202\arcsec{} $\times$ 202\arcsec{} field of view with a pixel scale of 0\farcs049\,~px$^{-1}$, although our final data products have been drizzled to a scale of 0\farcs05\,~px$^{-1}$. Each galaxy was observed for a single orbit with two exposures per filter, plus an addition 90\,s exposure in F850LP to correct any central saturation. Total exposure times were 750\,s in F475W and 1210\,s in F850LP. The center of each object was initially positioned on the WFC1 detector, one of the WFC's two 2048 $\times$ 4096 detectors, roughly 15--20\arcsec{} from the chip gap, depending on galaxy brightness.

After correcting for any small offsets ($\lesssim 0.2$ px) between exposures, the images were drizzled and cosmic ray corrected using \textit{multidrizzle} in \texttt{PyRAF}. Sky subtraction was omitted in the drizzling process because many target galaxies in the full ACSVCS sample dominate the field of view. The drizzling process also applies a kernel to the images when distributing flux onto the final science image. The ACSVCS reduction created science images with both the ``Gaussian" and ``Lanczos3" kernels. For this work, we use images created with the ``Gaussian" kernel, which allows for more effective bad pixel repair and therefore better estimates of the light profile in the central galaxy regions where the nucleus dominates. Point spread functions (PSFs) were generated using DAOPHOT II \citep{Stetson:1987aa, Stetson:1993aa} and archival observations of the globular cluster 47 Tucanae, and were allowed to have second order variations across the field. For each galaxy we retrieved a PSF at the nucleus' position on the chip. Additional details of the observational techniques and data reduction are available in \citet{Jordan:2004aa}.

With its excellent resolution, high signal-to-noise ratio (SNR) and comparatively wide field of view, the ACSVCS data are the clear choice for reference images in the 2D decompositions of our program galaxies. Figure~\ref{fig:subcolor} shows 20\arcsec{} $\times$ 20\arcsec{} ($1.6 \times 1.6$\,kpc) cutouts of the nuclear regions for the dataset. The nuclei are prominent and resolved in most of the galaxies, which aids in modeling and separating the nucleus and galaxy components. We therefore use the ACS F475W image to measure one set of structural parameters that are then applied to the full dataset. This procedure is described fully in \S\ref{subsec:galfit}.

\subsection{CFHT Imaging: MegaCam and WIRCam} \label{subsec:ngvs}

Full details on the NGVS observing strategy and data reduction procedures can be found in \citet{Ferrarese:2012aa} and \citet{Munoz:2014aa}. Here, we briefly explain the salient details of the observations.

NGVS was allocated ${\sim}$900 hours between 2008 and 2013 with the MegaCam \citep{Boulade:2003aa} instrument on CFHT. The survey was designed to cover 104 $\deg^2$ of the Virgo Cluster in the $u^*$, $g$, $r$, $i$, and $z$ bands --- an area that fully covers the region within the virial radii of the Virgo A and B subclusters (which are centered on the galaxies M87 and M49, respectively; see Figure~\ref{fig:footprint}). Unfortunately, bad weather and dome shutter problems made it impossible to complete the $r$-band imaging; therefore, full-depth $r$-band exposures are available for only ${\approx}~9\deg^2$. Complete coverage of the survey region is available in the $u^*giz$ bands.

Each MegaCam exposure covers $0\fdg96 \times 0\fdg94$ on the sky using a mosaic of 36 CCDs arranged in a $4 \times 9$ grid. With a pixel scale of 0\farcs187\,px$^{-1}$ and typical seeing ${\sim}$0\farcs7, the PSF is well sampled. The data are of good quality, with a median seeing of 0\farcs88, 0\farcs80, 0\farcs54, and 0\farcs75 in the $u^*$, $g$, $i$ and $z$ bands, respectively. The long exposure data used in this work have exposure times between 2055\,s in $i$ and 6402\,s in $u^*$, and reach limiting surface brightnesses of 29.3, 29.0, 27.4, and 26.0 AB mag\,arcsec$^{-2}$ in the $u^*$, $g$, $i$ and $z$ bands, respectively.

All NGVS data were reduced using the \textit{Elixir} pipeline which carries out bias subtraction, flat fielding, bad pixel masking, and applies a fringing correction to the $i$- and $z$-band images. A number of stacked science images were then produced using the \textit{MegaPipe} pipeline \citep{Gwyn:2008aa}. In this pipeline, all frames are matched to the Sloan Digital Sky Survey (SDSS) DR7 astrometric and photometric catalogs to produce astrometric corrections and photometric zeropoints. For this paper, we use the ``global background subtraction" stacks. In these stacks, a background map, estimated from median-combined archival MegaCam imaging processed with \textit{Elixir}, is scaled to each frame and then subtracted. While these stacks tend to have higher sky residuals (a few percent of the background level) compared to those created using \textit{Elixir-LSB}, we nevertheless use them for this analysis because they are available for all filters --- \textit{Elixir-LSB} requires the input images to be acquired in a specific dither pattern, which was not possible for the $r$ band. Despite the sky residuals, our chosen stacks have superior photometric accuracy compared to the other stacking techniques. PSFs were created using DAOPHOT and stars in each frame detected by both DAOPHOT and SExtractor. As with the ACS PSFs, second order variations were permitted, and a PSF in each filter was generated for the position of the nucleus.

The NGVS includes $K_s$-band imaging from the Wide-Field InfraRed Camera \citep[WIRCam;][]{Puget:2004aa} over the $2^\circ \times 2^\circ$ region centered on M87 \citep{Munoz:2014aa}. WIRCam has a $21\arcmin \times 21\arcmin$ field of view covered by four detectors with a pixel scale of 0\farcs3\,px$^{-1}$. A total of 36 pointings were made between December 2009 and July 2010. Each pointing was built from a series of 25\,s exposures observed in specific dither patterns designed to cover the 45\arcsec{} chip gap between detectors, ensure that each pixel covers a different sky region for each exposure, and enable precise sky subtraction. Total exposure times were 2700\,s per pointing. The raw images were processed for dark subtraction, nonlinearity correction, flat fielding, and bad pixel masking using the {\tt 'I'iwi} 2.0 pipeline\footnote{\url{http://www.cfht.hawaii.edu/Instruments/Imaging/WIRCam/IiwiVersion2Doc.html}} . After removing cosmic rays and satellite trails, sky subtraction was performed in two steps. First a median sky for each science frame was calculated using designated sky pointings, and then subtracted from the target frames. The sky-subtracted frames were then stacked and all sources identified in the stacked image were masked in each sky frame. New median skies were created with the masked sky frames and subtracted from the stacked target images. Additional corrections for variations in the amplifiers of each detector and large-scale sky fluctuations were also applpixel scaleied. Astrometric and photometric calibrations were performed by comparing to 2MASS, resulting in an astrometric accuracy better than 0\farcs02 and zeropoint uncertainty lower than 0.02 mag.

Images with seeing better than 0\farcs7 were selected to create four stacked frames that mirror the NGVS MegaCam positions and field of view for the $2 \deg \times 2\deg$ region around M87. These final images have a pixel scale 0\farcs186 and median seeing 0\farcs54. The limiting surface brightness is ${\sim}$24.0 AB mag\,arcsec$^{-2}$ in $K_s$. PSFs were generated separately for each of these four fields using PSFex \citep{Bertin:2011aa}. Spatial variations in the PSFs were modeled with a seventh-order polynomial. As with the other datasets, we extracted PSFs at the pixel position of each nucleus.

\subsection{HST/WFPC2 and HST/NICMOS Imaging} \label{subsec:redux}

A follow-up program to the ACSVCS, {\it Virgo Redux}, re-imaged the sample of ACSVCS galaxies at the UV and IR wavelengths (Program ID = 11083). UV observations were carried out using the F300W filter on the Wide Field Planetary Camera 2 (WFPC2). Galaxies were centered on the Planetary Camera (PC) chip, which has a finer pixel scale (0\farcs046\,px$^{-1}$) than the adjacent Wide Field (WF) chips (0\farcs1\,px$^{-1}$). Each galaxy was observed for one orbit; three exposures were collected in order to aid in cosmic ray rejection, for a total exposure times of 2100\,s. Our analysis uses only the PC chip, which measures 35\arcsec $\times$ 35\arcsec, since it contains the nucleus and most of the galaxy signal for all our targets. The data were retrieved from the Hubble Legacy Archive, which provides PC exposures that have been combined and scaled to 0\farcs05\,px$^{-1}$ using \textit{multidrizzle} in \texttt{PyRAF}. Flat-fielding, bias and dark subtraction, removal of saturated or bad pixels, and shutter shading correction were performed using the \textit{calwp2} software as part of the standard WFPC2 calibration pipeline. No sky subtraction was performed since the sky has a negligible contribution to the nuclei counts.

{\it Virgo Redux} also includes IR imaging taken in the F160W filter (${\sim}H$) with the Near-Infrared Camera and Multi-Object Spectrometer \citep[NICMOS;][]{Thompson:1994aa}. Images were acquired using the NIC1 detector which has a native pixel scale of 0\farcs043\,px$^{-1}$ and a field of view of 11\arcsec $\times$ 11\arcsec. Each galaxy was observed in a series of 12 exposures, in a spiral dither pattern, for a total exposure time of 1920\,s. One additional image was taken offset 2\arcmin{} from the galaxy center to aid in background sky measurements. For some of the larger, brighter galaxies in the survey, this offset pointing is likely contaminated with galaxy light, so we excluded it from our background measurements. An additional 13 exposures, offset ${\sim}$5\arcmin{} from the galaxy center, were taken in parallel with the WFPC2 imaging to serve as blank sky fields. Flat-fielding, bias and dark corrections and other reduction steps were performed using the \textit{calnica} pipeline in \texttt{IRAF}. NICMOS images also have a pedestal effect, appearing as a residual flat-field signature that differs for each quadrant on the detector and varies with time. We have corrected for this effect using \textit{pedsub}. Our final science images were created using \textit{multidrizzle} and have a pixel scale of 0\farcs03\,px$^{-1}$. Note that the drizzling task was run without sky subtraction. Instead, we later subtracted a sky level based on the mean sky measured from the 13 exposures with 5\arcmin{} offsets. PSFs for all Virgo Redux images were created using the Tiny Tim package\footnote{\url{http://www.stsci.edu/hst/observatory/focus/TinyTim}}.

\subsection{Ground-Based Spectroscopy} \label{subsec:spectra}

For 19 of the nuclei in our sample, high-quality optical spectroscopy is available from three different ground-based instruments. While these data are only available for a subset of the objects, they provide an important point of comparison for the photometric results, allowing us to evaluate the robustness of the photometrically-determined masses, ages and metallicities (and vice versa). In a few cases, spectroscopy is available from multiple sources which allows us to assess the level of agreement among parameters derived from the different spectroscopic datasets.

Five nuclei were observed with the Integral Field Unit \citep[IFU;][]{Allington-Smith:2002aa} within the Gemini Multi-Object Spectrograph \citep[GMOS;][]{Hook:2004aa} on the Gemini South telescope during the 2008A and 2009A observing seasons. The IFU's 7\arcsec $\times$ 5\arcsec{} field of view, containing 1000 fibers, was centered on each object and rotated to align with the galaxy's semi-major axis. Another 500 fibers were configured in a 5\arcsec $\times$ 3\arcsec{} field, offset by 1\arcmin{} from the center of the science field. Observations were performed in the two-slit mode using the B600 grating (600 l\,mm$^{-1}$) and $g^{\prime}$ filter, although data from one slit were excluded because the key spectral features --- H$_\beta$, Mgb, and Fe Lick indices --- fell on a CCD with a number of bad columns and pattern noise. Four exposures for each nucleus were acquired, giving total exposure times between 1600\,s and 8000\,s. The final binned science spectra have a dispersion of 0.9\,\AA\,px$^{-1}$. Full details on the instrumental configuration and data reduction procedures are given in \citet{Liu:2016aa}.

Another 17 galaxies were observed with the Echellette Spectrograph and Imager \citep[ESI;][]{Sheinis:2002aa} on the Keck II telescope during the 2003A and 2004A observing seasons. In its echelle mode, ESI offers 10 spectral orders, with complete wavelength coverage from 3900 to 10900\,\AA\ at a dispersion ranging from 0.15\,\AA\,px$^{-1}$ (for $\lambda$ = 3900--4400\,\AA~in order No.~15) to 0.39\,\AA\,px$^{-1}$ (for $\lambda$ = 9500--11000\,\AA~in order No.~6). The spectral dispersion, in units of velocity, is a nearly constant 11.5 km~s$^{-1}$~px$^{-1}$. Object were observed with either a 0\farcs75 $\times$ 20\arcsec~or a 1\farcs0 $\times$ 20\arcsec{} slit, giving an instrumental velocity resolution between 1.0 and 1.4\,\AA\ at ${\sim}$5200\,\AA. The processing of the raw data involved bias subtraction, finding and tracing the apertures, flat normalization, cosmic-ray removal, arc extraction, and spectral calibration. Reductions were carried out using the Mauna Kea Echelle Extraction package \citep[MAKEE;][]{Barlow:1997aa}. 

Lastly, seven nuclei were observed using the DEep Imaging Multi-Object Spectrograph \citep[DEIMOS;][]{Faber:2003aa} on the Keck II telescope during the 2012A observing season. The observations, which were optimized for radial velocity and chemical abundance studies of star clusters in these galaxies, were carried out using the 600 l\,mm$^{-1}$ grating centered at 7000\,\AA{}. When combined with the GG4455 filter, this set-up provided a wavelength coverage of 4800--9500\,\AA{} at a dispersion of 0.52\,\AA\,px$^{-1}$. Slit lengths were kept somewhat short, typically ${\sim}$\,4--8\arcsec, in order to place as many globular clusters, stars, and other point-like objects, including nuclei, as possible on each slit mask. A slit width of 0\farcs8 was used in all cases. Exposure times varied between 3600\,s and 4800\,s, with the different exposure times meant to account for variations in the observing conditions (i.e., the seeing varied between 0\farcs6 and 0\farcs9). Additional observational details are presented in \citet{Toloba:2016aa} and Guhatakurta et al.~(2017, in preparation).

\begin{figure}[htbp]
   \centering
   \epsscale{1.1}
   \plotone{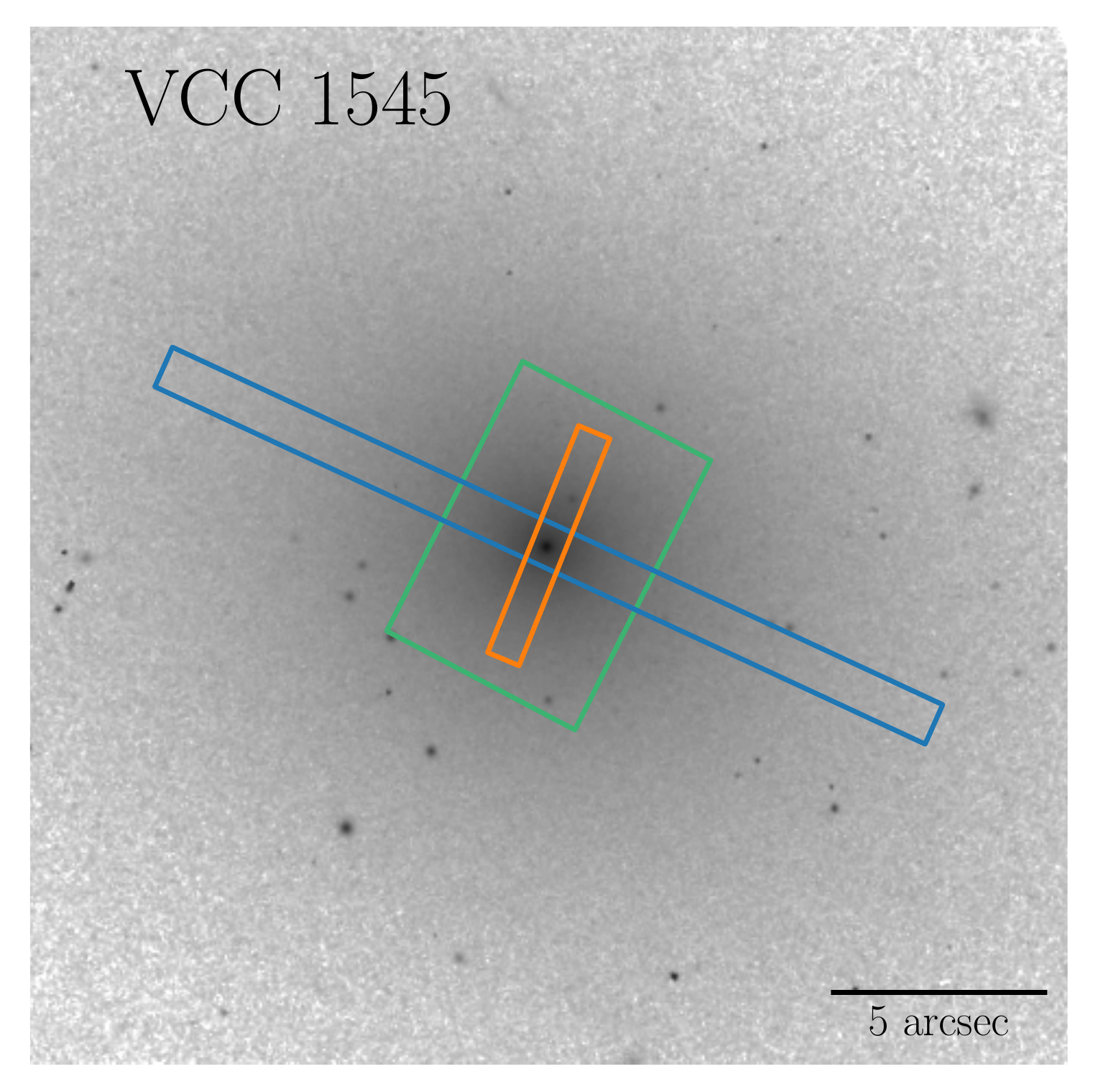} 
   \caption{NGVS $g$-band image for VCC~1545 with the location of the Keck/ESI and Keck/DEIMOS slits shown in blue and red, respectively. The green rectangle indicates the location of the Gemini/GMOS IFU.}
   \label{fig:slitshow}
\end{figure}

A summary of the spectroscopic observations is provided in Table~\ref{tab:spec}. In Figure~\ref{fig:slitshow}, we show slit and IFU orientations for VCC~1545, one of two galaxies included in all three spectral datasets.

\begin{deluxetable*}{clcccrcrc}
\tablecaption{Summary of Spectroscopic Observations \label{tab:spec}}
\tablehead{
\colhead{VCC} & \colhead{Telescope} & \colhead{Instrument} & \colhead{Grating} & \colhead{Dimenions} & \colhead{$\Theta$} & \colhead{$\lambda$ Range} & \colhead{Date} & \colhead{Exposure Time} \\
\colhead{} & \colhead{} & \colhead{} & \colhead{} & \colhead{(arcsec)} & \colhead{(deg)} & \colhead{(\AA{})} & \colhead{(dd/mm/yyyy)} & \colhead{(sec)}
}
\colnumbers
\startdata
~~33 & Gemini-S & GMOS-IFU & $g^\prime$ filter + B600 & 7\arcsec $\times$ 5\arcsec & 115.0 & $3980-5520$ & 22/2/2009 & 4400\\
~~33 & Keck II & ESI & echelle & 1\arcsec $\times$ 20\arcsec & 29.7 & 4020--7200 & 26/02/2003 & 1800\\
~200 & Keck II & ESI & echelle & 1\arcsec $\times$ 20\arcsec & 165.0 & 4020--7200 & 17/03/2004 & 1800\\
~230 & Keck II & ESI & echelle & 1\arcsec $\times$ 20\arcsec & 27.4 & 4020--7200 & 27/02/2003 & 2400\\
~538 & Keck II & ESI & echelle & 1\arcsec $\times$ 20\arcsec & 60.6 & 4020--7200 & 26/02/2003 & 2100\\
1075 & Keck II & DEIMOS & 600ZD & 0\farcs8 $\times$ 7\farcs2 & -1.9 & 4800--9500 & 23/4/2012 & 4500\\
1075 & Keck II & ESI & echelle & 0\farcs75 $\times$ 20\arcsec & 45.0 & 4020--7200 & 01/05/2000 & 1200\\
1185 & Gemini-S & GMOS-IFU & $g^\prime$ filter + B600 & 7\arcsec $\times$ 5\arcsec & 77.0 & $3980-5520$ & 25/2/2009 & 8400\\
1185 & Keck II & ESI & echelle & 1\arcsec $\times$ 20\arcsec & 28.2 & 4020--7200 & 26/02/2003 & 2700\\
1192 & Keck II & ESI & echelle & 1\arcsec $\times$ 20\arcsec & 37.6 & 4020--7200 & 26/02/2003 & 1800\\
1199 & Keck II & ESI & echelle & 1\arcsec $\times$ 20\arcsec & 45.0 & 4020--7200 & 26/02/2003 & 2100\\
1355 & Gemini-S & GMOS-IFU & $g^\prime$ filter + B600 & 7\arcsec $\times$ 5\arcsec & 127.0 & $3980-5520$ & 5/3/2009 & 8400\\
1407 & Keck II & DEIMOS & 600ZD & 0\farcs8 $\times$ 4\farcs39 & -173.6 & 4800--9500 & 22/4/2012 & 3730\\
1407 & Keck II & ESI & echelle & 0\farcs75 $\times$ 20\arcsec & 68.0 & 4020--7200 & 01/05/2000 & 1800\\
1440 & Keck II & ESI & echelle & 1\arcsec $\times$ 20\arcsec & 28.9 & 4020--7200 & 26/02/2003 & 2100\\
1489 & Keck II & ESI & echelle & 1\arcsec $\times$ 20\arcsec & 60.5 & 4020--7200 & 27/02/2003 & 3600\\
1539 & Gemini-S & GMOS-IFU & $g^\prime$ filter + B600 & 7\arcsec $\times$ 5\arcsec & 345.0 & $3980-5520$ & 2/3/2009 & 8800\\
1539 & Keck II & DEIMOS & 600ZD & 0\farcs8 $\times$ 4\farcs06 & 119.2 & 4800--9500 & 21/4/2012 & 4800\\
1539 & Keck II & ESI & echelle & 1\arcsec $\times$ 20\arcsec & 46.0 & 4020--7200 & 17/03/2004 & 2100\\
1545 & Gemini-S & GMOS-IFU & $g^\prime$ filter + B600 & 7\arcsec $\times$ 5\arcsec & 335.0 & $3980-5520$ & 9/4/2008 & 3600\\
1545 & Keck II & DEIMOS & 600ZD & 0\farcs8 $\times$ 5\farcs66 & -20.0 & 4800--9500 & 21/4/2012 & 3600\\
1545 & Keck II & ESI & echelle & 1\arcsec $\times$ 20\arcsec & 67.8 & 4020--7200 & 27/02/2003 & 2400\\
1627 & Keck II & ESI & echelle & 1\arcsec $\times$ 20\arcsec & 93.6 & 4020--7200 & 26/02/2003 & 2100\\
1826 & Keck II & ESI & echelle & 1\arcsec $\times$ 20\arcsec & 131.2 & 4020--7200 & 26/02/2003 & 2100\\
1828 & Keck II & DEIMOS & 600ZD & 0\farcs8 $\times$ 6\farcs45 & -50.7 & 4800--9500 & 23/4/2012 & 4499\\
1828 & Keck II & ESI & echelle & 0\farcs75 $\times$ 20\arcsec & 75.0 & 4020--7200 & 30/04/2000 & 1800\\
1861 & Keck II & DEIMOS & 600ZD & 0\farcs8 $\times$ 3\farcs86 & 14.9 & 4800--9500 & 22/4/2012 & 3599\\
1871 & Keck II & DEIMOS & 600ZD & 0\farcs8 $\times$ 4\farcs62 & 14.9 & 4800--9500 & 22/4/2012 & 3599\\
2050 & Keck II & ESI & echelle & 1\arcsec $\times$ 20\arcsec & 127.0 & 4020--7200 & 26/02/2003 & 2400\\
\enddata
\tablecomments{Key to columns: (1) VCC identification number, (2) telescope, (3) spectrograph, (4) grating, (5) slit or IFU dimensions, (6) position angle, $\Theta$, of the slit or major axis of the IFU, (7) wavelength range, (8) date of observation, and (9) total exposure time. For DEIMOS, slit lengths vary from galaxy to galaxy, with values in the range ${\sim}$\,4\arcsec to~8\arcsec{}.}
\end{deluxetable*}

\section{Photometric and Structural Measurements} \label{sec:photometry}

There are a number of challenges involved in the measurement of photometric and structural parameters for nuclei, and, ultimately, in the characterization of their stellar populations. Nucleus-galaxy decompositions can be uncertain due to such factors as the number of components used in the modeling of the light distribution, the PSF used for the model convolution, and the presence of complex or nonparametric structures that may skew the model fit \citep[see, e.g,][for a discussion of these issues]{Turner:2012aa}. When deriving stellar population properties, one familiar difficulty is the age-metallicity degeneracy. This is undoubtedly important for nuclei in early-type hosts, as both the galaxies and the nuclei are likely to contain old to intermediate age ($t>5$ Gyr) populations. Age sensitivity can be improved by extending the photometric coverage into the UV and IR regions \citep{Worthey:1994aa, Puzia:2002aa, Hempel:2003aa, de-Meulenaer:2014aa} --- a prime motivation for our study. Of course, the derived stellar population parameters can also vary with the choice of comparison models, which often rely on different isochrones, spectral libraries and stellar evolution treatments \citep{Conroy:2010aa, Powalka:2016aa}. 

To test the robustness of our results with these uncertainties in mind, we use multiple procedures to understand any possible systematics. This includes using different nucleus-galaxy decomposition techniques (described in \S\ref{subsec:ellipse} and \S\ref{subsec:galfit}) as well as deriving stellar population parameters from various population synthesis models.

\subsection{\texttt{ELLIPSE}-Based Analysis} \label{subsec:ellipse}

Nuclei and galaxy parameters were derived from S\'ersic component fits to radial profiles created using the \texttt{IRAF} task \texttt{ELLIPSE} which fits elliptical isophotes to each object using the method of \citet{Jedrzejewski:1987aa}. The image intensity is sampled along each ellipse's path, creating an intensity distribution as a function of azimuthal angle, $\phi$. If the ellipse's parameters are well matched to the galaxy's shape, then the intensity should be constant at all values of $\phi$. Any deviations from the isophote can be expressed as higher order moments of a Fourier series:
\begin{equation} \label{eq:ellip}
  I(\phi) = I_0 + \sum_{k}[A_k\sin(k\phi) + B_k\cos(k\phi)]
  \end{equation}
$A_k$ and $B_k$ represent the amplitude of each moment. A pure ellipse can be described by the first two moments of the series, while any deviations (e.g., a disky or boxy isophote) can be expressed with terms for $k\geq3$. \texttt{ELLIPSE} determines the best-fit parameters by least-squares minimization of the residuals between the sampled image intensity and Eq.~\ref{eq:ellip}. This isophotal fitting process is described in full in \citet{Ferrarese:2006ab} and \citet{Cote:2006aa}. 

To capitalize simultaneously on HST's superior resolution and the depth of the wide-field NGVS imaging, composite surface brightness profiles were created by combining ACS F475W and MegaCam $g$-band profiles, as well as F850LP and $z$-band profiles. The two filter pairs are nearly identical, although a small ($\lesssim0.01$ mag) zeropoint correction is required to transform each set to a common system. In addition, the HST profiles must have the sky removed. To accomplish both these tasks, the $g$ and $z$ profiles were first transformed to the SDSS photometric system using the color transformations provided by the MegaPipe webpages\footnote{\url{http://www.cadc-ccda.hia-iha.nrc-cnrc.gc.ca/en/megapipe/docs/filt.html}}. Next, we estimated the zeropoint shifts and sky levels of the ACS images simultaneously. For the ACS profiles, a corrected surface brightness profile can be calculated using the equation
\begin{equation} \label{eq:shift}
   \mu_{AB}(r)=-2.5\log_{10}(f(r)+f_{sky})+z+\Delta z,
   \end{equation}
   where $f(r)$ is the measured flux at each radial step, $f_{sky}$ is the estimated sky level, $z$ is the zeropoint for initial photometric system, and $\Delta z$ is the zeropoint correction.
 
\begin{figure}[htbp]
   \centering
   \epsscale{1.1}
   \plotone{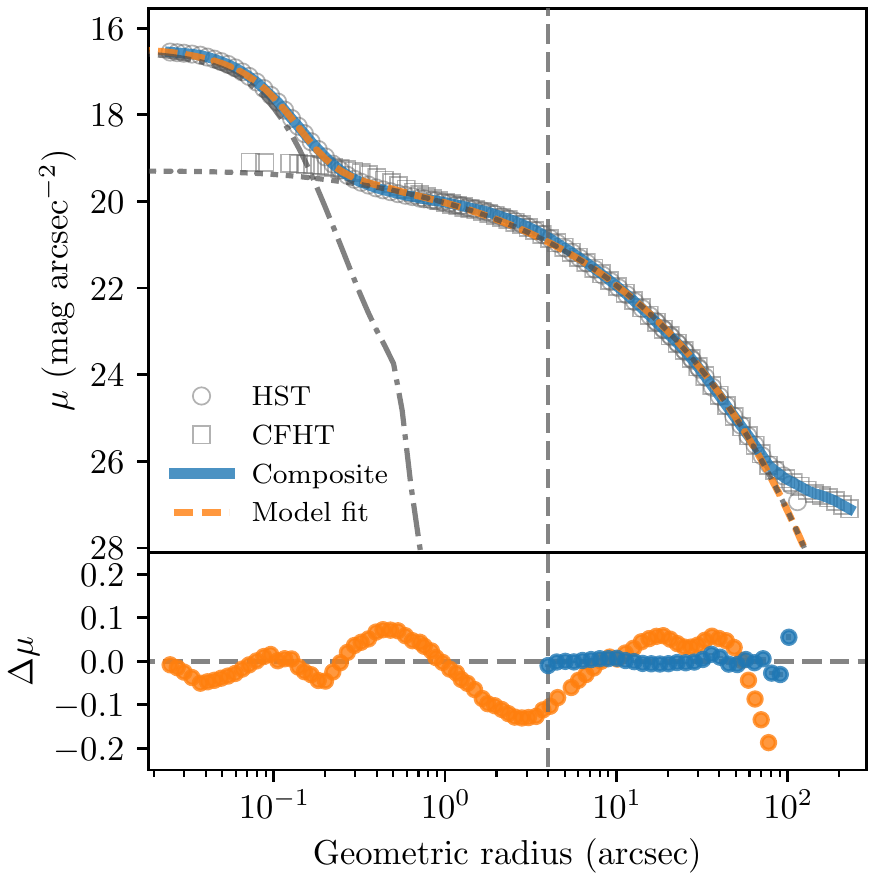} 
   \caption{\textit{(Top panel)}. Matched HST/ACS (gray circles) and CFHT (gray squares) $g$-band surface brightness profiles for VCC~1422. The final composite profile is plotted in blue. The vertical dashed line indicates the inner boundary of the region used to match the space- and ground-based profiles. The dashed orange line shows a fit to the profile using two S\'ersic components; the nucleus component is shown by the dot-dashed gray curve, and the galaxy component by the dotted gray curve. \textit{(Bottom panel)}. Blue points show the residuals between the HST and CFHT profiles in the region used to match the profiles. Orange points show the residuals between the composite profile and the best-fit model. Error bars are smaller than the data points.}
   \label{fig:profs}
\end{figure}

These corrections were determined on a galaxy-by-galaxy basis by matching the ACS profile produced by Eq.~\ref{eq:shift} to the CFHT profile using the orthogonal distance regression (ODR) package within SciPy. For the fit, only the profile regions beyond 4\arcsec{} were considered. This is roughly five times the seeing of the NGVS data, which should safely avoid any blurring of the nucleus and galaxy profiles \citep{Schweizer:1979aa}. Figure~\ref{fig:profs} shows the matched profiles for galaxy VCC~1422 once the ACS component has been zeropoint-corrected and sky-subtracted. Residuals between the two original profiles in the fitted regions, unaffected by smearing, are shown as well.

Parametric fits to the composite profiles were produced following a method similar to that in \citet{Cote:2006aa}, \citet{Ferrarese:2006aa}, and \citet{Turner:2012aa}. However, the approach here differs in a few ways, most notably in that we fitted the nucleus light using a S\'ersic profile rather than a King profile (i.e., the entire profile nominally contains two S\'ersic components). The intensity in a S\'ersic profile is described by the equation
\begin{equation} \label{eq:sersic}
   I(r) = I_e \exp\left\{-b_n\left[\left(\frac{r}{r_e}\right)^{1/n}-1\right]\right\}
   \end{equation}
   where the free parameters are $I_e$, the intensity at the effective radius $r_e$, and the S\'ersic index $n$. The constant $b_n$ is defined by complete and incomplete gamma functions, $\Gamma(n)$ and $\gamma(n, x)$, respectively, such that \(\Gamma(2n)=2\gamma(2n,b_n)\). For the nuclear component, $n$ was fixed to \(n=2\) to diminish the likelihood of the nucleus component fitting non-nucleus light in galaxies with complex substructure.

The best-fit model was determined using $\chi^2$ minimization with equal weight applied to all points in the profile after convolution with the appropriate PSF. Fits to the $g$ and $z$ profiles were performed both independently for each filter as well as simultaneously. We found that the structural parameters derived with independent and simultaneous fits were consistent, although small variations between the $g$ and $z$ parameters arose in the independent fits. However, each fit remained well-behaved. We adopt the results from the independent fits for our analysis.
      
\subsection{\textsc{galfit} Analysis} \label{subsec:galfit}

\begin{figure*}
\gridline{\fig{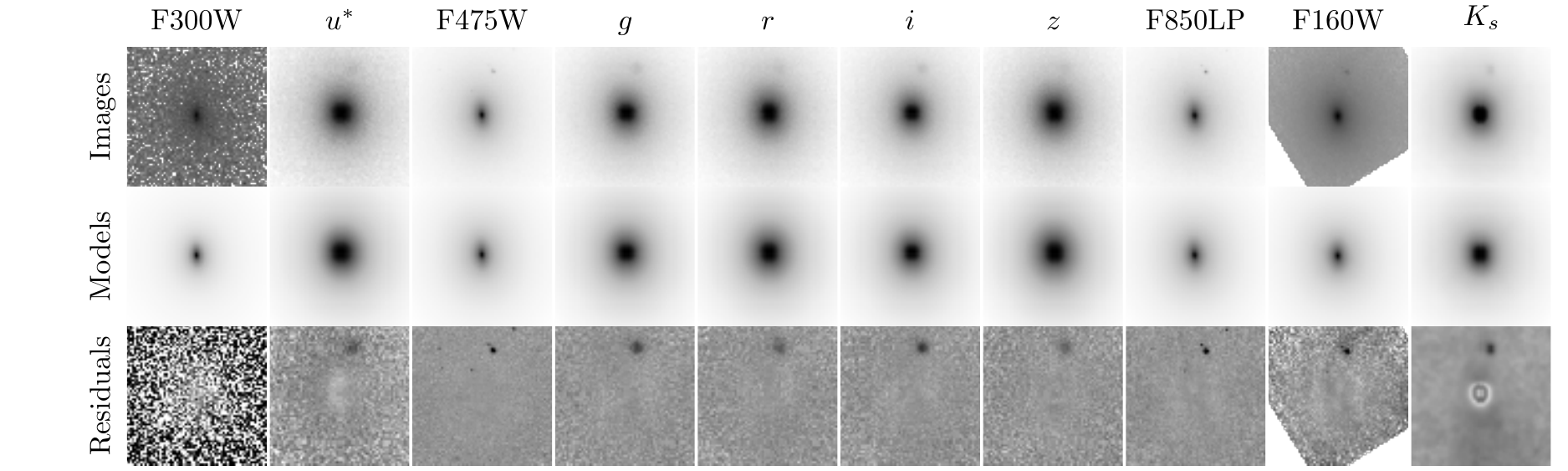}{\textwidth}{a) VCC~1146.}}
\gridline{\fig{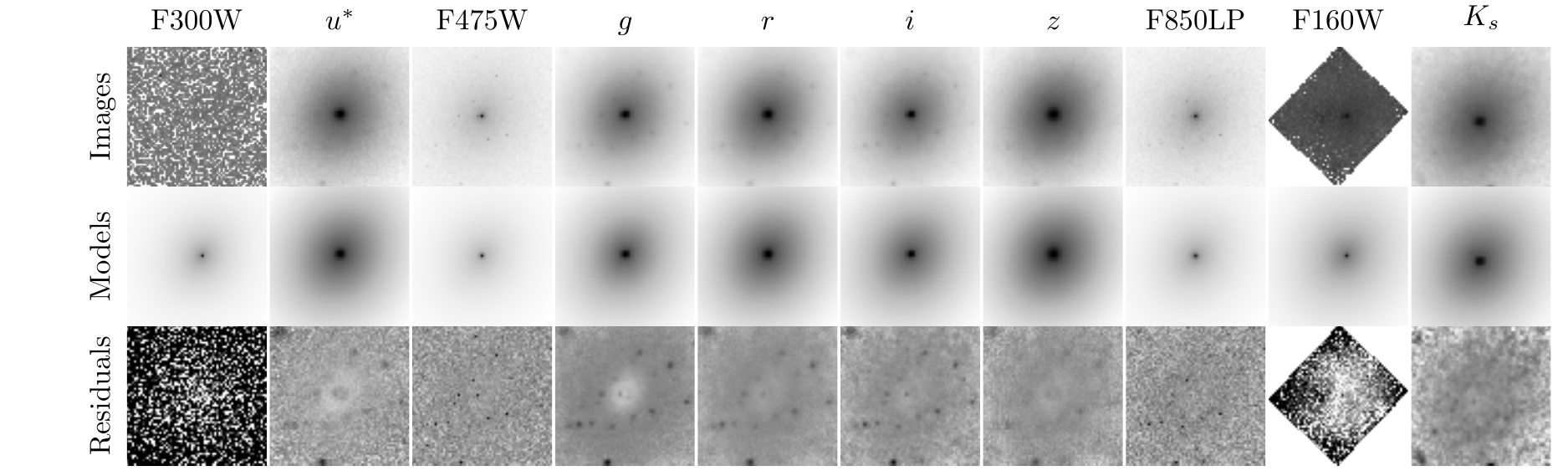}{\textwidth}{b) VCC~1407.}}
\gridline{\fig{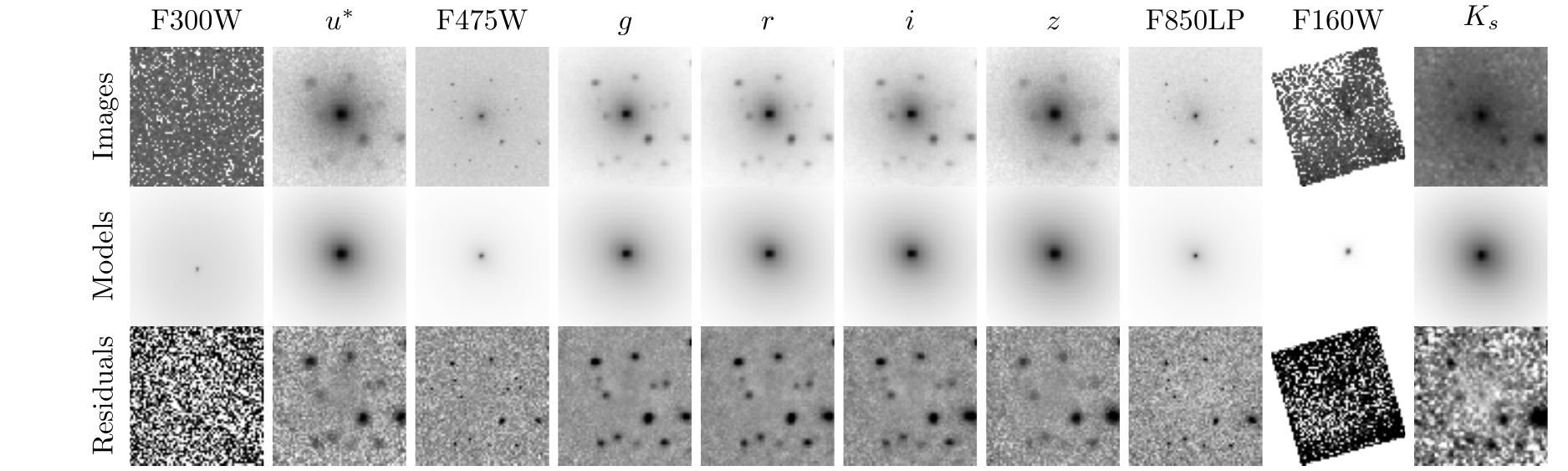}{\textwidth}{c) VCC~1539.}}
\caption{Final images, best-fit \textsc{galfit} models, and model residuals for three of our program galaxies: VCC~1146 (a), VCC~1407 (b), and VCC~1539 (c).}
\label{fig:galfit}
\end{figure*}

Galaxy and nucleus magnitudes were measured simultaneously using \textsc{galfit}, a familiar algorithm that fits two-dimensional (2D) parametric models to images \citep{Peng:2002aa,Peng:2010aa}. A galaxy model can be composed of an arbitrary number of components (e.g., exponential disk, S\'ersic profile, point source) that are combined to best fit the 2D galaxy image. As inputs, \textsc{galfit} requires the original image, a PSF image to convolve with the model component(s), and either a sigma map containing the errors for each pixel, or the gain of the instrument in order to estimate errors from the Poisson noise.

Final magnitude measurements for the various components were obtained using an iterative process. Initially, each object was fitted with a basic model containing (1) a S\'ersic component for the galaxy and (2) a S\'ersic component for the nucleus. These components were defined by a single ellipticity and position angle (i.e., no isophote twisting), as well as an effective radius $r_e$, S'ersic index $n$, axis ratio $b/a$, and magnitude. Although \textsc{galfit} can estimate and fit the sky level, its brightness was fixed to a predetermined value because the fitted images are often object, not sky, dominated, and so \textsc{galfit}'s sky level estimates were found to be consistently high. For the $u^*griz$, F300W and F160W images, sky subtraction had already been applied to the data, so sky levels were fixed to zero. For the ACS F475W and F850LP images, sky levels were determined from matching the F475W and F850LP one-dimensional (1D) profiles to the corresponding CFHT $g$ and $z$ profiles (see \S\ref{subsec:ellipse}). The output parameters of the fit obtained with the basic model then were inspected for each object. The model was subsequently refined with additional S\'ersic components for the galaxy if the fit met either, or both, of the following conditions: (1) $n$ for any component was outside the range $0.5 < n < 4$; and (2) the nucleus' $r_e$ was more than 10\% of the galaxy's $r_e$. This second condition, a quite conservative criterion, was imposed to catch only the most obvious outliers: i.e., typically, the nucleus radius is $\sim$2\% of the galaxy half-light radius. In all cases, a single component was used to fit the nucleus, resulting in an average $n = 1.57 \pm 0.64$, and average $r_e = 5.74 \pm 1.74$\,pc. The models were also refined if a visual inspection of the fit residuals indicated an incomplete fit, even if the above conditions were not met. Even dwarf galaxies can display multiple structural components \citep[see e.g.,][and references therein]{Janz:2012aa}, so most of the galaxies in our sample have been modeled with more than one S\'ersic component, typically requiring two or three. The maximum number of components required was seven for VCC~784, one of the brightest galaxies in the sample.

To ensure consistent measurements across all bands, the same physical parameters were held fixed and fitted to all images, with the only free parameters being the magnitudes of the model components (the sky level, however, was always held fixed). To determine the values of the fixed parameters, a completely free fit was performed on the F475W images. These are the obvious choice as reference images due to the combination of high SNR, high resolution and relatively wide field coverage. All nuclei in this sample were resolved in the HST data, enabling better measurement of nuclei parameters. Once the fits to the F475W images were completed, the best-fit parameters for each object were extracted and fitted to the remaining images with only the magnitudes being allow to vary from their input values. Using this technique, we were able to measure component magnitudes in a homogeneous way. This is particularly important for the NICMOS and WFPC2 data, where a limited field of view, or marginal SNR, can present challenges in the fitting process. Figure~\ref{fig:galfit} shows the \textsc{galfit} results for three program galaxies at a range of magnitudes and structural complexities. With the flexibility of adding multiple galaxy components, even a bright, complex galaxy such as VCC~1146 can be well-modeled. For the faintest galaxies in the sample, such as VCC~1539, the fixed structural parameters ensure consistent fits, even in filters where the detection is limited. Tables~\ref{tab:nucs}~and~\ref{tab:gals} list the extracted magnitudes for the nuclei and galaxies, respectively. Galaxy magnitudes are defined as the total magnitude of all components \emph{excluding} the nucleus component.

\begin{longrotatetable}
\begin{deluxetable*}{ccccccccccc}
\tablecaption{Photometric Measurements for Program Nuclei \label{tab:nucs}}
\tablehead{
\colhead{VCC} & \colhead{F300W} & \colhead{$u^*$} & \colhead{F475W} & \colhead{$g$} & \colhead{$r$} & \colhead{$i$} & \colhead{$z$} & \colhead{F850LP} & \colhead{F160W} & \colhead{$K_s$} \\
\colhead{} & \colhead{(AB mag)} & \colhead{(AB mag)} & \colhead{(AB mag)} & \colhead{(AB mag)} & \colhead{(AB mag)} & \colhead{(AB mag)} & \colhead{(AB mag)} & \colhead{(AB mag)} & \colhead{(AB mag)} & \colhead{(AB mag)}
}
\colnumbers
\startdata
~~33 & 25.79 $\pm$ 0.30 & 23.43 $\pm$ 0.37 & 22.20 $\pm$ 0.37 & 22.27 $\pm$ 0.37 & \ldots & 21.45 $\pm$ 0.37 & 22.06 $\pm$ 0.37 & 21.34 $\pm$ 0.37 & 21.04 $\pm$ 0.37 & \ldots\\
~140 & 24.49 $\pm$ 0.30 & 23.72 $\pm$ 0.37 & 22.20 $\pm$ 0.37 & 21.87 $\pm$ 0.37 & \ldots & 21.27 $\pm$ 0.37 & 21.61 $\pm$ 0.37 & 21.27 $\pm$ 0.37 & 20.53 $\pm$ 0.37 & \ldots\\
~200 & 22.86 $\pm$ 0.30\tablenotemark{a} & 23.20 $\pm$ 0.35\tablenotemark{a} & 22.97 $\pm$ 0.35 & 23.18 $\pm$ 0.35 & \ldots & 21.29 $\pm$ 0.35 & 21.12 $\pm$ 0.35 & 21.75 $\pm$ 0.35 & 20.76 $\pm$ 0.35 & \ldots\\
~230 & 22.55 $\pm$ 0.30 & 21.31 $\pm$ 0.30 & 20.16 $\pm$ 0.30 & 19.95 $\pm$ 0.30 & \ldots & 19.14 $\pm$ 0.30 & 19.01 $\pm$ 0.30 & 19.13 $\pm$ 0.30 & 19.06 $\pm$ 0.30 & \ldots\\
~538 & 23.50 $\pm$ 0.30 & 21.84 $\pm$ 0.27 & 20.84 $\pm$ 0.27 & 20.96 $\pm$ 0.27 & \ldots & 19.80 $\pm$ 0.27 & 19.36 $\pm$ 0.27 & 19.75 $\pm$ 0.27 & 19.09 $\pm$ 0.27 & \ldots\\
~698 & 22.43 $\pm$ 0.30 & 21.11 $\pm$ 0.29 & 20.05 $\pm$ 0.29 & 19.27 $\pm$ 0.29 & \ldots & 18.95 $\pm$ 0.29 & 18.68 $\pm$ 0.29 & 18.78 $\pm$ 0.29 & 18.37 $\pm$ 0.29 & \ldots\\
~784 & 21.92 $\pm$ 0.30 & 20.59 $\pm$ 0.21 & 18.75 $\pm$ 0.21 & 18.34 $\pm$ 0.21 & \ldots & 17.21 $\pm$ 0.21 & 16.84 $\pm$ 0.21 & 17.26 $\pm$ 0.21 & 16.31 $\pm$ 0.21 & \ldots\\
~828 & 21.77 $\pm$ 0.30 & 19.37 $\pm$ 0.20 & 18.54 $\pm$ 0.20 & 18.45 $\pm$ 0.20 & \ldots & 17.10 $\pm$ 0.20 & 17.28 $\pm$ 0.20 & 16.96 $\pm$ 0.20 & 16.22 $\pm$ 0.20 & \ldots\\
~856 & 22.02 $\pm$ 0.30 & 20.57 $\pm$ 0.32 & 19.56 $\pm$ 0.32 & 19.38 $\pm$ 0.32 & \ldots & 18.51 $\pm$ 0.32 & 18.37 $\pm$ 0.32 & 18.49 $\pm$ 0.32 & 17.90 $\pm$ 0.32 & \ldots\\
1075 & 23.54 $\pm$ 0.30 & 22.16 $\pm$ 0.27 & 21.08 $\pm$ 0.27 & 21.13 $\pm$ 0.27 & \ldots & 20.24 $\pm$ 0.27 & 20.20 $\pm$ 0.27 & 20.19 $\pm$ 0.27 & 19.76 $\pm$ 0.27 & \ldots\\
1087 & 23.23 $\pm$ 0.30 & 22.12 $\pm$ 0.30 & 20.16 $\pm$ 0.30 & 17.52 $\pm$ 0.30\tablenotemark{a} & 17.32 $\pm$ 0.30\tablenotemark{a} & 16.77 $\pm$ 0.30\tablenotemark{a} & 16.00 $\pm$ 0.30\tablenotemark{a} & 18.93 $\pm$ 0.30 & 18.30 $\pm$ 0.30 & 18.17 $\pm$ 0.30\\
1146 & 19.32 $\pm$ 0.30 & 17.49 $\pm$ 0.12 & 16.60 $\pm$ 0.12 & 16.58 $\pm$ 0.12 & 16.05 $\pm$ 0.12 & 15.42 $\pm$ 0.12 & 15.37 $\pm$ 0.12 & 15.10 $\pm$ 0.12 & 14.24 $\pm$ 0.12 & 15.67 $\pm$ 0.12\\
1185 & \ldots & 21.88 $\pm$ 0.28 & 20.76 $\pm$ 0.28 & 20.67 $\pm$ 0.28 & 20.12 $\pm$ 0.28 & 19.81 $\pm$ 0.28 & 19.70 $\pm$ 0.28 & 19.80 $\pm$ 0.28 & 19.42 $\pm$ 0.28 & 19.43 $\pm$ 0.28\\
1192 & 21.12 $\pm$ 0.30 & 19.70 $\pm$ 0.20 & 18.47 $\pm$ 0.20 & 18.44 $\pm$ 0.20 & 17.73 $\pm$ 0.20 & 17.61 $\pm$ 0.20 & 17.14 $\pm$ 0.20 & 17.23 $\pm$ 0.20 & 16.43 $\pm$ 0.20 & \ldots\\
1199 & 22.41 $\pm$ 0.30 & 20.63 $\pm$ 0.32 & 19.43 $\pm$ 0.32 & 19.08 $\pm$ 0.32 & 18.35 $\pm$ 0.32 & 18.25 $\pm$ 0.32 & 17.82 $\pm$ 0.32 & 18.01 $\pm$ 0.32 & 17.09 $\pm$ 0.32 & \ldots\\
1242 & 20.32 $\pm$ 0.30 & 19.12 $\pm$ 0.11 & 17.61 $\pm$ 0.11 & 17.37 $\pm$ 0.11 & 16.63 $\pm$ 0.11 & 16.25 $\pm$ 0.11 & 16.03 $\pm$ 0.11 & 16.09 $\pm$ 0.11 & 15.33 $\pm$ 0.11 & \ldots\\
1261 & 22.55 $\pm$ 0.30 & 21.39 $\pm$ 0.29 & 19.87 $\pm$ 0.29 & 20.33 $\pm$ 0.29 & \ldots & 18.96 $\pm$ 0.29 & 18.59 $\pm$ 0.29 & 18.69 $\pm$ 0.29 & 18.69 $\pm$ 0.29 & \ldots\\
1283 & 22.75 $\pm$ 0.30 & 20.88 $\pm$ 0.29 & 19.94 $\pm$ 0.29 & 19.50 $\pm$ 0.29 & 18.89 $\pm$ 0.29 & 18.61 $\pm$ 0.29 & 18.32 $\pm$ 0.29 & 18.47 $\pm$ 0.29 & 17.50 $\pm$ 0.29 & \ldots\\
1355 & 23.48 $\pm$ 0.30 & 22.39 $\pm$ 0.26 & 21.10 $\pm$ 0.26 & 21.23 $\pm$ 0.26 & 20.67 $\pm$ 0.26 & 20.20 $\pm$ 0.26 & 20.06 $\pm$ 0.26 & 20.12 $\pm$ 0.26 & 19.68 $\pm$ 0.26 & \ldots\\
1407 & 23.36 $\pm$ 0.30 & 21.48 $\pm$ 0.29 & 20.76 $\pm$ 0.29 & 20.37 $\pm$ 0.29 & 19.94 $\pm$ 0.29 & 19.67 $\pm$ 0.29 & 19.38 $\pm$ 0.29 & 19.78 $\pm$ 0.29 & 19.31 $\pm$ 0.29 & 18.83 $\pm$ 0.29\\
1422 & 22.86 $\pm$ 0.30 & 21.31 $\pm$ 0.29 & 20.09 $\pm$ 0.29 & 19.99 $\pm$ 0.29 & \ldots & 18.98 $\pm$ 0.29 & 18.76 $\pm$ 0.29 & 18.87 $\pm$ 0.29 & 18.32 $\pm$ 0.29 & \ldots\\
1431 & 22.16 $\pm$ 0.30 & 21.05 $\pm$ 0.30 & 19.89 $\pm$ 0.30 & 19.62 $\pm$ 0.30 & \ldots & 18.83 $\pm$ 0.30 & 18.61 $\pm$ 0.30 & 18.79 $\pm$ 0.30 & 18.32 $\pm$ 0.30 & \ldots\\
1440 & 22.75 $\pm$ 0.30 & 21.09 $\pm$ 0.31 & 20.24 $\pm$ 0.31 & 19.80 $\pm$ 0.31 & \ldots & 19.39 $\pm$ 0.31 & 18.36 $\pm$ 0.31 & 19.01 $\pm$ 0.31 & 18.54 $\pm$ 0.31 & \ldots\\
1489 & 24.38 $\pm$ 0.30 & 23.51 $\pm$ 0.38 & 22.41 $\pm$ 0.38 & 22.29 $\pm$ 0.38 & \ldots & 21.63 $\pm$ 0.38 & 21.66 $\pm$ 0.38 & 21.62 $\pm$ 0.38 & 21.25 $\pm$ 0.38 & \ldots\\
1539 & 23.67 $\pm$ 0.30\tablenotemark{a} & 22.02 $\pm$ 0.27 & 21.20 $\pm$ 0.27 & 21.01 $\pm$ 0.27 & 20.53 $\pm$ 0.27 & 20.34 $\pm$ 0.27 & 20.36 $\pm$ 0.27 & 20.38 $\pm$ 0.27 & 19.74 $\pm$ 0.27 & 20.21 $\pm$ 0.27\\
1545 & 24.22 $\pm$ 0.30 & 22.50 $\pm$ 0.28 & 21.81 $\pm$ 0.28 & 21.29 $\pm$ 0.28 & 20.20 $\pm$ 0.28 & 20.28 $\pm$ 0.28 & 20.10 $\pm$ 0.28 & 20.61 $\pm$ 0.28 & 21.36 $\pm$ 0.28 & 19.05 $\pm$ 0.28\\
1619 & 23.30 $\pm$ 0.30 & 21.58 $\pm$ 0.19 & 18.99 $\pm$ 0.19 & 19.06 $\pm$ 0.19 & \ldots & 17.86 $\pm$ 0.19 & 17.35 $\pm$ 0.19 & 17.78 $\pm$ 0.19 & 16.33 $\pm$ 0.19 & \ldots\\
1627 & \ldots & 21.37 $\pm$ 0.30 & 20.20 $\pm$ 0.30 & 19.56 $\pm$ 0.30 & \ldots & 20.03 $\pm$ 0.30 & 18.26 $\pm$ 0.30 & 18.81 $\pm$ 0.30 & \ldots & \ldots\\
1630 & 21.80 $\pm$ 0.30 & 19.93 $\pm$ 0.20 & 18.67 $\pm$ 0.20 & 18.45 $\pm$ 0.20 & \ldots & 17.35 $\pm$ 0.20 & 16.65 $\pm$ 0.20 & 17.04 $\pm$ 0.20 & 16.04 $\pm$ 0.20 & \ldots\\
1661 & 24.01 $\pm$ 0.30 & 22.12 $\pm$ 0.27 & 20.87 $\pm$ 0.27 & 20.23 $\pm$ 0.27 & \ldots & 19.92 $\pm$ 0.27 & 19.36 $\pm$ 0.27 & 21.27 $\pm$ 0.27 & 19.28 $\pm$ 0.27 & \ldots\\
1826 & 22.69 $\pm$ 0.30 & 21.13 $\pm$ 0.29 & 20.05 $\pm$ 0.29 & 19.97 $\pm$ 0.29 & \ldots & 18.99 $\pm$ 0.29 & 18.84 $\pm$ 0.29 & 18.91 $\pm$ 0.29 & 18.67 $\pm$ 0.29 & \ldots\\
1828 & 23.59 $\pm$ 0.30 & 22.81 $\pm$ 0.25 & 21.49 $\pm$ 0.25 & 21.44 $\pm$ 0.25 & \ldots & 20.54 $\pm$ 0.25 & 20.39 $\pm$ 0.25 & 20.48 $\pm$ 0.25 & 20.17 $\pm$ 0.25 & \ldots\\
1861 & 21.95 $\pm$ 0.30 & 21.09 $\pm$ 0.29 & 20.01 $\pm$ 0.29 & 19.84 $\pm$ 0.29 & \ldots & 19.02 $\pm$ 0.29 & 18.93 $\pm$ 0.29 & 18.92 $\pm$ 0.29 & 18.43 $\pm$ 0.29 & \ldots\\
1871 & 21.86 $\pm$ 0.30 & 20.32 $\pm$ 0.19 & 19.04 $\pm$ 0.19 & 19.31 $\pm$ 0.19 & \ldots & 18.42 $\pm$ 0.19 & 17.58 $\pm$ 0.19 & 17.77 $\pm$ 0.19 & 17.15 $\pm$ 0.19 & \ldots\\
1883 & 20.03 $\pm$ 0.30 & 18.30 $\pm$ 0.13 & 17.61 $\pm$ 0.13 & 17.74 $\pm$ 0.13 & \ldots & 29.30 $\pm$ 0.13\tablenotemark{a} & 16.09 $\pm$ 0.13 & 17.42 $\pm$ 0.13 & 15.53 $\pm$ 0.13 & \ldots\\
1886 & 24.27 $\pm$ 0.30\tablenotemark{a} & 23.06 $\pm$ 0.37 & 22.11 $\pm$ 0.37 & 22.02 $\pm$ 0.37 & \ldots & 21.24 $\pm$ 0.37 & 21.30 $\pm$ 0.37 & 21.20 $\pm$ 0.37 & 21.10 $\pm$ 0.37 & \ldots\\
1910 & 22.36 $\pm$ 0.30 & 20.84 $\pm$ 0.31 & 19.74 $\pm$ 0.31 & 19.62 $\pm$ 0.31 & 19.01 $\pm$ 0.31 & 18.69 $\pm$ 0.31 & 18.53 $\pm$ 0.31 & 18.62 $\pm$ 0.31 & 18.30 $\pm$ 0.31 & \ldots\\
2019 & 23.01 $\pm$ 0.30 & 21.45 $\pm$ 0.32 & 20.27 $\pm$ 0.32 & 20.25 $\pm$ 0.32 & \ldots & 19.27 $\pm$ 0.32 & 19.07 $\pm$ 0.32 & 19.18 $\pm$ 0.32 & 18.61 $\pm$ 0.32 & \ldots\\
2050 & 21.92 $\pm$ 0.30\tablenotemark{a} & 29.35 $\pm$ 0.37\tablenotemark{a} & 22.47 $\pm$ 0.37 & 23.08 $\pm$ 0.37 & \ldots & 21.46 $\pm$ 0.37 & 22.11 $\pm$ 0.37 & 21.39 $\pm$ 0.37 & 20.56 $\pm$ 0.37 & \ldots\\
\enddata
\tablecomments{Magnitudes have not been corrected for Milky Way foreground extinction.}
\tablenotetext{a}{Magnitude excluded from analysis due to image artifacts, failure of \textsc{galfit} to converge on a model fit, or, in the F300W, a non-detection.}
\end{deluxetable*}
\end{longrotatetable}

\begin{longrotatetable}
\begin{deluxetable*}{ccccccccccc}
\tablecaption{Photometric Measurements for Program Galaxies \label{tab:gals}}
\tablehead{
\colhead{VCC} & \colhead{F300W} & \colhead{$u^*$} & \colhead{F475W} & \colhead{$g$} & \colhead{$r$} & \colhead{$i$} & \colhead{$z$} & \colhead{F850LP} & \colhead{F160W} & \colhead{$K_s$} \\
\colhead{} & \colhead{(AB mag)} & \colhead{(AB mag)} & \colhead{(AB mag)} & \colhead{(AB mag)} & \colhead{(AB mag)} & \colhead{(AB mag)} & \colhead{(AB mag)} & \colhead{(AB mag)} & \colhead{(AB mag)} & \colhead{(AB mag)}
}
\colnumbers
\startdata
~~33 & 16.76 $\pm$ 0.30 & 16.05 $\pm$ 0.02 & 14.98 $\pm$ 0.02 & 14.90 $\pm$ 0.02 & \ldots & 14.08 $\pm$ 0.02 & 13.83 $\pm$ 0.02 & 13.91 $\pm$ 0.02 & 13.50 $\pm$ 0.02 & \ldots\\
~140 & 16.68 $\pm$ 0.30 & 15.26 $\pm$ 0.03 & 14.01 $\pm$ 0.03 & 13.95 $\pm$ 0.03 & \ldots & 13.05 $\pm$ 0.03 & 12.83 $\pm$ 0.03 & 12.83 $\pm$ 0.03 & 12.52 $\pm$ 0.03 & \ldots\\
~200 & 15.96 $\pm$ 0.30\tablenotemark{a} & 15.87 $\pm$ 0.02 & 14.69 $\pm$ 0.02 & 14.59 $\pm$ 0.02 & \ldots & 13.70 $\pm$ 0.02 & 13.46 $\pm$ 0.02 & 13.50 $\pm$ 0.02 & 13.23 $\pm$ 0.02 & \ldots\\
~230 & 17.41 $\pm$ 0.30 & 16.59 $\pm$ 0.02 & 15.47 $\pm$ 0.02 & 15.39 $\pm$ 0.02 & \ldots & 14.50 $\pm$ 0.02 & 14.27 $\pm$ 0.02 & 14.32 $\pm$ 0.02 & 14.09 $\pm$ 0.02 & \ldots\\
~538 & 18.89 $\pm$ 0.30 & 16.97 $\pm$ 0.02 & 15.87 $\pm$ 0.02 & 15.81 $\pm$ 0.02 & \ldots & 14.91 $\pm$ 0.02 & 14.75 $\pm$ 0.02 & 14.74 $\pm$ 0.02 & 14.55 $\pm$ 0.02 & \ldots\\
~698 & 15.51 $\pm$ 0.30 & 14.16 $\pm$ 0.01 & 12.98 $\pm$ 0.01 & 12.85 $\pm$ 0.01 & \ldots & 11.91 $\pm$ 0.01 & 11.64 $\pm$ 0.01 & 11.65 $\pm$ 0.01 & 10.83 $\pm$ 0.01 & \ldots\\
~784 & 15.81 $\pm$ 0.30 & 13.45 $\pm$ 0.10 & 12.23 $\pm$ 0.10 & 12.12 $\pm$ 0.10 & \ldots & 11.12 $\pm$ 0.10 & 10.83 $\pm$ 0.10 & 10.83 $\pm$ 0.10 & 10.17 $\pm$ 0.10 & \ldots\\
~828 & 15.68 $\pm$ 0.30 & 13.24 $\pm$ 0.10 & 12.36 $\pm$ 0.10 & 12.15 $\pm$ 0.10 & 11.76 $\pm$ 0.10 & 11.12 $\pm$ 0.10 & 10.78 $\pm$ 0.10 & 10.92 $\pm$ 0.10 & 10.25 $\pm$ 0.10 & \ldots\\
~856 & 16.22 $\pm$ 0.30 & 15.38 $\pm$ 0.03 & 14.28 $\pm$ 0.03 & 14.19 $\pm$ 0.03 & \ldots & 13.30 $\pm$ 0.03 & 13.09 $\pm$ 0.03 & 13.08 $\pm$ 0.03 & 12.85 $\pm$ 0.03 & \ldots\\
1075 & 17.00 $\pm$ 0.30 & 16.04 $\pm$ 0.02 & 14.87 $\pm$ 0.02 & 14.83 $\pm$ 0.02 & \ldots & 13.92 $\pm$ 0.02 & 13.67 $\pm$ 0.02 & 13.68 $\pm$ 0.02 & 13.44 $\pm$ 0.02 & \ldots\\
1087 & 15.69 $\pm$ 0.30 & 15.46 $\pm$ 0.02 & 14.41 $\pm$ 0.02 & 14.11 $\pm$ 0.02\tablenotemark{a} & 13.41 $\pm$ 0.02\tablenotemark{a} & 13.18 $\pm$ 0.02\tablenotemark{a} & 13.05 $\pm$ 0.02\tablenotemark{a} & 13.06 $\pm$ 0.02 & 12.57 $\pm$ 0.02 & 12.95 $\pm$ 0.02\\
1146 & 15.18 $\pm$ 0.30 & 13.60 $\pm$ 0.04 & 12.48 $\pm$ 0.04 & 12.36 $\pm$ 0.04 & 11.68 $\pm$ 0.04 & 11.35 $\pm$ 0.04 & 11.16 $\pm$ 0.04 & 11.14 $\pm$ 0.04 & 10.56 $\pm$ 0.04 & 10.82 $\pm$ 0.04\\
1185 & \ldots & 16.31 $\pm$ 0.03 & 15.20 $\pm$ 0.03 & 15.12 $\pm$ 0.03 & 14.56 $\pm$ 0.03 & 14.26 $\pm$ 0.03 & 13.97 $\pm$ 0.03 & 14.02 $\pm$ 0.03 & 13.30 $\pm$ 0.03 & 14.43 $\pm$ 0.03\\
1192 & 17.60 $\pm$ 0.30 & 16.18 $\pm$ 0.02 & 14.76 $\pm$ 0.02 & 14.66 $\pm$ 0.02 & 14.00 $\pm$ 0.02 & 13.61 $\pm$ 0.02 & 13.34 $\pm$ 0.02 & 13.29 $\pm$ 0.02 & 12.57 $\pm$ 0.02 & \ldots\\
1199 & 18.97 $\pm$ 0.30 & 17.68 $\pm$ 0.01 & 16.13 $\pm$ 0.01 & 16.02 $\pm$ 0.01 & 15.31 $\pm$ 0.01 & 14.90 $\pm$ 0.01 & 14.64 $\pm$ 0.01 & 14.57 $\pm$ 0.01 & 13.91 $\pm$ 0.01 & \ldots\\
1242 & 15.55 $\pm$ 0.30 & 13.33 $\pm$ 0.11 & 12.11 $\pm$ 0.11 & 12.01 $\pm$ 0.11 & 11.35 $\pm$ 0.11 & 10.95 $\pm$ 0.11 & 10.73 $\pm$ 0.11 & 10.73 $\pm$ 0.11 & 10.14 $\pm$ 0.11 & \ldots\\
1261 & 15.64 $\pm$ 0.30 & 14.50 $\pm$ 0.02 & 13.42 $\pm$ 0.02 & 13.31 $\pm$ 0.02 & \ldots & 12.45 $\pm$ 0.02 & 12.23 $\pm$ 0.02 & 12.22 $\pm$ 0.02 & 11.75 $\pm$ 0.02 & \ldots\\
1283 & 14.30 $\pm$ 0.30 & 14.24 $\pm$ 0.01 & 13.05 $\pm$ 0.01 & 12.92 $\pm$ 0.01 & 12.25 $\pm$ 0.01 & 11.92 $\pm$ 0.01 & 11.63 $\pm$ 0.01 & 11.65 $\pm$ 0.01 & 10.81 $\pm$ 0.01 & \ldots\\
1355 & 16.84 $\pm$ 0.30 & 15.27 $\pm$ 0.03 & 14.17 $\pm$ 0.03 & 14.09 $\pm$ 0.03 & 13.53 $\pm$ 0.03 & 13.22 $\pm$ 0.03 & 13.01 $\pm$ 0.03 & 13.00 $\pm$ 0.03 & 12.73 $\pm$ 0.03 & \ldots\\
1407 & 16.62 $\pm$ 0.30 & 15.97 $\pm$ 0.02 & 14.94 $\pm$ 0.02 & 14.82 $\pm$ 0.02 & 14.26 $\pm$ 0.02 & 13.91 $\pm$ 0.02 & 13.70 $\pm$ 0.02 & 13.73 $\pm$ 0.02 & 13.14 $\pm$ 0.02 & 13.65 $\pm$ 0.02\\
1422 & 15.79 $\pm$ 0.30 & 14.48 $\pm$ 0.02 & 13.43 $\pm$ 0.02 & 13.34 $\pm$ 0.02 & \ldots & 12.45 $\pm$ 0.02 & 12.17 $\pm$ 0.02 & 12.24 $\pm$ 0.02 & 11.73 $\pm$ 0.02 & \ldots\\
1431 & 17.06 $\pm$ 0.30 & 15.49 $\pm$ 0.03 & 14.27 $\pm$ 0.03 & 14.19 $\pm$ 0.03 & 13.53 $\pm$ 0.03 & 13.17 $\pm$ 0.03 & 12.92 $\pm$ 0.03 & 12.91 $\pm$ 0.03 & 12.50 $\pm$ 0.03 & \ldots\\
1440 & 17.96 $\pm$ 0.30 & 15.75 $\pm$ 0.02 & 14.78 $\pm$ 0.02 & 14.67 $\pm$ 0.02 & \ldots & 13.74 $\pm$ 0.02 & 13.53 $\pm$ 0.02 & 13.53 $\pm$ 0.02 & 12.91 $\pm$ 0.02 & \ldots\\
1489 & 17.45 $\pm$ 0.30 & 16.84 $\pm$ 0.02 & 15.80 $\pm$ 0.02 & 15.76 $\pm$ 0.02 & 15.22 $\pm$ 0.02 & 14.94 $\pm$ 0.02 & 14.73 $\pm$ 0.02 & 14.76 $\pm$ 0.02 & 14.65 $\pm$ 0.02 & \ldots\\
1539 & 14.68 $\pm$ 0.30\tablenotemark{a} & 16.81 $\pm$ 0.02 & 15.68 $\pm$ 0.02 & 15.62 $\pm$ 0.02 & 15.07 $\pm$ 0.02 & 14.74 $\pm$ 0.02 & 14.56 $\pm$ 0.02 & 14.47 $\pm$ 0.02 & 16.09 $\pm$ 0.02 & 14.52 $\pm$ 0.02\\
1545 & 18.30 $\pm$ 0.30 & 15.93 $\pm$ 0.02 & 14.82 $\pm$ 0.02 & 14.69 $\pm$ 0.02 & 14.09 $\pm$ 0.02 & 13.75 $\pm$ 0.02 & 13.56 $\pm$ 0.02 & 13.55 $\pm$ 0.02 & \ldots & 13.53 $\pm$ 0.02\\
1619 & 15.14 $\pm$ 0.30 & 13.36 $\pm$ 0.10 & 12.10 $\pm$ 0.10 & 12.03 $\pm$ 0.10 & \ldots & 11.02 $\pm$ 0.10 & 10.69 $\pm$ 0.10 & 10.71 $\pm$ 0.10 & 10.55 $\pm$ 0.10\tablenotemark{a} & \ldots\\
1627 & \ldots & 16.33 $\pm$ 0.02 & 15.04 $\pm$ 0.02 & 14.94 $\pm$ 0.02 & \ldots & 13.90 $\pm$ 0.02 & 13.65 $\pm$ 0.02 & 13.63 $\pm$ 0.02 & \ldots & \ldots\\
1630 & 15.24 $\pm$ 0.30 & 13.77 $\pm$ 0.10 & 12.41 $\pm$ 0.10 & 12.30 $\pm$ 0.10 & \ldots & 11.27 $\pm$ 0.10 & 10.91 $\pm$ 0.10 & 10.93 $\pm$ 0.10 & 10.36 $\pm$ 0.10 & \ldots\\
1661 & 17.56 $\pm$ 0.30 & 16.88 $\pm$ 0.13 & 16.78 $\pm$ 0.13 & 15.58 $\pm$ 0.13 & \ldots & 14.62 $\pm$ 0.13 & 14.40 $\pm$ 0.13 & \ldots & 14.90 $\pm$ 0.13 & \ldots\\
1826 & 18.30 $\pm$ 0.30 & 16.74 $\pm$ 0.02 & 15.52 $\pm$ 0.02 & 15.47 $\pm$ 0.02 & \ldots & 14.59 $\pm$ 0.02 & 14.37 $\pm$ 0.02 & 14.38 $\pm$ 0.02 & 14.18 $\pm$ 0.02 & \ldots\\
1828 & 17.65 $\pm$ 0.30 & 16.17 $\pm$ 0.02 & 15.08 $\pm$ 0.02 & 14.99 $\pm$ 0.02 & \ldots & 14.07 $\pm$ 0.02 & 13.83 $\pm$ 0.02 & 13.85 $\pm$ 0.02 & 13.27 $\pm$ 0.02 & \ldots\\
1861 & 16.10 $\pm$ 0.30 & 15.28 $\pm$ 0.03 & 14.17 $\pm$ 0.03 & 14.03 $\pm$ 0.03 & \ldots & 13.09 $\pm$ 0.03 & 12.86 $\pm$ 0.03 & 12.87 $\pm$ 0.03 & 12.76 $\pm$ 0.03 & \ldots\\
1871 & 16.88 $\pm$ 0.30 & 15.41 $\pm$ 0.03 & 14.14 $\pm$ 0.03 & 14.02 $\pm$ 0.03 & \ldots & 12.98 $\pm$ 0.03 & 12.72 $\pm$ 0.03 & 12.73 $\pm$ 0.03 & 12.21 $\pm$ 0.03 & \ldots\\
1883 & 14.53 $\pm$ 0.30 & 12.91 $\pm$ 0.10 & 12.06 $\pm$ 0.10 & 11.68 $\pm$ 0.10 & \ldots & 17.99 $\pm$ 0.10\tablenotemark{a} & 10.71 $\pm$ 0.10 & 10.49 $\pm$ 0.10 & 10.19 $\pm$ 0.10 & \ldots\\
1886 & 15.71 $\pm$ 0.30\tablenotemark{a} & 16.24 $\pm$ 0.03 & 15.26 $\pm$ 0.03 & 15.19 $\pm$ 0.03 & \ldots & 14.40 $\pm$ 0.03 & 14.23 $\pm$ 0.03 & 14.25 $\pm$ 0.03 & 13.51 $\pm$ 0.03 & \ldots\\
1910 & 16.93 $\pm$ 0.30 & 15.32 $\pm$ 0.03 & 14.15 $\pm$ 0.03 & 13.99 $\pm$ 0.03 & 13.35 $\pm$ 0.03 & 13.02 $\pm$ 0.03 & 12.74 $\pm$ 0.03 & 12.78 $\pm$ 0.03 & 12.05 $\pm$ 0.03 & \ldots\\
2019 & 16.43 $\pm$ 0.30 & 15.63 $\pm$ 0.02 & 14.46 $\pm$ 0.02 & 14.42 $\pm$ 0.02 & \ldots & 13.51 $\pm$ 0.02 & 13.30 $\pm$ 0.02 & 13.29 $\pm$ 0.02 & 12.89 $\pm$ 0.02 & \ldots\\
2050 & 17.57 $\pm$ 0.30\tablenotemark{a} & 14.33 $\pm$ 0.02 & 14.95 $\pm$ 0.02 & 14.82 $\pm$ 0.02 & \ldots & 14.01 $\pm$ 0.02 & 13.74 $\pm$ 0.02 & 13.83 $\pm$ 0.02 & 13.71 $\pm$ 0.02 & \ldots\\
\enddata
\tablecomments{Magnitudes have not been corrected for Milky Way foreground extinction.}
\tablenotetext{a}{Magnitude excluded from analysis due to image artifacts, failure of \textsc{galfit} to converge on a model fit, or, in the F300W, a non-detection.}
\end{deluxetable*}
\end{longrotatetable}

\subsection{Comparison of Results} \label{subsec:1d2d}

\begin{figure}[htbp]
   \centering
   \epsscale{1.1}
   \plotone{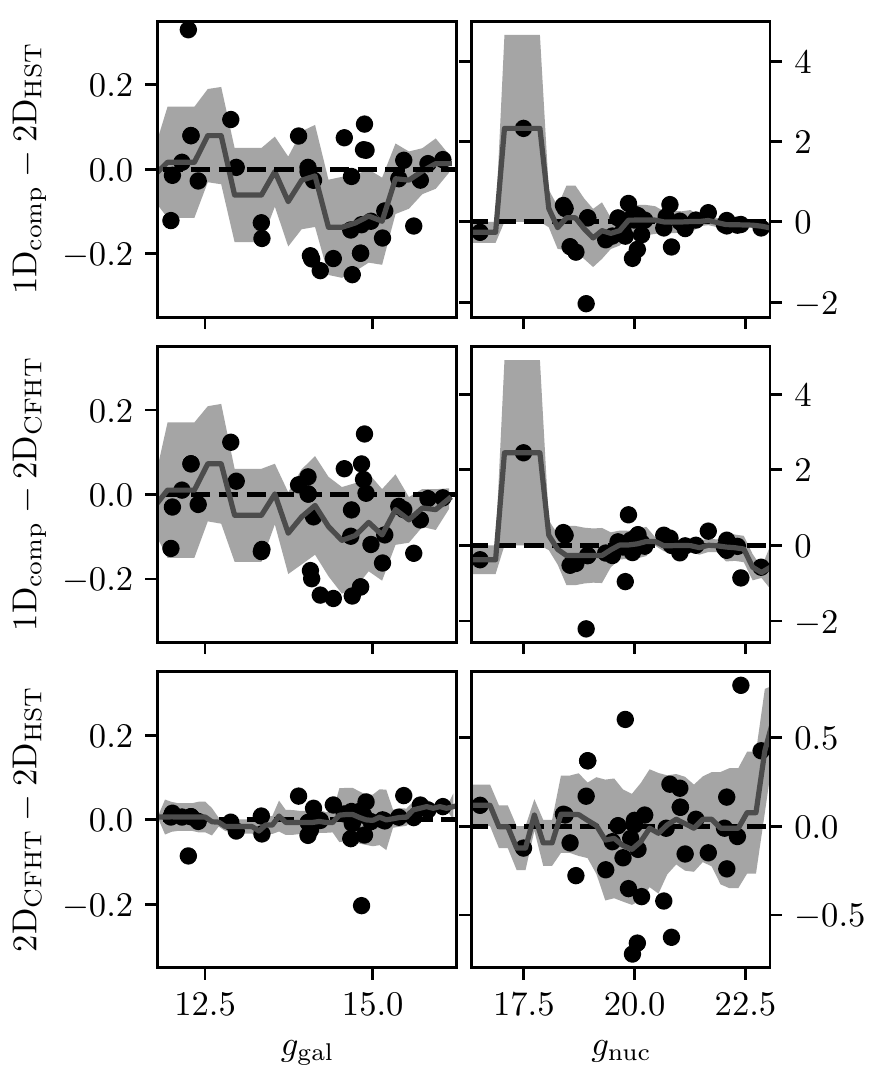} 
   \caption{Differences in magnitudes measured, measured with different methods, plotted as a function of magnitude. The left column shows galaxy magnitudes, while the right shows nuclear magnitudes. The top row shows differences between the 1D HST+CFHT composite profile and 2D HST magnitudes. The middle row compares the 1D HST+CFHT composite profile and 2D CFHT magnitudes, while the bottom row compares the 2D CFHT and HST magnitudes. In each panel, the gray curve shows the average offset as a function of magnitude, with the shaded region showing the associated $1\,\sigma$ scatter.}
   \label{fig:errors}
\end{figure}

Before proceeding, we pause to consider the robustness of our magnitude measurements. For the F475W and F850LP images, it is possible to compare magnitudes derived using the two methods described in \S\ref{subsec:ellipse} and \S\ref{subsec:galfit}. We can also compare the space- and ground-based 2D magnitudes from \S\ref{subsec:galfit} to investigate the effect of resolution on the resulting values. All \textsc{galfit} F475W, F850LP, $g$-band and $z$-band magnitudes were first transformed to the SDSS photometric system following the same procedure outlined in \S\ref{subsec:ellipse}. Comparisons among the various methods in the $g$ bandpass are shown in Figure~\ref{fig:errors}. For the galaxies, we find that both ground- and space-based 2D magnitudes differ slightly from the 1D composite profile magnitudes, with a typical scatter of $\sim$0.1 mag. The 1D magnitudes tend to be systematically brighter in galaxies with $g \gtrsim 14$~mag. This is not surprising as the 1D and 2D models differ in complexity. We find excellent agreement between the space- and ground-based 2D galaxy magnitudes, with a typical scatter of just $\sim$0.02 mag.

For the nuclei, the scatter among all measurement methods is understandably larger. Adjusting the nucleus model, even dramatically, will generally have a negligible effect on the derived galaxy magnitude (or on the total magnitude of the system). When comparing the 1D and 2D magnitudes, the scatter is largest for the brightest nuclei, which have $g \lesssim 19.5$~mag. In this regime, the nuclei are embedded in the brightest, and most structurally complex galaxies in our sample and are thus the most difficult to model well \citep[see][]{Turner:2012aa}. These objects are also expected to be most affected by the number of components used in the model fit, which is a factor when comparing the 1D profile measurements (made with two S\'ersic components) to the 2D method (which may include as many components as needed). However, despite the fairly large scatter of 0.2--0.5 mag, we see no evidence of systematic offsets among the various measurement methods. The good overall agreement between the space- and ground-based 2D magnitudes is especially notable: the nuclei are unresolved in the CFHT imaging, but we are nevertheless able to measure consistent total magnitudes.

\subsection{Adopted Errors} \label{subsec:errors}

The uncertainties provided by \textsc{galfit} are purely statistical and do not account for systematic effects such as the deviation of real galaxies from parametric models. As a result, the errors on \textsc{galfit} parameters are unrealistically small and not well determined \citep{Haussler:2007aa, Lange:2016aa}, so we do not apply these errors to our results. Instead we estimate the errors based on the comparisons in \S\ref{subsec:1d2d}. Our largest source of error for the nuclei is almost certainly due to the modeling process, as even subtle adjustments to the overall galaxy model may affect the distribution of light in its center. As a result, the nucleus parameters can vary significantly. In addition, with only a few pixels in each image providing information on the nucleus, the $\chi^2$ values calculated by \textsc{galfit} are dominated by the quality of the fit of the galaxy components, so determining the best-fit nucleus model can prove challenging. Comparing the results from multiple fitting methods can help quantify the uncertainties in our component magnitudes.

Magnitude differences in the $g$ and $z$ bands appear similar, so we do not expect a strong wavelength dependence on our estimated errors. In fact, the errors should be quite correlated, as all structural parameters have been held fixed at those measured from the F475W images. For the $u^*griz$, $K_s$, F475W, F850LP and F160W images, we adopt errors based on the typical scatter in the bottom left and right panels of Figure~\ref{fig:errors} at each object's $g$ magnitude. However, other factors can contribute to the error budget in case of the WFPC2 imaging: e.g., the low SNR and the limited field of view. In these images, we estimate errors directly from annuli on the images. We treat any signal outside 20 pixels of the nucleus to be noise that dominates the uncertainty on the nucleus measurement. We estimate this uncertainty from an annulus 5 pixels wide and with an inner radius of 20 pixels, centered on the nucleus.

\section{Spectroscopic Analysis} \label{sec:spectrometry}

The spectroscopic observations for our target nuclei are summarized in Table~\ref{tab:spec}. For our spectroscopic analysis, we focus entirely on the nuclei spectra because coverage of the galaxy is usually quite limited. For all three datasets (GMOS, ESI and DEIMOS), we have employed as homogeneous an analysis as possible in order to minimize any differences arising from different techniques. 

\subsection{Data Reduction and Calibration} \label{subsec:specreduction}

Full details on the reduction of the GMOS spectra are given in \citet{Liu:2016aa}. In brief, cosmic rays were removed from the spectra before reducing them using the standard GMOS-IFU pipeline in \texttt{IRAF}, which performs bias subtraction, dark correction, flat fielding, sky subtraction, and wavelength calibration. The spectra were then continuum normalized and stacked. For each galaxy, the galaxy light was modeled as a S\'ersic profile using the signal outside one FWHM of the galaxy center. This profile was then extrapolated into the central region and subtracted to isolate the nucleus spectrum.

The ESI spectra were reduced using the MAuna Kea Echelle Extraction (MAKEE) pipeline \citep{Barlow:1997aa}. MAKEE is designed to extract isolated and unresolved sources, subtract a sky spectrum from the source, and perform wavelength calibrations using a sixth order polynomial fit to each echelle order. While the nuclei are unresolved, they clearly are not isolated. Therefore, MAKEE was adapted to treat the adjacent galaxy spectrum as the ``sky" component during the sky removal step of the pipeline. As before, the nuclei spectra were continuum normalized and then shifted to rest-frame wavelengths.

\begin{figure}[htbp]
   \centering
   \epsscale{1.1}
   \plotone{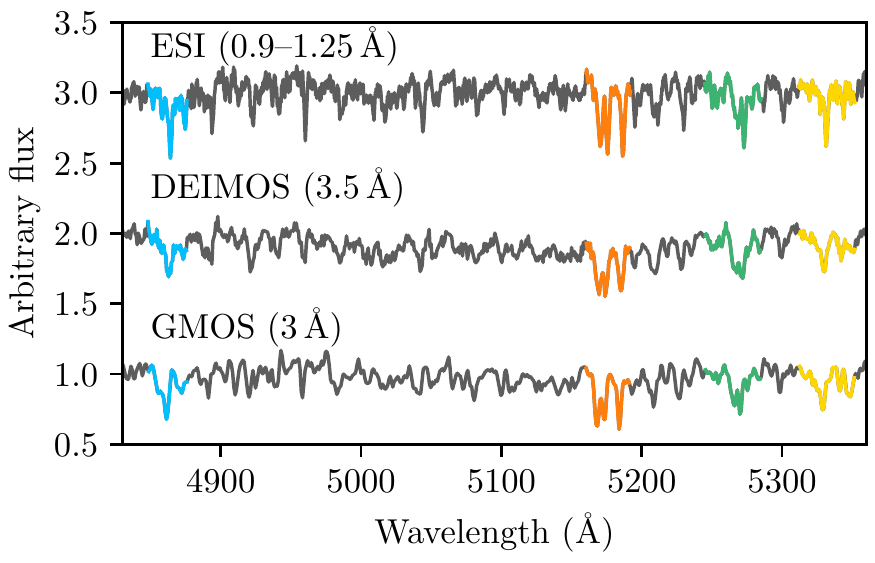} 
   \caption{Wavelength-calibrated, continuum-normalized, rest-frame spectra for VCC~1545 obtained using the ESI, DEIMOS and GMOS instruments. Four indices used to measure ages, metallicities, and $\alpha$-element abundances have been highlighted: H$_\beta$ (light blue), Mgb (orange), Fe5270 (green), and Fe5335 (yellow). Values in parentheses after each instrument name indicate the spectral resolution at 5000\,\AA{}.}
   \label{fig:allspec}
\end{figure}

The \texttt{spec2D} pipeline was used on the DEIMOS spectra to reduce 1D and 2D spectra corrected for flat fielding, sky subtraction, cosmic ray removal, and wavelength calibration. The nucleus light in each spectrum was extracted from the galaxy light by collapsing the 2D spectrum in the wavelength direction and fitting a Gaussian distribution to the resulting light profile. The width of this Gaussian defines an extraction window. Each pixel within this window is weighted by the value of the Gaussian distribution at that position before being added to the final 1D spectrum. Complete details of this reduction process are provided in \citet{Toloba:2016aa}. Once again, the nuclei spectra were then continuum normalized and shifted to rest-frame wavelengths. As an illustration of the data quality, the final, wavelength-calibrated, continuum-normalized spectra for the nucleus of VCC~1545 are shown in Figure~\ref{fig:allspec}.

\subsection{Line Index Measurements} \label{subsec:indices}

Line indices were measured using the IDL script \texttt{Lick-EW}, provided as part of the \texttt{EZ-AGES} code package \citep{Schiavon:2007aa, Graves:2008aa}. \texttt{Lick-EW} measures equivalent widths on the Lick system by broadening the input spectra to Lick/IDS resolution and following the method described in \citet{Worthey:1994aa}. For this sample, we omit corrections for the velocity dispersion, as the low dispersions (${\sim}$50\,km~s$^{-1}$) for dwarf early-type galaxies do not significantly affect the line widths \citep{Kuntschner:2004aa}. While \texttt{Lick-EW} will measure every available Lick index, we selected only the H$_\beta$, Mgb, Fe5270 and Fe5335 lines to estimate ages, metallicities and $\alpha$-element abundances due to their strong features and presence in the wavelength coverage of all three datasets. The measured Lick indices for each nucleus are listed in Table~\ref{tab:indices}. The tabulated values are the mean values of the measurements from each dataset when multiple observations are available.

\begin{table*}[htbp]
   \centering
   \caption{Mean Measured Lick Indices}
   \begin{tabular}{r l c c c c c c c}
      \hline \hline
      VCC & Instruments  & H$_\beta$ & Mgb & Fe5270 & Fe5335 \\
               &                      & (\AA) & (\AA) & (\AA) & (\AA) \\
      (1) & (2) & (3) & (4) & (5) & (6)\\
      \hline
33 & ESI,GMOS & $3.13 \pm 0.41$ & $1.38 \pm 0.42$ & $1.91 \pm 0.47$ & $1.12 \pm 0.57$ \\

200 & ESI & $2.09 \pm 0.90$ & $2.27 \pm 0.70$ & $2.38 \pm 1.45$ & $2.12 \pm 0.81$ \\

230 & ESI & $2.30 \pm 0.61$ & $1.56 \pm 0.79$ & $1.91 \pm 0.59$ & $1.58 \pm 0.86$ \\

538 & ESI & $2.22 \pm 0.73$ & $2.09 \pm 0.82$ & $2.19 \pm 0.67$ & $2.21 \pm 1.03$ \\

1075 & ESI,DEIMOS & $2.26 \pm 0.58$ & $1.17 \pm 0.48$ & $1.69 \pm 0.61$ & $1.52 \pm 0.61$\tablenotemark{a} \\

1185 & ESI,GMOS & $2.29 \pm 0.42$ & $1.70 \pm 0.41$ & $1.96 \pm 0.54$ & $1.71 \pm 0.77$ \\

1192 & ESI & $1.76 \pm 0.71$ & $4.06 \pm 0.97$ & $2.75 \pm 0.79$ & $2.43 \pm 0.97$ \\

1199 & ESI & $1.86 \pm 0.83$ & $4.05 \pm 0.70$ & $3.12 \pm 0.81$ & $3.17 \pm 0.77$ \\

1355 & GMOS & $2.81 \pm 0.42$ & $1.42 \pm 0.48$ & $0.80 \pm 0.58$ & $0.63 \pm 0.66$ \\

1407 & ESI,DEIMOS & $2.66 \pm 0.84$ & $2.04 \pm 0.54$ & $1.55 \pm 0.63$ & $1.22 \pm 1.00$ \\

1440 & ESI & $1.90 \pm 0.76$ & $2.63 \pm 0.92$ & $2.49 \pm 0.62$ & $2.29 \pm 0.96$ \\

1489 & ESI & $2.84 \pm 0.39$ & $0.71 \pm 0.54$ & $1.11 \pm 0.85$ & $1.86 \pm 0.92$ \\

1539 & ESI,DEIMOS,GMOS & $1.98 \pm 0.47$ & $1.76 \pm 0.43$ & $1.33 \pm 0.55$\tablenotemark{c} & $0.91 \pm 0.57$ \\

1545 & ESI,DEIMOS,GMOS & $2.09 \pm 0.26$ & $2.81 \pm 0.32$ & $2.66 \pm 0.44$ & $2.53 \pm 0.31$ \\

1627 & ESI & $1.87 \pm 0.72$ & $3.24 \pm 0.73$ & $2.81 \pm 0.86$ & $2.82 \pm 0.79$ \\

1826 & ESI & $2.26 \pm 0.66$ & $1.96 \pm 0.71$ & $2.25 \pm 1.11$ & $1.55 \pm 1.15$ \\

1828 & ESI,DEIMOS & $2.32 \pm 0.49$\tablenotemark{b} & $1.94 \pm 0.86$ & $2.47 \pm 0.71$ & $1.48 \pm 0.60$ \\

1861 & DEIMOS & $2.33 \pm 0.83$ & $1.22 \pm 0.55$ & $1.79 \pm 0.57$ & $1.60 \pm 0.63$ \\

1871 & DEIMOS & $1.80 \pm 0.45$ & $3.61 \pm 0.20$ & $2.92 \pm 0.21$ & $2.73 \pm 0.24$ \\

2050 & ESI & $2.33 \pm 0.76$ & $1.45 \pm 1.06$ & $2.37 \pm 1.23$ & $1.23 \pm 0.97$ \\
\hline
  \end{tabular}
   \\ \tablenotetext{a}{ESI data excluded from measurement due to non-detection.} \tablenotetext{b}{DEIMOS data excluded from measurement due to a gap in the spectrum.} \tablenotetext{c}{GMOS data excluded from measurement due to non-detection.}
   \label{tab:indices}
\end{table*}

\begin{figure}[htbp]
   \centering
   \epsscale{1.1}
   \plotone{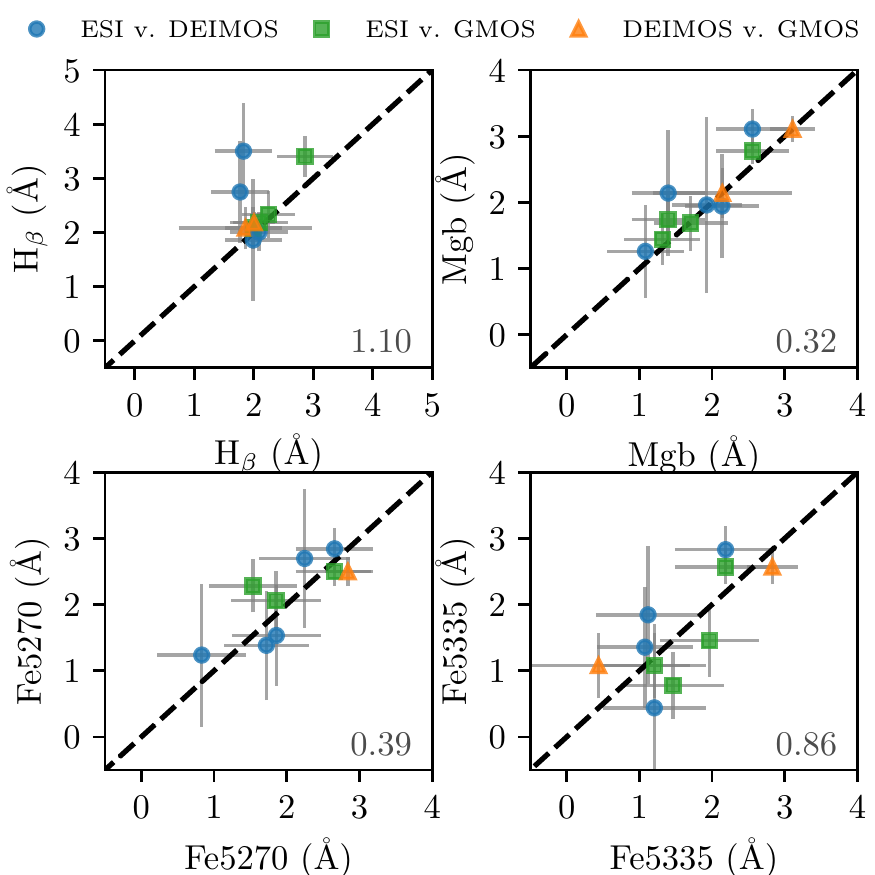} 
   \caption{Comparison of Lick index measurements among the ESI, GMOS and DEIMOS datasets. The blue points show the value measured with ESI data compared to that measured with the DEIMOS data. Similarly, green squares show the ESI values against the GMOS values. In both cases, ESI measurements are plotted along the abscissa. The orange triangles are DEIMOS values compared to GMOS values for the two nuclei (VCC~1539 and VCC~1545) that appear in both datasets. The dashed line in each panel shows the one-to-one relation, while numbers in the bottom right corner show the RMS scatter in \AA{}.}
   \label{fig:indexcompare}
\end{figure}

To investigate the robustness of our line index measurements, we compare results from the three spectroscopic datasets for all nuclei that appear in more than one dataset. These comparisons are shown in Figure~\ref{fig:indexcompare} for the H$_\beta$, Mgb, Fe5270 and Fe5335 indices. There is generally very good agreement among the index measurements, particularly for Fe5270. The H$_\beta$ agreement for DEIMOS data is somewhat poorer for two objects; however, we note that the H$_\beta$ feature is at the extreme blue end of the DEIMOS wavelength range where the detector's efficiency drops quickly, leading to low SNR. Overall, the agreement among datasets suggests that our line index measurements are reliable.

\section{Results} \label{sec:results}

\subsection{Nucleus and Galaxy Colors} \label{subsec:colors}

\begin{figure*}[htbp]
   \centering
   \epsscale{1.1}
   \plotone{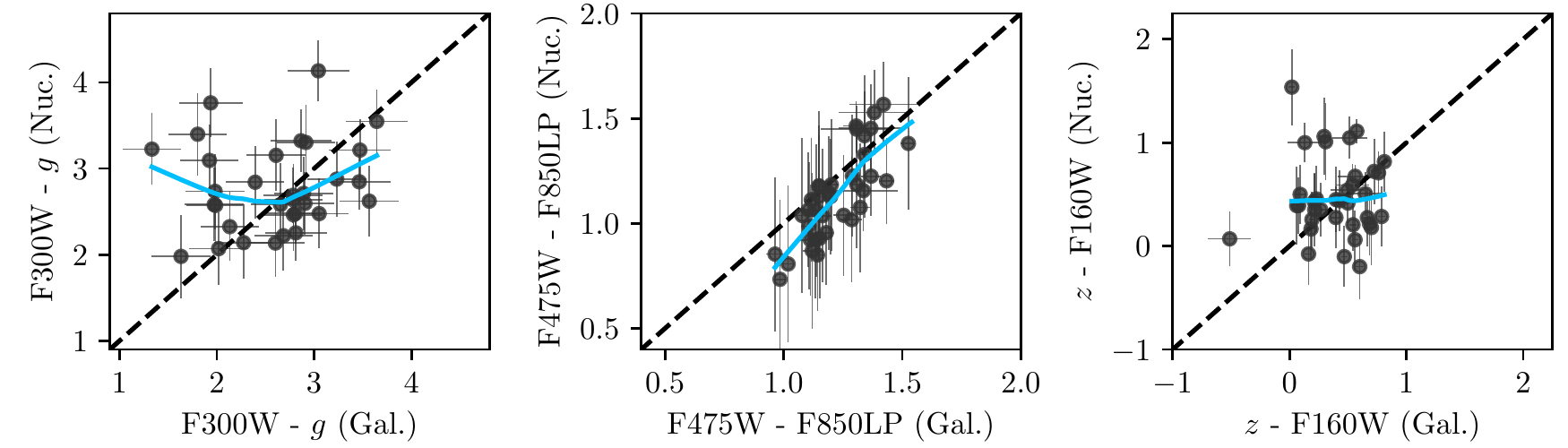} 
   \caption{Nuclei UV-optical (left panel), optical (middle panel) and optical-IR (right panel) colors as a function of galaxy color. In each panel, one to one lines are shown to guide the eye. The blue lines in each panel are LOWESS \citep{Cleveland:1979aa} curves for the data to illustrate the typical color trend for the sample.}
   \label{fig:colors}
\end{figure*}

Many studies have noted that nuclei are typically bluer than their hosts in optical colors \citep[e.g.,][]{Lotz:2004aa,Cote:2006aa,Turner:2012aa}. This is consistent with our findings here. In Figure~\ref{fig:colors} we present (F475W -- F850LP) colors, as well as (F300W -- $g$) and ($z$ -- F160W) colors. We confirm that nuclei are bluer than their hosts in optical colors. This trend does not persist in infrared colors, with no clear color offsets between nuclei and their host galaxies. Unfortunately, given the uncertainties in our UV data it is difficult to draw any strong conclusions about how nuclei compare to their hosts' UV colors. If the nuclei are truly bluer, this could be indicative of some fraction of the stellar content consisting of a young ($\lesssim 2$ Gyr) population. Alternatively, redder UV colors in the nuclei could be a sign of internal dust extinction. Although dwarf early-type galaxies are not expected to have a {\it substantial} dust content, any dust that is present tends to be more concentrated than the stellar content \citep{di-Serego-Alighieri:2013aa}. In that case, nuclei may be more affected by centrally-concentrated galactic dust. Moreover, nuclei may well show larger dust fractions if they have recently formed stars. In this work we have assumed zero internal extinction for both galaxies and nuclei, which may present a bias in the derived population parameters. In the future, high-resolution UV and MIR data could be used to investigate the dust content and extinction in these objects. Further details on dust content and its effect on the results are included in \S\ref{subsec:dust}.

\subsection{SED Fitting and Parameter Estimation} \label{subsec:sedfits}

The results of \S\ref{subsec:1d2d} suggest that the 2D image decomposition technique yields the most homogeneous photometry for each nucleus and galaxy. We therefore use these UV, optical and IR measurements as the basis of our SED analysis for all objects.

The overarching goal of our SED analysis is to estimate the masses, ages and metallicities for each nucleus and its galaxy in a consistent way. To do this, we adopt a Markov Chain Monte Carlo (MCMC) approach. MCMC methods are designed to select $N$ samples from the parameter space in a random walk such that, as $N$ increases, the sample distribution approaches the true probability distribution. We constructed an SED fitting program in Python based on the \texttt{emcee} package \citep{Foreman-Mackey:2013aa}. The \texttt{emcee} algorithm is the affine-invariant MCMC ensemble sampler, which has a few substantial differences from the more common Metropolis-Hastings algorithm. The Metropolis-Hastings method relies on a single Markov chain to probe the parameter space, and each time the chain attempts to jump to a new region of parameter space, the move is rejected or accepted based only on the likelihood of the proposed position relative to the likelihood of the current position. In contrast, the affine-invariant method uses multiple \emph{walkers} to probe parameter space, and the proposed jump for each walker is based on the likelihood of that walker's current position as well as the likelihood of one other randomly selected walker's position. As a result, this method should require fewer tuning parameters and be less sensitive to the initial choices of model parameters.

Predicted stellar population properties can vary substantially based on the models used for comparison in the SED fit \citep{Kannappan:2007aa, Muzzin:2009aa}. When creating a model population, a number of components must be included, and the choices for these ingredients naturally will affect the resulting population. Such components include, for example: an initial mass function (IMF), spectral libraries, stellar isochrones, and the treatment of post-main sequence phases. The last point is especially important for poorly-understood evolutionary phases such as the thermally-pulsating asymptotic giant branch (TP-AGB). Depending on how TP-AGB stars are modeled, the population spectrum for ages $0.3\leq t \leq 2$ Gyr can change dramatically \citep{Maraston:2005aa}. A thorough overview of the uncertainties among models is given in \citet{Conroy:2010aa}.

To minimize the effect of model-specific features, we fit our data to an assortment of model SEDs and look for any results that remain consistent regardless of the adopted model. We also have chosen similar or matching IMFs whenever possible: i.e., a Chabrier or Kroupa IMF, both of which are appropriate for dwarf, low-$\sigma$ early-type galaxies \citep{Cappellari:2012aa, Mentz:2016aa}. We used simple stellar population (SSP) spectra from \citet[hereafter BC03]{Bruzual:2003aa}, \citet[\textsc{p\'egase.2}]{Fioc:1997aa, Fioc:1999aa}, and \citet[hereafter M05]{Maraston:2005aa}. We summarize the features of each model set in Table~\ref{tab:models}.

\begin{table*}[htbp]
   \centering
   \caption{Properties of Population Synthesis Models}
   \begin{tabular}{l c c c c c c}
      \hline \hline
      Model Set & Stellar Tracks & Spectral Libraries & IMF & Metallicities \\
      (1) & (2) & (3) & (4) & (5) \\
      \hline
      BC03 & Padova 1994 & STELIB and BaSeL 3.1 & Chabrier & $0.0001\leq Z \leq0.05$ \\
      M05 & Cassisi and Geneva & BaSeL 3.1 & Kroupa & $0.0004\leq Z \leq0.04$ \\
      \textsc{p\'egase.2} & Padova 1994 & BaSeL 2.2 & Kroupa & $0.0001\leq Z \leq 0.1$ \\
\hline
   \end{tabular}
   \\ 
   \label{tab:models}
\end{table*}

In all cases, the model SEDs are purely stellar in nature: i.e., we assume zero dust content and no nebular emission. Therefore, only three free parameters are needed: stellar mass $M_{\star}$, metallicity $Z$, and age $t$. We assume a flat prior in the mass range $1 \leq M_{\star} \leq 10^{14}\,M_{\odot}$, and across the full metallicity range of each model set (see Table~\ref{tab:models}). We also apply flat priors to ages in the range $0 \leq t \leq 14$ Gyr, eliminating ages older than that of the universe. Model grids were created with 50 metallicity steps and 100 age steps, regularly spaced across the full space of $\log_{10}Z$ and $\log_{10}t$ covered by each model family. To evaluate how well the model reproduces the observed data, we use the log-likelihood equation:
\begin{equation} 
   \ln \mathcal{L} = -\frac{1}{2}\displaystyle\sum_{i=1}^{N}\left[\ln(2\pi\sigma_{obs,i}^2)+\frac{\left(\frac{M_{\star}}{m(t, Z)}F_i(t, Z) - F_{obs,i}\right)^2}{\sigma_{obs,i}^2}\right]
\end{equation}
Here $F_{obs,i}$ is the observed flux in each filter, $\sigma_{obs,i}$ is the flux error, $m(t, Z)$ is the stellar mass of the model, and $F_i(t, Z)$ is the model flux in each filter. Because the model SEDs are normalized to one solar mass at $t=0$, $M_{\star}/m(t, Z)$ is effectively a scale factor that is applied to best match the model to the observed flux.

Before fitting each object, we omitted any magnitudes that were obvious outliers based on visual inspection of the SED. These outliers represent cases in which the \textsc{galfit} model clearly failed to converge on a reasonable fit, and affected, at most, a single data point for each observed SED. These values were often at least an order of magnitude brighter or fainter than the surrounding measurements and could be confidently excluded.

\begin{figure}[htbp]
   \centering
   \epsscale{1.1}
   \plotone{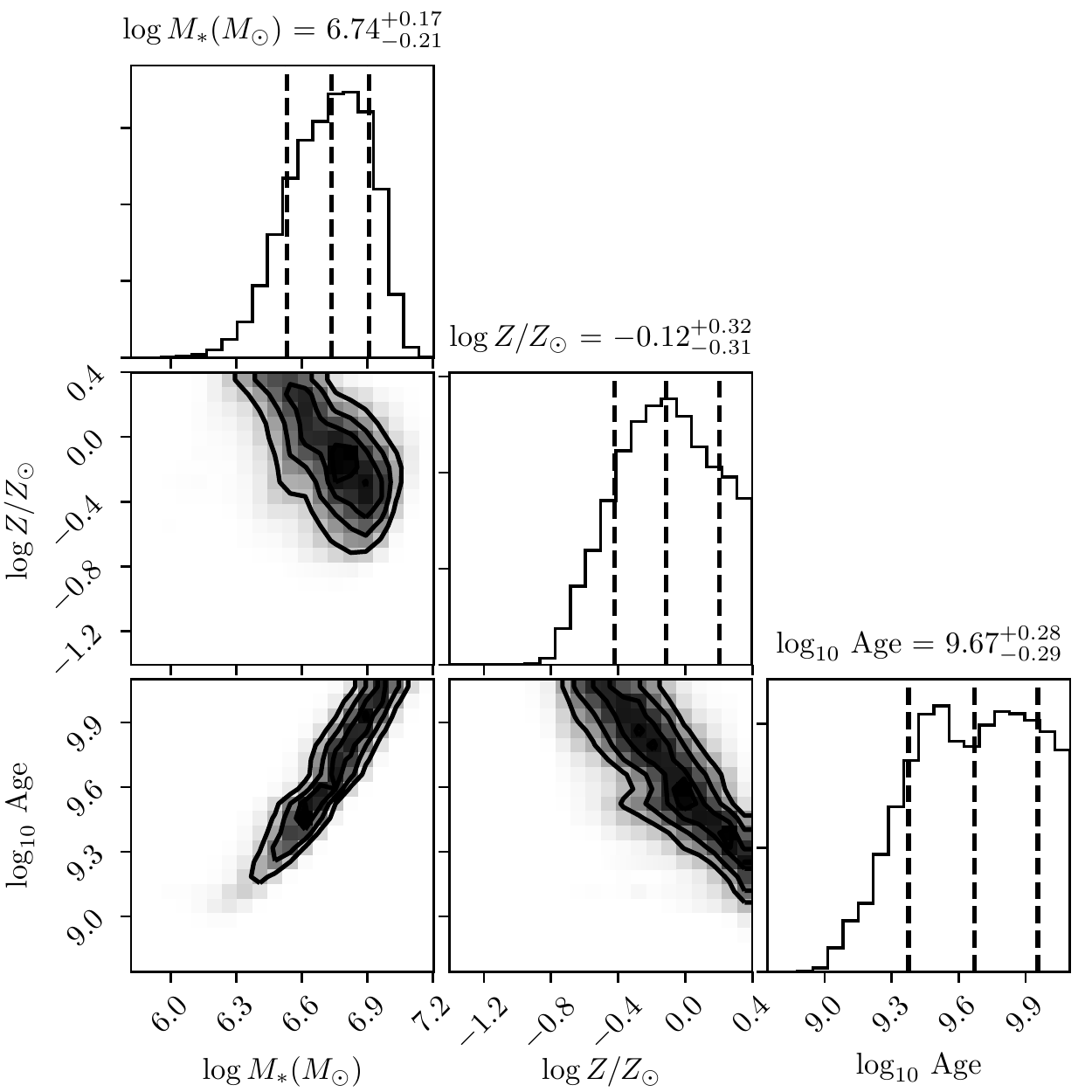} 
   \caption{Stellar masses, metallicities, and ages derived using the BC03 models with a Chabrier IMF for the nucleus in VCC~1422. Plots along the diagonal show the collapsed individual posterior probability distributions for the mass (top left), metallicity (middle), and age (bottom right), while other panels show joint probability distributions. The median value for each parameter is quoted along the top of the diagonal, with error bars determined from the 16$^{th}$ and 84$^{th}$ percentiles.}
   \label{fig:1422pdfs}
\end{figure}

\begin{figure}[htbp]
   \centering
   \epsscale{1.1}
   \plotone{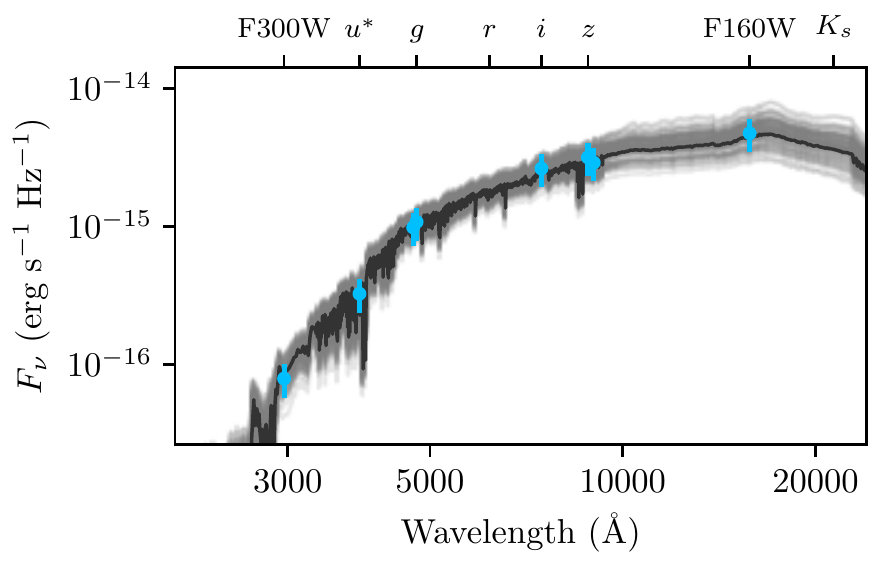} 
   \caption{Best-fit model (in black) compared to the observed SED for the nucleus of VCC~1422 (blue points). The model SED has been generated using the best-fit parameters shown in Figure~\ref{fig:1422pdfs}. 100 models have been randomly extracted from the posterior distribution and plotted in gray to demonstrate the uncertainty on the best-fit model.}
   \label{fig:1422sed}
\end{figure}

For each nucleus and galaxy, 500 walkers with 1000 steps through parameter space (after a burn-in period of 500 steps) provided a total of 500,000 samplings to generate posterior probability distributions for each parameter. Figure~\ref{fig:1422pdfs} shows an example of the joint and individual posteriors for the VCC~1422 nucleus and the selected best-fit parameters. We extracted the median of each distribution as the best-fit value, and use the 16$^{th}$ and 84$^{th}$ percentiles as 1\,$\sigma$ uncertainties. The best-fit SED using these values is shown in Figure~\ref{fig:1422sed}, again for the VCC~1422 nucleus. For comparison, we also calculate the $\chi^2$ value for each parameter combination, and determine an additional set of best-fit parameters based on $\chi^2$ minimization. This technique is more consistent with previous work, but produces systematically different values from the median best-fit parameters, which are generally older and less metal rich. We find that the $\chi^2$ value changes very minimally due to the age-metallicity degeneracy, such that the $\chi^2$ value of each set of median parameters is only marginally larger than the minimum $\chi^2$. With this in mind, we adopt the median parameters as our final best-fit parameters, as we believe that the MCMC technique and error estimation better accounts for the degeneracy. The resulting masses, metallicities and ages estimated using the BC03 models for nuclei and galaxies are listed in Table~\ref{tab:sedagez}.

\begin{figure*}[htbp]
   \centering
   \epsscale{1.1}
   \plotone{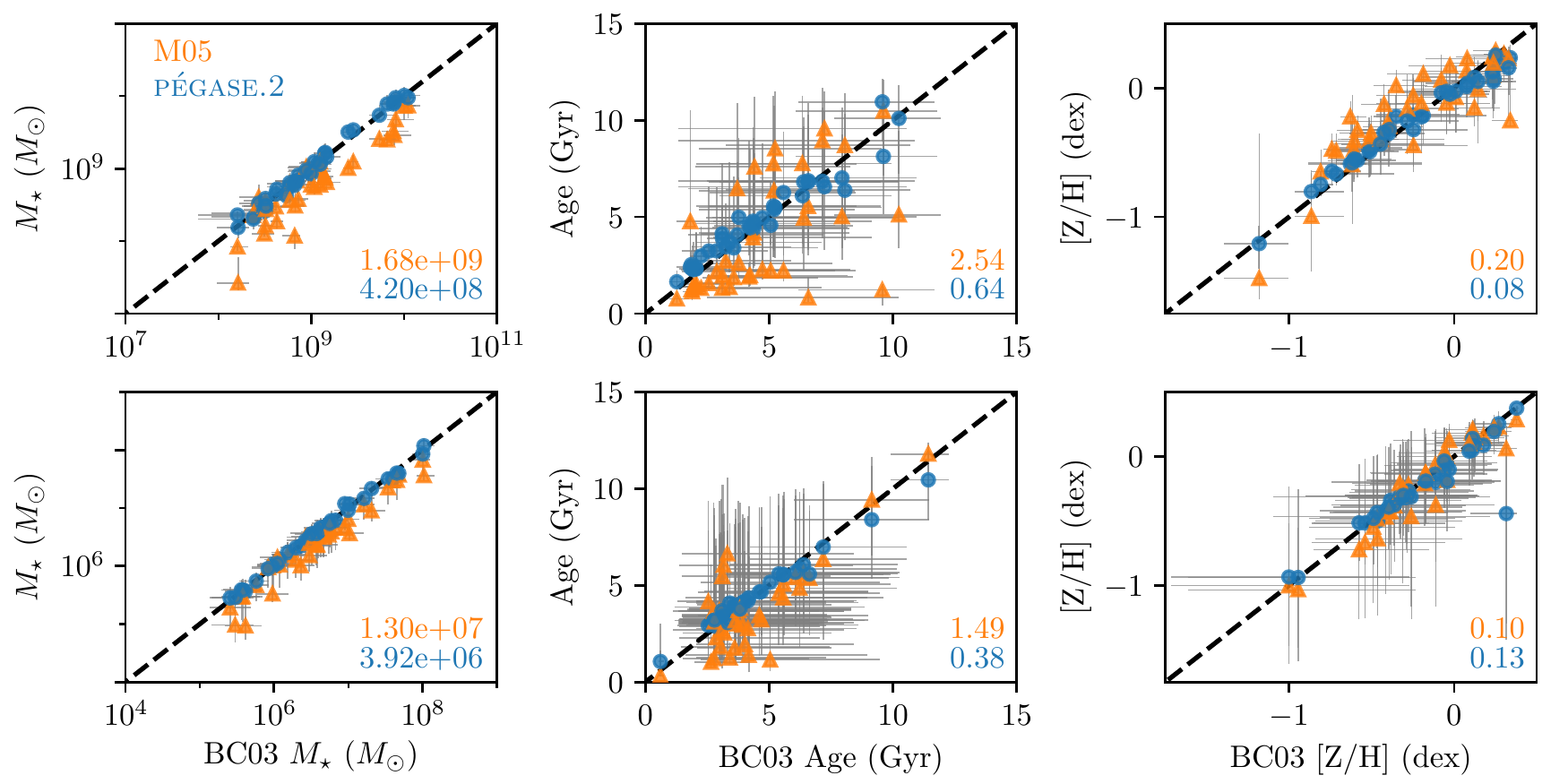} 
   \caption{Comparison of stellar masses (left column), ages (middle column), and metallicities (right column) found using different SSP models in the SED fitting analysis. Values from the BC03 models (plotted along the abscissa, listed in Table~\ref{tab:sedagez}, and adopted as our preferred values) are plotted against the M05 \citep{Maraston:2005aa} and \textsc{p\'egase.2} \citep{Fioc:1997aa, Fioc:1999aa} values (orange triangles and blue circles, respectively). Results are shown for both the galaxies (top row) and nuclei (bottom row). The dashed line in each panel shows the one-to-one relation, while the numbers in the bottom right corner show the RMS scatter, in the same units as each panel, between M05 and BC03 (in orange) and \textsc{p\'egase.2} and BC03 (in blue).}
   \label{fig:compare_sedfits}
\end{figure*}

By comparing results from different models, we can explore possible systematic differences among the calculated parameters. A comparison of the stellar population properties for galaxies and nuclei derived from our three models is shown in Figure~\ref{fig:compare_sedfits}. The BC03 values are adopted as a baseline on the abscissa, with the M05 or \textsc{p\'egase.2} value on the ordinate. There are no strong systematic differences among the derived parameters. All models produce masses and metallicities that are in very good agreement within the uncertainties. The BC03 and \textsc{p\'egase.2} results are quite consistent with each other, although the M05 models do tend to predict a different range of ages. While the BC03 and \textsc{p\'egase.2} models tend to produce ages between 5 and 12 Gyr, the M05 ages can be as young as 1 Gyr. This is likely an effect of the treatment of the TP-AGB population and other post-main sequence evolutionary stages. The unique fuel consumption model employed by M05 for these stages of stellar evolution means that the contribution of these red stars to the population's total becomes highly significant at ages $\sim$1--3 Gyr. As a result, the M05 models can match relatively redder observed colors with younger populations compared to the other models.

\begin{table*}[htbp]
   \centering
   \caption{Masses, Metallicities and Ages Derived from SED Fitting Using BC03}
   \begin{tabular}{r c c c c c c}
      \hline \hline
      VCC & $\log_{10}M_{\star,\textrm{gal}}$ & $\log_{10}M_{\star,\textrm{nuc}}$ & T$_{\textrm{gal}}$ & T$_{\textrm{nuc}}$  & $\log_{10}Z/Z_{\odot,\textrm{gal}}$ & $\log_{10}Z/Z_{\odot,\textrm{nuc}}$ \\
       & ($M_\odot$) &  ($M_\odot$) & (Gyr) & (Gyr) & (dex) & (dex) \\
      (1) & (2) & (3) & (4) & (5) & (6) & (7)\\
      \hline
33 & $8.72_{-0.27}^{+0.15}$ & $5.67_{-0.24}^{+0.15}$ & $6.63_{-4.28}^{+4.86}$ & $6.32_{-3.94}^{+4.91}$ & $-0.85_{-0.27}^{+0.33}$ & $-0.32_{-0.27}^{+0.47}$ \\

140 & $9.19_{-0.20}^{+0.13}$ & $5.68_{-0.21}^{+0.19}$ & $7.14_{-3.89}^{+4.21}$ & $4.00_{-2.14}^{+4.89}$ & $-0.48_{-0.19}^{+0.31}$ & $-0.50_{-0.25}^{+0.38}$ \\

200 & $8.98_{-0.17}^{+0.10}$ & $5.75_{-0.18}^{+0.13}$ & $8.19_{-4.19}^{+3.75}$ & $7.35_{-3.60}^{+4.04}$ & $-0.38_{-0.20}^{+0.29}$ & $-0.16_{-0.36}^{+0.30}$ \\

230 & $8.56_{-0.22}^{+0.15}$ & $6.67_{-0.25}^{+0.13}$ & $6.63_{-3.70}^{+4.73}$ & $7.97_{-4.84}^{+4.18}$ & $-0.61_{-0.17}^{+0.31}$ & $-0.78_{-0.16}^{+0.25}$ \\

538 & $8.48_{-0.15}^{+0.09}$ & $6.48_{-0.20}^{+0.13}$ & $8.90_{-4.05}^{+3.34}$ & $6.70_{-3.74}^{+4.35}$ & $-0.42_{-0.17}^{+0.27}$ & $-0.33_{-0.26}^{+0.34}$ \\

698 & $9.68_{-0.19}^{+0.14}$ & $6.86_{-0.19}^{+0.12}$ & $6.11_{-3.22}^{+4.36}$ & $7.18_{-3.82}^{+4.12}$ & $-0.19_{-0.28}^{+0.33}$ & $-0.36_{-0.23}^{+0.29}$ \\

784 & $10.23_{-0.11}^{+0.07}$ & $7.72_{-0.10}^{+0.07}$ & $10.69_{-3.20}^{+2.18}$ & $10.68_{-2.92}^{+2.12}$ & $0.16_{-0.14}^{+0.14}$ & $0.20_{-0.14}^{+0.12}$ \\

828 & $10.15_{-0.13}^{+0.09}$ & $7.65_{-0.15}^{+0.10}$ & $9.25_{-3.63}^{+3.13}$ & $8.46_{-3.69}^{+3.62}$ & $0.12_{-0.17}^{+0.17}$ & $0.09_{-0.20}^{+0.19}$ \\

856 & $9.02_{-0.25}^{+0.15}$ & $6.93_{-0.23}^{+0.16}$ & $6.37_{-3.82}^{+4.94}$ & $5.58_{-3.16}^{+5.02}$ & $-0.72_{-0.20}^{+0.32}$ & $-0.50_{-0.22}^{+0.38}$ \\

1075 & $8.78_{-0.24}^{+0.15}$ & $6.22_{-0.27}^{+0.16}$ & $6.04_{-3.46}^{+5.01}$ & $6.13_{-3.97}^{+5.54}$ & $-0.63_{-0.20}^{+0.32}$ & $-0.73_{-0.21}^{+0.39}$ \\

1087 & $9.14_{-0.16}^{+0.11}$ & $6.94_{-0.15}^{+0.10}$ & $7.98_{-3.83}^{+3.75}$ & $8.68_{-3.59}^{+3.46}$ & $-0.34_{-0.20}^{+0.25}$ & $0.09_{-0.18}^{+0.17}$ \\

1146 & $9.93_{-0.18}^{+0.11}$ & $8.04_{-0.23}^{+0.16}$ & $7.36_{-3.73}^{+3.93}$ & $6.16_{-3.44}^{+5.12}$ & $-0.23_{-0.21}^{+0.25}$ & $-0.61_{-0.16}^{+0.30}$ \\

1185 & $8.70_{-0.18}^{+0.11}$ & $6.38_{-0.25}^{+0.17}$ & $8.81_{-4.48}^{+3.53}$ & $5.29_{-3.06}^{+5.34}$ & $-0.79_{-0.15}^{+0.17}$ & $-0.57_{-0.21}^{+0.35}$ \\

1192 & $9.09_{-0.16}^{+0.10}$ & $7.44_{-0.19}^{+0.13}$ & $7.77_{-3.58}^{+4.04}$ & $6.32_{-3.41}^{+4.26}$ & $-0.02_{-0.23}^{+0.23}$ & $-0.25_{-0.27}^{+0.34}$ \\

1199 & $8.60_{-0.14}^{+0.10}$ & $7.30_{-0.15}^{+0.10}$ & $8.31_{-3.41}^{+3.69}$ & $8.49_{-3.70}^{+3.67}$ & $0.04_{-0.21}^{+0.20}$ & $0.03_{-0.20}^{+0.21}$ \\

1242 & $10.04_{-0.19}^{+0.13}$ & $8.11_{-0.11}^{+0.08}$ & $6.42_{-3.44}^{+4.21}$ & $9.70_{-3.21}^{+2.81}$ & $-0.30_{-0.26}^{+0.32}$ & $0.08_{-0.14}^{+0.16}$ \\

1261 & $9.37_{-0.23}^{+0.16}$ & $6.31_{-0.29}^{+0.30}$ & $5.20_{-2.94}^{+5.23}$ & $3.25_{-2.01}^{+7.16}$ & $-0.53_{-0.24}^{+0.38}$ & $-0.71_{-0.46}^{+0.64}$ \\

1283 & $9.64_{-0.19}^{+0.15}$ & $7.13_{-0.15}^{+0.10}$ & $5.85_{-3.09}^{+4.32}$ & $8.42_{-3.74}^{+3.68}$ & $-0.26_{-0.29}^{+0.35}$ & $0.05_{-0.20}^{+0.20}$ \\

1355 & $9.10_{-0.22}^{+0.14}$ & $6.25_{-0.23}^{+0.15}$ & $6.81_{-3.94}^{+4.41}$ & $5.92_{-3.29}^{+4.96}$ & $-0.52_{-0.18}^{+0.32}$ & $-0.59_{-0.19}^{+0.33}$ \\

1407 & $8.83_{-0.20}^{+0.14}$ & $6.41_{-0.26}^{+0.18}$ & $6.97_{-3.76}^{+4.29}$ & $4.14_{-2.55}^{+5.10}$ & $-0.51_{-0.17}^{+0.29}$ & $-0.37_{-0.29}^{+0.49}$ \\

1422 & $9.39_{-0.22}^{+0.15}$ & $6.86_{-0.17}^{+0.11}$ & $5.59_{-3.09}^{+4.85}$ & $7.57_{-3.93}^{+3.99}$ & $-0.49_{-0.23}^{+0.36}$ & $-0.27_{-0.23}^{+0.29}$ \\

1431 & $9.21_{-0.17}^{+0.11}$ & $6.81_{-0.24}^{+0.16}$ & $7.77_{-3.98}^{+3.82}$ & $5.56_{-3.11}^{+5.16}$ & $-0.32_{-0.22}^{+0.28}$ & $-0.59_{-0.20}^{+0.35}$ \\

1440 & $8.97_{-0.20}^{+0.12}$ & $6.69_{-0.06}^{+0.05}$ & $7.19_{-3.92}^{+4.20}$ & $13.26_{-0.79}^{+0.40}$ & $-0.21_{-0.25}^{+0.31}$ & $0.38_{-0.01}^{+0.01}$ \\

1489 & $8.39_{-0.25}^{+0.13}$ & $5.65_{-0.27}^{+0.17}$ & $8.15_{-4.74}^{+3.88}$ & $6.81_{-4.22}^{+4.79}$ & $-1.08_{-0.43}^{+0.27}$ & $-1.19_{-0.56}^{+0.38}$ \\

1539 & $8.23_{-0.21}^{+0.18}$ & $6.19_{-0.29}^{+0.18}$ & $5.69_{-2.92}^{+4.14}$ & $6.30_{-4.21}^{+5.61}$ & $-1.81_{-0.33}^{+0.41}$ & $-0.79_{-0.22}^{+0.34}$ \\

1545 & $8.85_{-0.25}^{+0.15}$ & $6.18_{-0.18}^{+0.17}$ & $6.46_{-3.75}^{+4.80}$ & $4.34_{-2.12}^{+4.59}$ & $-0.69_{-0.18}^{+0.28}$ & $-0.06_{-0.35}^{+0.33}$ \\

1619 & $9.98_{-0.21}^{+0.14}$ & $7.55_{-0.09}^{+0.07}$ & $6.44_{-3.47}^{+4.36}$ & $11.46_{-2.53}^{+1.68}$ & $-0.51_{-0.21}^{+0.31}$ & $0.31_{-0.10}^{+0.06}$ \\

1627 & $8.90_{-0.17}^{+0.12}$ & $6.35_{-0.34}^{+0.25}$ & $7.51_{-3.85}^{+4.04}$ & $4.78_{-3.51}^{+6.05}$ & $-0.29_{-0.27}^{+0.33}$ & $-1.23_{-0.67}^{+0.79}$ \\

1630 & $10.04_{-0.16}^{+0.10}$ & $7.78_{-0.11}^{+0.07}$ & $8.17_{-3.93}^{+3.78}$ & $10.51_{-3.34}^{+2.32}$ & $-0.11_{-0.22}^{+0.26}$ & $0.24_{-0.14}^{+0.10}$ \\

1661 & $8.40_{-0.21}^{+0.19}$ & $6.43_{-0.14}^{+0.09}$ & $6.66_{-3.29}^{+4.29}$ & $9.80_{-3.94}^{+2.76}$ & $-0.05_{-0.51}^{+0.30}$ & $-0.54_{-0.13}^{+0.21}$ \\

1826 & $8.50_{-0.24}^{+0.15}$ & $6.72_{-0.23}^{+0.15}$ & $6.48_{-3.74}^{+4.82}$ & $6.54_{-3.77}^{+4.77}$ & $-0.72_{-0.18}^{+0.29}$ & $-0.60_{-0.18}^{+0.33}$ \\

1828 & $8.79_{-0.20}^{+0.14}$ & $6.11_{-0.23}^{+0.15}$ & $6.40_{-3.55}^{+4.55}$ & $6.18_{-3.44}^{+5.00}$ & $-0.40_{-0.24}^{+0.33}$ & $-0.68_{-0.18}^{+0.28}$ \\

1861 & $9.10_{-0.23}^{+0.15}$ & $6.71_{-0.24}^{+0.16}$ & $6.31_{-3.59}^{+4.77}$ & $5.24_{-3.03}^{+5.51}$ & $-0.62_{-0.19}^{+0.31}$ & $-0.61_{-0.22}^{+0.38}$ \\

1871 & $9.27_{-0.17}^{+0.12}$ & $7.17_{-0.19}^{+0.13}$ & $7.30_{-3.69}^{+3.96}$ & $6.38_{-3.41}^{+4.39}$ & $-0.26_{-0.23}^{+0.28}$ & $-0.19_{-0.27}^{+0.33}$ \\

1883 & $10.09_{-0.20}^{+0.14}$ & $4.89_{-0.46}^{+0.21}$ & $6.71_{-3.68}^{+4.24}$ & $5.76_{-4.91}^{+5.34}$ & $-0.47_{-0.22}^{+0.31}$ & $-1.27_{-0.73}^{+0.98}$ \\

1886 & $8.56_{-0.27}^{+0.17}$ & $5.87_{-0.20}^{+0.10}$ & $5.92_{-3.87}^{+5.33}$ & $9.37_{-4.74}^{+3.14}$ & $-0.91_{-0.47}^{+0.42}$ & $-1.16_{-0.40}^{+0.29}$ \\

1910 & $9.19_{-0.20}^{+0.15}$ & $6.89_{-0.20}^{+0.14}$ & $5.85_{-3.14}^{+4.48}$ & $7.04_{-3.83}^{+4.36}$ & $-0.32_{-0.28}^{+0.35}$ & $-0.55_{-0.17}^{+0.30}$ \\

2019 & $8.94_{-0.22}^{+0.15}$ & $6.69_{-0.20}^{+0.13}$ & $5.58_{-3.05}^{+5.08}$ & $6.89_{-3.77}^{+4.35}$ & $-0.59_{-0.22}^{+0.34}$ & $-0.40_{-0.23}^{+0.32}$ \\

2050 & $8.94_{-0.10}^{+0.07}$ & $5.65_{-0.21}^{+0.15}$ & $11.04_{-3.99}^{+2.07}$ & $6.70_{-3.74}^{+4.59}$ & $-0.11_{-0.16}^{+0.19}$ & $-0.23_{-0.48}^{+0.41}$ \\
\hline
   \end{tabular}
   \\ 
   \label{tab:sedagez}
\end{table*}

\subsection{A Note on Dust Effects}\label{subsec:dust}

Given the potential impact of internal dust on the measured colors and population parameters, we have investigated the degree to which dust could alter the estimated stellar population parameters. We have done so by carrying out an independent set of SED fits with the addition of a free parameter, $E(B-V)$. The model fluxes were reddened according to the attenuation law in \citet{Calzetti:2000aa}. On average, the best-fit $E(B -V)$ for the nuclei is $\sim$0.3~mag, and $\sim$0.1~mag for the galaxies. There are, however, large uncertainties on these values, as well as substantial degeneracies with the best-fit ages. The galaxy results appear to be relatively unchanged compared to the dust-free models, with generally only slight decreases (1-2 Gyr) in age after including the reddening parameter. The galaxy metallicities remain roughly the same, around solar values, but lower metallicity objects become even more metal poor once reddening is included. The nucleus metallicities exhibit a similar change, while the shift in nucleus ages is more dramatic with reddening: i.e, the nuclei as a population become clustered around 0.5-1.5 Gyr. However, the uncertainties on these results do not rule out the possibility of older ages (3-6 Gyr) --- consistent with the dust-free results.

In a parallel approach, we explored using FIR data as a means of constraining the range of plausible $E(B-V)$ values for our sample objects. In the Herschel Virgo Cluster Survey \citep[HeViCS;][]{di-Serego-Alighieri:2013aa}, only one of our program galaxies, VCC~1619, had a dust detection, with an estimated dust mass (25.5$\pm$5.5) $\times$ $10^4~M_\odot$ and dust temperature 21.7$\pm$1.0~K. For the remainder of the sample, we assume an upper limit on the dust mass of 2.44 $\times$ $10^3~M_\odot$, based on stacking the images of 227 early type dwarfs with non-detections in HeViCS \citep{De-Looze:2010aa}.

For the simple scenario of a foreground dust screen in front of a stellar point source (much like a nucleus with foreground galactic dust), the dust optical depth at a given wavelength, $\tau_{\lambda}$, is proportional to the reddening $E(B-V)$, following the equation
$$\tau_{\lambda} = 0.921E(B-V){\kappa}({\lambda}),$$
where $\kappa({\lambda})$ is value of the attenuation curve at a given wavelength. The optical depth depends on the dust mass density, $\rho_d$, the path length through the dust, $L$, and the dust absorption coefficient, $k_\lambda$, such that $\tau_\lambda = {\rho_d}Lk_\lambda$. The dust mass density is determined by assuming various areas and values of $L$ to create volumes in which the upper limit dust mass can be distributed. In the following calculations, we adopt the extinction curve from \citet{Calzetti:2000aa} and the absorption coefficients tabulated in \citet{Li:2001aa}.

To estimate $E(B-V)$ in our program objects, we consider two cases. In the simplest scenario, we assume that all of the dust is evenly distributed across the galaxy in some sort of foreground screen. Assuming a typical dE effective radius of $\sim$1~kpc, and therefore a screen of 1~kpc $\times$ 1~kpc, this produces a negligibly small reddening of $E(B-V)$ = 0.0014~mag. On the other hand, we find an (extreme) upper limit on $E(B-V)$ if we assume that all of the dust is contained in a foreground cylinder with a radius of $\sim$5~pc (i.e., the size of a typical nucleus). This leads to an upper limit of $E(B-V)$ = 18\,mag, which is obviously not a useful constraint on the plausible reddening values for these objects.

In a more realistic approach, we assume that the dust has uniform densities throughout the galaxy and nucleus. To estimate this density, we assume that $0.3\%$ of the total dust mass is contained within a spherical volume with a nucleus-sized radius of 5~pc --- in other words, the dust mass follows the same nucleus-galaxy stellar mass relation. The value of $E(B-V)$ can be calculated for various path lengths through a foreground screen with this density of dust. A dust screen 5~pc thick generates 0.04~mag of reddening, while a screen of 1~kpc (the typical $r_e$ for early-type Virgo dwarfs) causes 8~mag of reddening. To produce $E(B-V) \sim 0.3$~mag, as suggested by the SED fits for the nuclei, the dust screen must be only 30-40~pc thick.

To summarize, we find that dust can produce non-negligible amounts of reddening, even in objects considered to contain minimal dust, like our program objects. The severity of the effect on the resulting stellar population parameters depends strongly on the distribution of this dust. Unfortunately, without high-resolution, deep imaging in FIR bands, we can only guess at the intrinsic dust distribution. We note, however, that the good agreement between the dust-free metallicity and the spectroscopic metallicities (see \S\ref{subsec:specphotcompare} for details) suggests that the assumption of minimal dust is reasonable for these objects, as the Lick indices --- given the narrow wavelength coverage of each absorption feature --- should be relatively unaffected by reddening.

\subsection{Measurement of Spectroscopic Parameters}  \label{subsec:specpars}

We use the Lick indices measured in \S\ref{subsec:indices} to estimate an age, [Fe/H] and [Mg/Fe] for each nucleus using the code \texttt{EZ-AGES} \citep{Graves:2008aa}. This code uses the models from \citet{Schiavon:2007aa} which probe ages $0.1 \leq t \leq 15.8$ Gyr and metallicities $-1.3 \leq \mbox{[Fe/H]} \leq 0.2$ for the solar-scaled isochrone that we chose for our analysis. \texttt{EZ-AGES} uses a sequential grid inversion technique to determine varied abundance ratios for Fe, Mg, C, N and Ca (with options to specify ratios for O, Na, Si, Ti and Cr); however, with the four Lick indices we have available, we can only determine [Fe/H] and [Mg/Fe]. Following the default settings of \texttt{EZ-AGES}, C, N, Ca, O and Cr are fixed to solar values (\(\text{[X/Fe]} = 0\)), and Na, Si and Ti are fixed to Mg (\(\text{[X/Fe]} = \text{[Mg/Fe]}\)).

In brief, \texttt{EZ-AGES} first calculates an initial guess for the population age and [Fe/H] using a model grid of H$_\beta$ and $\langle$Fe$\rangle$, an average of Fe5270 and Fe5335. It then creates another model grid using $\langle$Fe$\rangle$ and Mgb to probe [Mg/Fe], and adjusts the [Mg/Fe] until this grid fits a model with age and [Fe/H] values sufficiently similar to the fiducial estimates. For our purposes, the code stops here because we provide no other indices to constrain other element ratios. Lastly, the code computes errors on age and [Fe/H] by shifting the H$_\beta$, Fe5270 and Fe5335 indices by their error bars and repeating the grid inversion. For other element abundances, errors are determined by uncertainties on the fiducial age and [Fe/H] as well as errors on the relevant Lick indices.
 
There are a few caveats to the results of this analysis. In some nuclei, low SNR can influence the pseudo-continuum estimates surrounding the Lick indices, introducing a bias to the measurements. Additional uncertainty arises due to smoothing the spectra to the Lick resolution, which may not be matched perfectly. Finally, even though there is evidence indicating these objects may be $\alpha$-enhanced \citep{Liu:2016aa}, we have used the solar-scaled isochrones in \texttt{EZ-AGES} rather than the $\alpha$-enhanced versions because those have been found to predict ages that are too old \citep{Weiss:2006aa, Schiavon:2007aa}. The fitting process in \texttt{EZ-AGES} can still produce super-solar enhancements for \emph{individual} elements such as Mg (and the other elements set to follow the [Mg/Fe] value); however, the abundances of the remaining elements will remain solar-scaled as we do not provide any index measurements for those elements. This may be a non-physical model for $\alpha$-enhanced objects and introduces uncertainty in our estimates.

For consistency with the photometric results, we quote [Z/H] based on the estimated [Fe/H] and [Mg/Fe] values using the equation
\begin{equation} \label{eq:zconv}
   \text{[Z/H]} = \text{[Fe/H]} + 0.94\,\text{[$\alpha$/Fe]},
   \end{equation}
where we use [Mg/Fe] as a proxy for [$\alpha$/Fe] \citep{Thomas:2003aa, Trager:2000aa}. Considering that most of the $\alpha$-elements are set to match the Mg abundance, this is a reasonable approximation. Our resulting age, [Z/H] and [$\alpha$/Fe] estimates are provided in Table~\ref{tab:specpars}. When multiple measurements are available for a nucleus, we quote the weighted median value.

\begin{figure*}[htbp]
   \centering
   \epsscale{1.1}
   \plotone{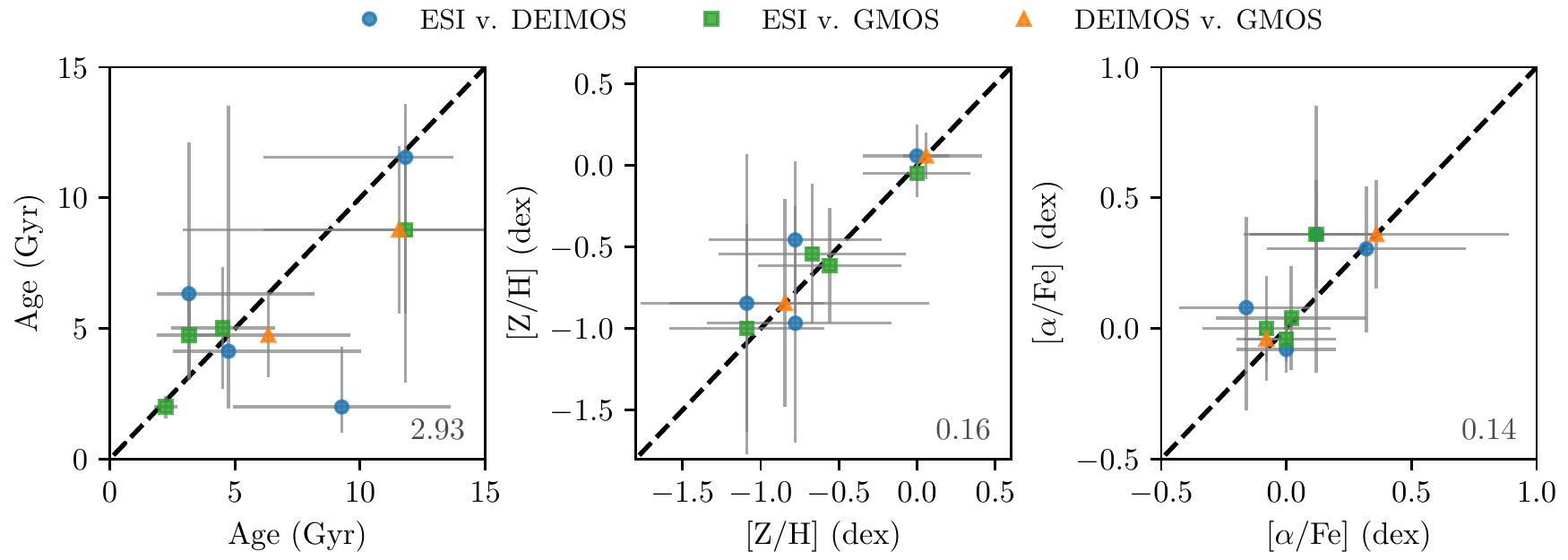} 
   \caption{Comparison of age, [Z/H], and [$\alpha$/Fe] estimates among the ESI, GMOS and DEIMOS datasets. Symbols are the same as in Figure~\ref{fig:indexcompare}. The dashed line in each panel shows the one-to-one relation, while numbers in the bottom right corner show the RMS scatter in the units of each panel.}
   \label{fig:agezcompare}
\end{figure*}

Just as we did for the indices in \S\ref{subsec:indices}, we now compare the age and abundance estimates for objects included in two or more of the spectroscopic datasets. These comparisons are shown in Figure~\ref{fig:agezcompare}. There are fewer data here compared to Figure~\ref{fig:indexcompare} because \texttt{EZ-AGES} could not always converge on a fit to the provided index measurements. We find excellent agreement for [$\alpha$/Fe] and [Z/H] among the datasets. Age estimates are less consistent and less certain than the abundance estimates, likely due to the inherent challenges of separating SSP ages for populations older than a few Gyr. 

\begin{table*}[htbp]
   \centering
   \caption{Best-fit SSP Parameters from Spectroscopy}
   \begin{tabular}{r l c c c}
      \hline \hline
      VCC & Instruments  & T & [Z/H] & [$\alpha$/Fe] \\
       & & (Gyr) & (dex) & (dex) \\
      (1) & (2) & (3) & (4) & (5)\\
      \hline
      
33 & ESI,GMOS & $2.12^{+0.32}_{-0.75}$ & $-0.61^{+0.37}_{-0.42}$& $0.03^{+0.18}_{-0.19}$ \\
200 & ESI & $3.40^{+1.29}_{-4.69}$ & $-0.15^{+0.38}_{-0.43}$& $0.00^{+0.20}_{-0.20}$ \\
230 & ESI & $3.68^{+1.46}_{-4.01}$ & $-0.56^{+0.47}_{-0.54}$& $0.00^{+0.28}_{-0.26}$ \\
538 & ESI & $3.26^{+1.18}_{-4.51}$ & $-0.22^{+0.39}_{-0.43}$& $-0.04^{+0.20}_{-0.20}$ \\
1075 & ESI,DEIMOS & $4.44^{+1.56}_{-5.40}$ & $-0.87^{+0.46}_{-0.47}$& $-0.04^{+0.24}_{-0.22}$ \\
1185 & ESI,GMOS & $4.77^{+1.56}_{-3.57}$ & $-0.59^{+0.29}_{-0.32}$& $-0.04^{+0.16}_{-0.16}$ \\
1192 & ESI & $6.51^{+3.77}_{-6.60}$ & $0.23^{+0.31}_{-0.39}$& $0.24^{+0.22}_{-0.20}$ \\
1199 & ESI & $3.78^{+1.89}_{-6.60}$ & $0.20^{+0.33}_{-0.33}$& $0.04^{+0.14}_{-0.14}$ \\
1407 & ESI,DEIMOS & $5.63^{+2.22}_{-2.46}$ & $-0.62^{+0.36}_{-0.37}$& $0.31^{+0.26}_{-0.23}$ \\
1440 & ESI & $5.59^{+2.89}_{-7.90}$ & $-0.11^{+0.36}_{-0.41}$& $0.04^{+0.23}_{-0.23}$ \\
1489 & ESI & $3.67^{+1.46}_{-2.90}$ & $-1.00^{+0.46}_{-0.47}$& $-0.22^{+0.14}_{-0.21}$ \\
1539 & ESI,DEIMOS,GMOS & $10.72^{+3.60}_{-2.27}$ & $-0.98^{+0.41}_{-0.40}$& $0.28^{+0.21}_{-0.20}$ \\
1545 & ESI,DEIMOS,GMOS & $4.75^{+1.29}_{-2.66}$ & $0.00^{+0.13}_{-0.16}$& $-0.04^{+0.08}_{-0.08}$ \\
1627 & ESI & $4.93^{+3.20}_{-7.95}$ & $0.19^{+0.33}_{-0.32}$& $0.04^{+0.15}_{-0.12}$ \\
1826 & ESI & $3.73^{+1.53}_{-4.68}$ & $-0.44^{+0.44}_{-0.51}$& $0.00^{+0.26}_{-0.26}$ \\
1828 & ESI & $5.20^{+2.59}_{-6.15}$ & $-0.56^{+0.49}_{-0.57}$& $0.12^{+0.29}_{-0.29}$ \\
1861 & DEIMOS & $5.47^{+3.33}_{-5.33}$ & $-0.80^{+0.43}_{-0.49}$& $-0.12^{+0.24}_{-0.26}$ \\
1871 & DEIMOS & $9.15^{+5.67}_{-4.40}$ & $0.04^{+0.12}_{-0.14}$& $0.00^{+0.08}_{-0.09}$ \\
2050 & ESI & $3.31^{+1.20}_{-3.99}$ & $-0.50^{+0.46}_{-0.51}$& $0.00^{+0.26}_{-0.26}$ \\

\hline
   \end{tabular}
   \\ 
   \label{tab:specpars}
\end{table*}

\subsection{Comparison to Previous Spectroscopic Studies} \label{subsec:speccompare}

\begin{figure}[htbp]
   \centering
   \epsscale{1.1}
   \plotone{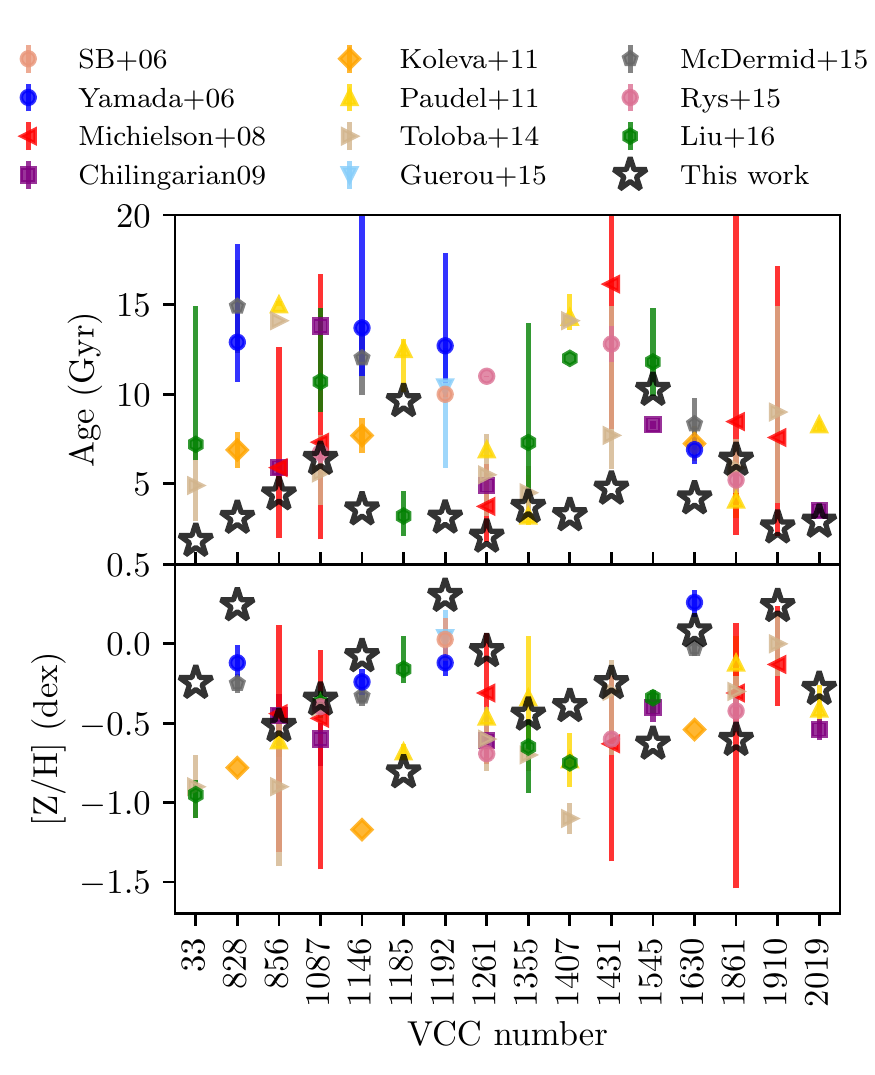} 
   \caption{Comparison of our spectroscopic ages (upper panel) and metallicities (lower panel) with estimates from literature for our program galaxies. Only objects with more than one literature estimate are included. Estimates derived from our SED analysis are plotted as black stars.}
   \label{fig:litspec_gal}
\end{figure}

In Figure~\ref{fig:litspec_gal}, we show published spectroscopic age and metallicity measurements for our sample galaxies along with our estimates from SED fitting of the photometry. In general, spectroscopic values are consistent within their uncertainties, although the average age differences among measurements are $4.3 \pm 2.2$\,Gyr for ages and $0.48 \pm 0.23$\,dex for metallicities. Some of these discrepancies can likely be explained by the use of different model sets when fitting the data, or by differences in the spatial coverage of the galaxies themselves. Consider, for example, the ATLAS3D measurements from \citet{McDermid:2015aa}, whose values are measured within one effective radius and use the \citet{Schiavon:2007aa} models for comparison to Lick index measurements. Their measurements often differ from those of \citet{Koleva:2011aa} who relied on full spectral fitting and compared to the \textsc{p\'egase.hr} models (although they too focused on the region inside one effective radius). Similarly, the discrepancies with \citet{Yamada:2006aa} are likely due to model differences (i.e., they used SSP models from \citet{Vazdekis:1999ab}), the large number of spectral indices used in their analysis, and differences in spatial sampling (i.e., the latter study focused on the galaxy spectrum within $\sim$0.1\,$R_e$). An additional hurdle in measuring ages is the difficulty in distinguishing between SSP models older than ${\sim}6$ Gyr \citep[see, e.g,][]{Powalka:2016aa}. Since most early-type galaxies contain a prominent old stellar population, it is clear that the estimation of accurate ages is quite challenging.

\begin{figure}[htbp]
   \centering
   \epsscale{1.1}
   \plotone{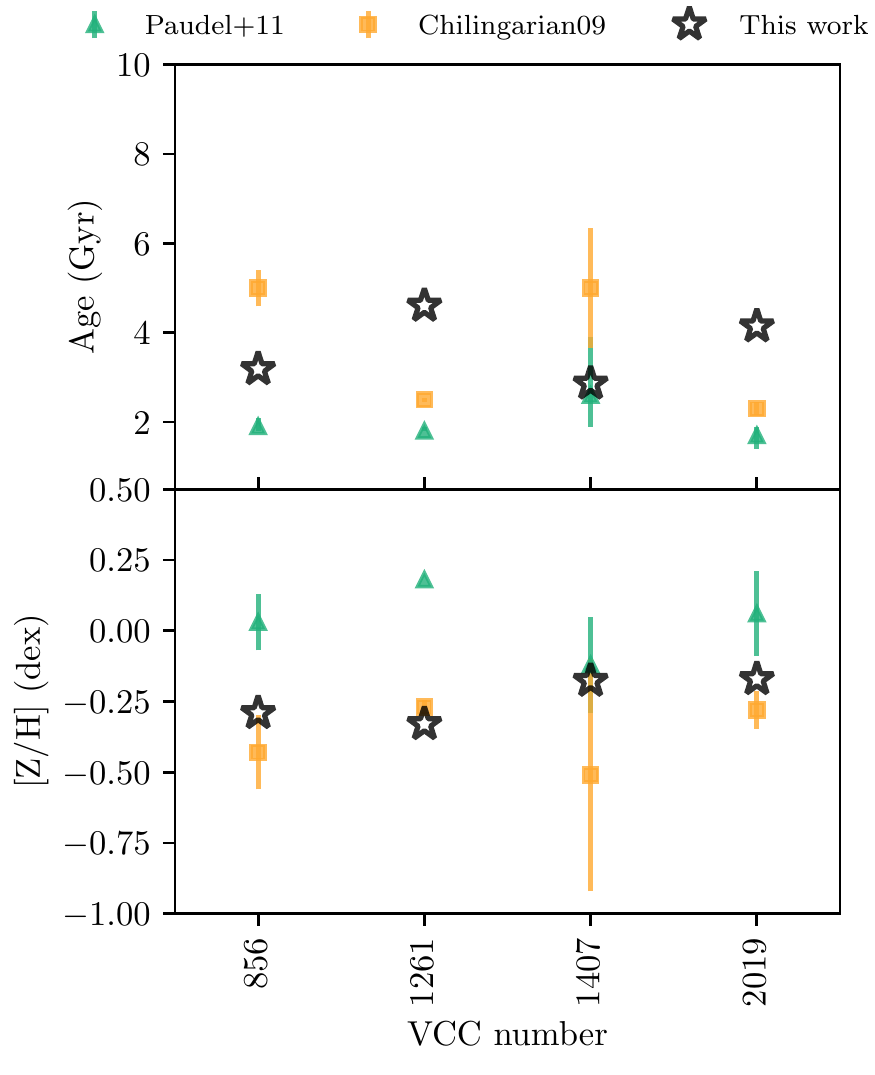} 
   \caption{The same comparison as Figure~\ref{fig:litspec_gal}, but for the nuclei.}
   \label{fig:litspec_nuc}
\end{figure}

In Figure~\ref{fig:litspec_nuc}, we show a similar comparison for the nuclei. Parameter estimates for the nuclei have an additional source of uncertainty --- possible contamination of the nucleus spectrum by the underlying galaxy which could affect the derived nuclei parameters. A comparison of independent spectroscopic measurements may therefore help us understand the importance of such possible systematic errors. Unfortunately, such measurements are available in the literature for only four nuclei in our sample, from two studies: \citet{Chilingarian:2009aa} and \citet{Paudel:2011aa}. The ages are generally in good agreement, with only one nucleus, VCC~856, showing discrepant spectral age estimates from the literature. Three nuclei (VCC~856, VCC~1261 and VCC~2019) have conflicting metalllicity estimates from the literature. We note that \citet{Paudel:2011aa} modeled the galaxy light profile and subtracted it from their nuclei estimates, while \citet{Chilingarian:2009aa} did not. This is likely a key factor in the overall discrepancy between the two sets of measurements. It is also interesting to note that our photometric metallicities seem more consistent with those from \citet{Chilingarian:2009aa}, even though our extraction methodology is more similar to that of \citet{Paudel:2011aa}. This is perhaps an effect of the different SSP models used in each analysis. Of course, it is difficult to draw firm conclusions with only four nuclei in common among the samples.

\subsection{Comparison of Spectroscopic and Photometric Results} \label{subsec:specphotcompare}

In this section, we compare the ages and metallicities derived for the nuclei using the ESI, GMOS and DEIMOS spectra to those found from SED fitting of the photometry. It is worth emphasizing that spectra have all been reduced in similar ways, with the galaxy light modeled and removed from the nucleus in all cases (see \S\ref{subsec:specreduction}. In addition, the spectra have been analyzed in an identical manner, using a single model set and methodology (as described in \S\ref{subsec:indices} and \S\ref{subsec:specpars}). This homogeneous analysis should reduce the possible sources of disagreement among the datasets, so that any scatter in the results should largely be attributable to the data.

\begin{figure}[htbp]
   \centering
   \epsscale{1.1}
   \plotone{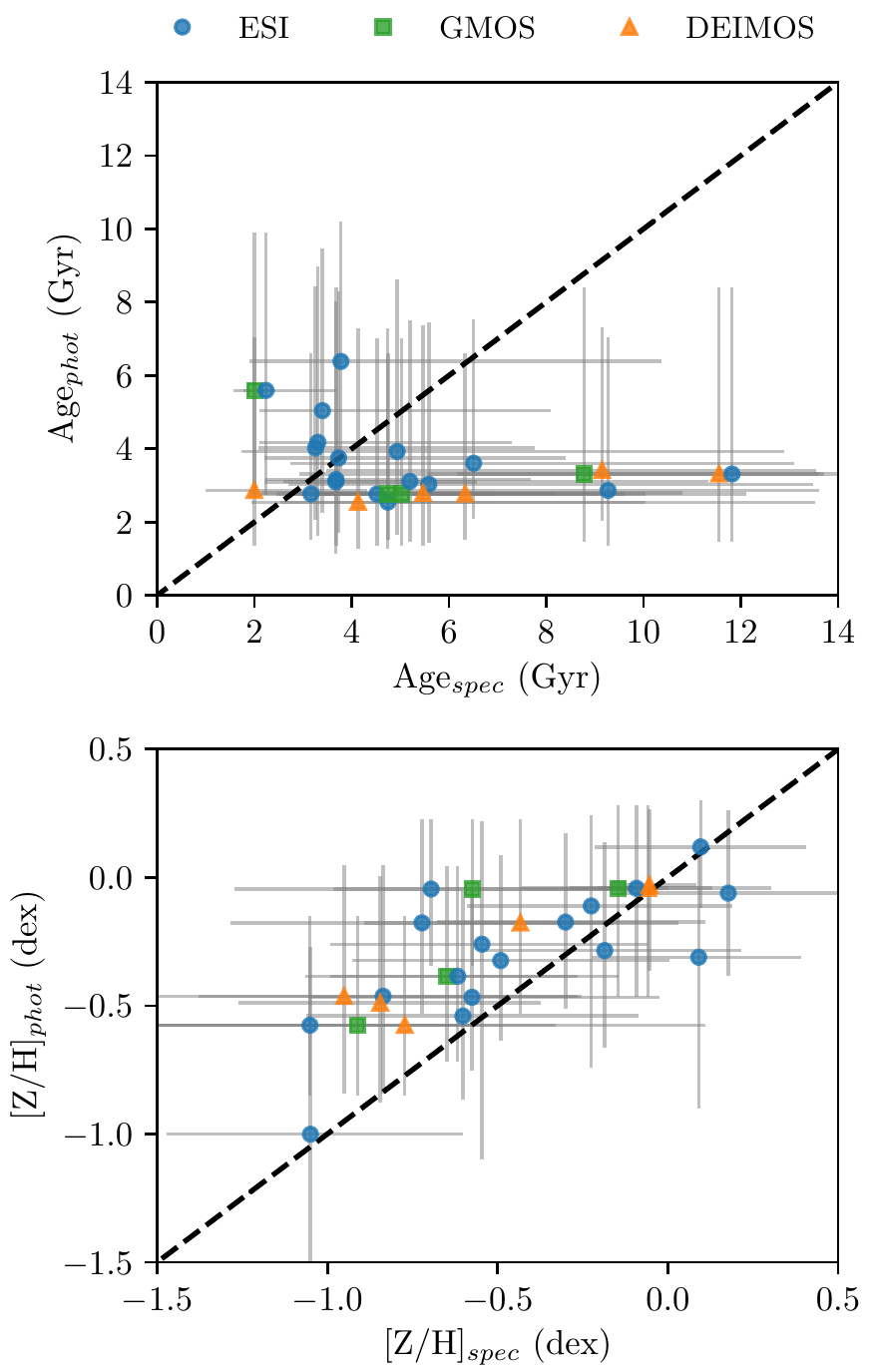} 
   \caption{Comparison of nuclei ages (top panel) and metallicities (bottom panel) derived from our photometric and spectroscopic analyses. The values from SED fitting have been calculated using the BC03 models. Typical uncertainties for the measurements are shown by the black errorbars in each panel. The dashed line in each panel shows the one-to-one relation.}
   \label{fig:specphot}
\end{figure}

Figure~\ref{fig:specphot} compares our photometric age and metallicity estimates (calculated using the BC03 models) to the corresponding spectroscopic estimates. In general, there is good agreement within the uncertainties. In particular, the derived metallicities appear to be robust; the RMS scatter is 0.3~dex with a Spearman rank correlation coefficient $\rho = 0.79$. The ages seem to be more uncertain, five nuclei having fairly old (9--12 Gyr) spectroscopically derived ages but younger ($\sim$3 Gyr) photometric ages. This discrepancy can be attributed to the similar broadband features of populations older than a few Gyr. In addition, the spectroscopic data only include four optical lines, which also limits their ability to discriminate in age. Overall, this comparison illustrates, once again, the challenges inherent in distinguishing between old ($\geq5$\,Gyr) and intermediate-age populations. Based on these comparisons, we conclude that photometry alone can provide accurate metallicity estimates for galaxies and nuclei. The age estimates, while less tightly constrained, can at least eliminate the presence of prominent young ($\lesssim2$\,Gyr) stellar populations in either system.

\section{Discussion} \label{sec:discussion}

\subsection{Masses and Relation to Host Galaxies} \label{subsec:masses}

In the SED fitting process, we measure stellar masses for both nuclei and their host galaxies. The uncertainties in the derived masses range from $\sim$15\% to $\sim$75\%, with most masses having a precision of $\sim$35\%. These uncertainties are dramatically improved over previous literature estimates \citep{Ferrarese:2006aa, Leigh:2012aa, Georgiev:2016aa}, even without imposing fixed values for the age, metallicity, or $M/L$. Our mass uncertainties are dominated by the relatively large uncertainties on age: i.e., at a given metallicity, a 10 Gyr population will require $\sim$30\% more mass to emit as much light as a 2 Gyr population.

\begin{figure}[htbp]
   \centering
   \epsscale{1.1}
   \plotone{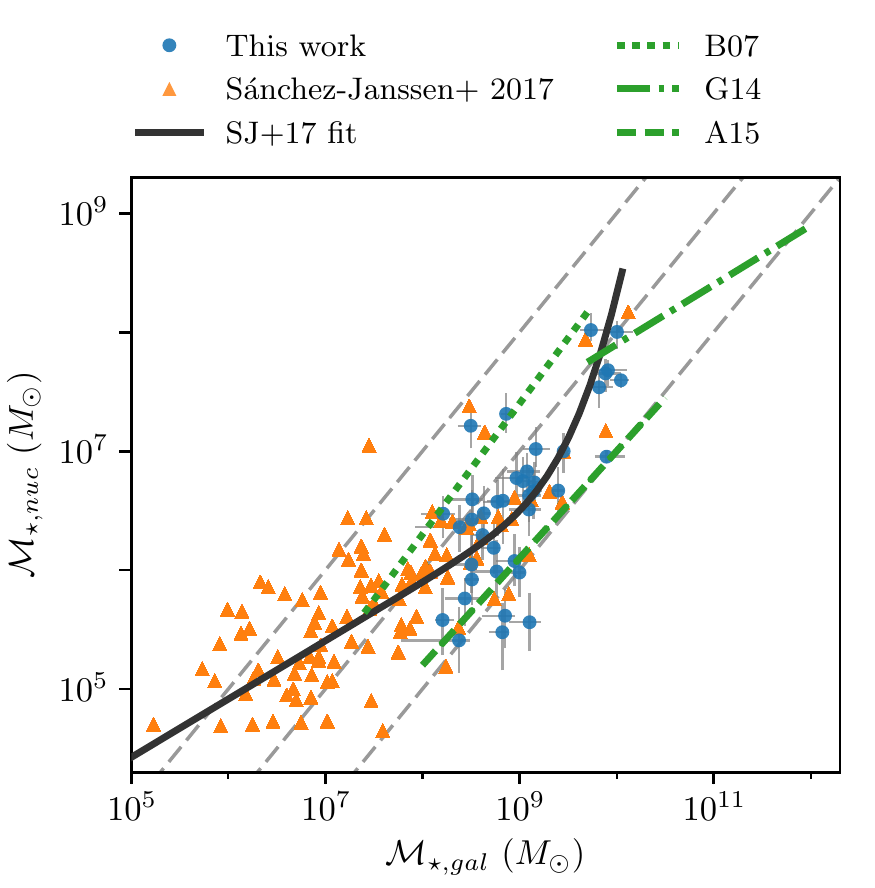} 
   \caption{Nucleus stellar mass plotted as a function of galaxy stellar mass. Blue points show the 39 galaxies from this work, while green points are estimates based on NGVS $u^*griz$ photometry from S\'anchez-Janssen et al.~(2017, in preparation) for 107 nucleated galaxies in a 4 $\deg^2$ region centered on M87. Dotted light gray lines indicate, from left to right, mass fractions of 10\%, 1\% and 0.1\%. The solid dark gray curve shows the mass relation derived by S\'anchez-Janssen et al.~(2017). We also include comparison to three nucleus formation models: \citet{Antonini:2015aa}, \citet{Gnedin:2014aa} and \citet{Bekki:2007aa}. A description of these models is given in the text.}
   \label{fig:masses}
\end{figure}

In Figure~\ref{fig:masses}, we show the nucleus-galaxy mass relation found using our median masses (which were computed using the BC03 SSP models and a Chabrier IMF). To extend the nucleus-galaxy relation to lower masses, we have included masses for 107 nucleated galaxies from the NGVS located in the 4 $\deg^2$ region surrounding M87 (S\'anchez-Janssen et al. 2017, in preparation). Note that these masses are approximate, being calculated from the observed $i$-band magnitudes and a mean $M/L$ computed from the NGVS optical colors. The gray curve shows the best-fit relation of S\'anchez-Janssen et al.~(2017).

We overplot the observed data with predictions from three different simulations: \citet[][hereafter A15]{Antonini:2015aa}, \citet[][hereafter G14]{Gnedin:2014aa} and \citet[][hereafter B07]{Bekki:2007aa}. A15 provide an analytic mass relation for their model \texttt{CliN}. This is a simple model of GC infall within an isolated spheroid. A15 also use a more complex model (\texttt{GxeV}) that allows for \textit{in situ} star formation as well as galaxy and black hole mergers in addition to GC infall. However, the resulting mass relations for both models look quite similar in the regime of our data, with \texttt{GxeV} producing a relation with larger scatter.

The G14 model is a numerical model based on pure cluster infall via dynamical friction. This model produces nucleus-to-galaxy mass fractions that are consistent with the observed fractions for Virgo nuclei (as well as other samples). However, these mass fractions are for more massive simulated galaxies (ranging from galaxies comparable to the Milky Way to M87-like systems). The results from G14 show that nuclei become slightly more prominent as galaxy mass decreases. Therefore, once their mass relation is extrapolated to the mass regime of this work, the predicted nuclei masses are somewhat over-massive. However, it is important to note that G14 quote the total stellar mass within 10 pc of the galaxy center as the nucleus mass; they also caution that their dynamical friction model might be too effective in migrating clusters to the galaxy center. The G14 predicted mass relation differs significantly from the \texttt{CliN} relation; this is likely due to the absence of black hole disruption in the G14 scenario.

Lastly, there is the prediction from the B07 model, which is a pure dissipative formation model accounting for nucleus growth regulation from stellar feedback and a central black hole. The model galaxy is a spheroid (usually about $10^9$ solar masses), within which is embedded a 1 kpc gas disk. Different iterations of the model assign between 2\% and 50\% of the mass to the gas disk. The models typically produce a nucleus with $\sim$4.6\% of the spheroid mass, a larger mass fraction than observed in this sample. This efficient formation scenario may be appropriate for nuclei in lower mass galaxies considered in the model ($M_{\star,gal}\approx10^8$), which tend to be more prominent within the host galaxy.

The B07 and A15 mass relations have slopes that are roughly consistent with the trend among the most massive nucleated galaxies ($M_{\star,gal} \geq 10^8$), although the B07 relation produces overmassive nuclei, while the A15 relation produces undermassive nuclei. The observed mass relation is effectively bounded by these two cases, suggesting that variation in mass fraction can be produced by varying the contribution of dissipative and dissipationless formation processes. This is supported by the results shown in A15, in which the \texttt{GxeV} model produces a wide range of nucleus masses at fixed galaxy mass.

\begin{figure}[htbp]
   \centering
   \epsscale{1.1}
   \plotone{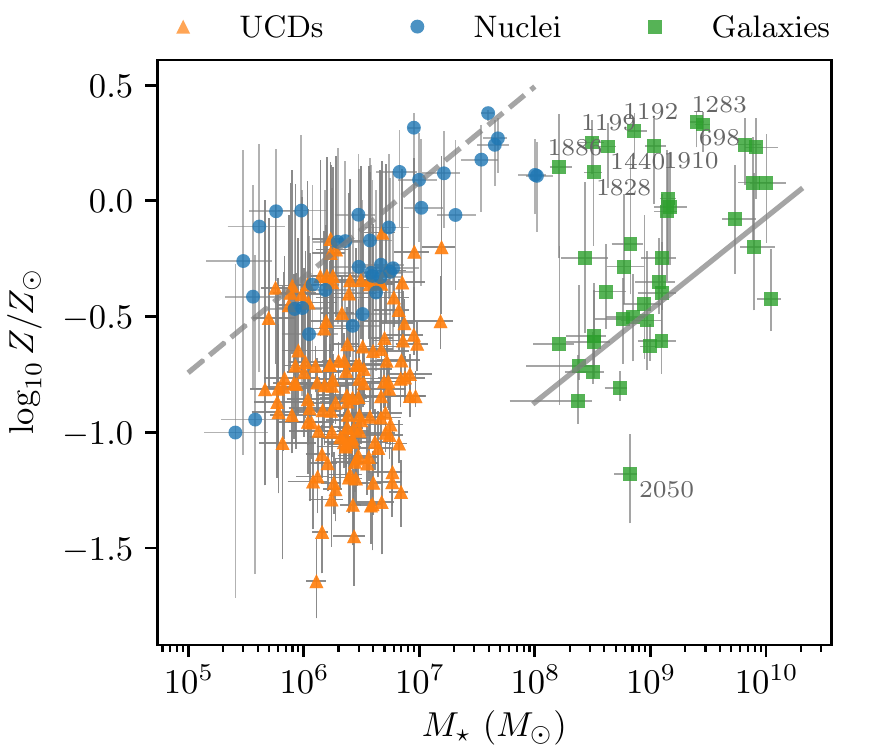} 
   \caption{Mass-metallicity relations for our sample nuclei and galaxies (blue circles and green squares, respectively). For comparison, orange triangles show UCDs in the M87 region from \citet{Liu:2015aa}. Masses and metallicities for the three different types of stellar systems were derived homogeneously using the BC03 SSP models with a Chabrier IMF. Note, however, that F300W and F160W photometry is unavailable for the UCDs, and roughly a third of the sample also does not have $r$ or $K_s$ imaging. The dotted line shows the fitted relation for the nuclei, while the solid line shows the galaxy mass-metallicity relation from the simulations in \citet{Ma:2016aa}, shifted to higher metallicities by 0.3\,dex. VCC numbers are labeled for galaxies that diverge from this mass-metallicity relation.}
   \label{fig:massz}
\end{figure}

\subsection{Abundances} \label{subsec:abundances}

\begin{figure*}[htbp]
   \centering
   \epsscale{1.1}
   \plotone{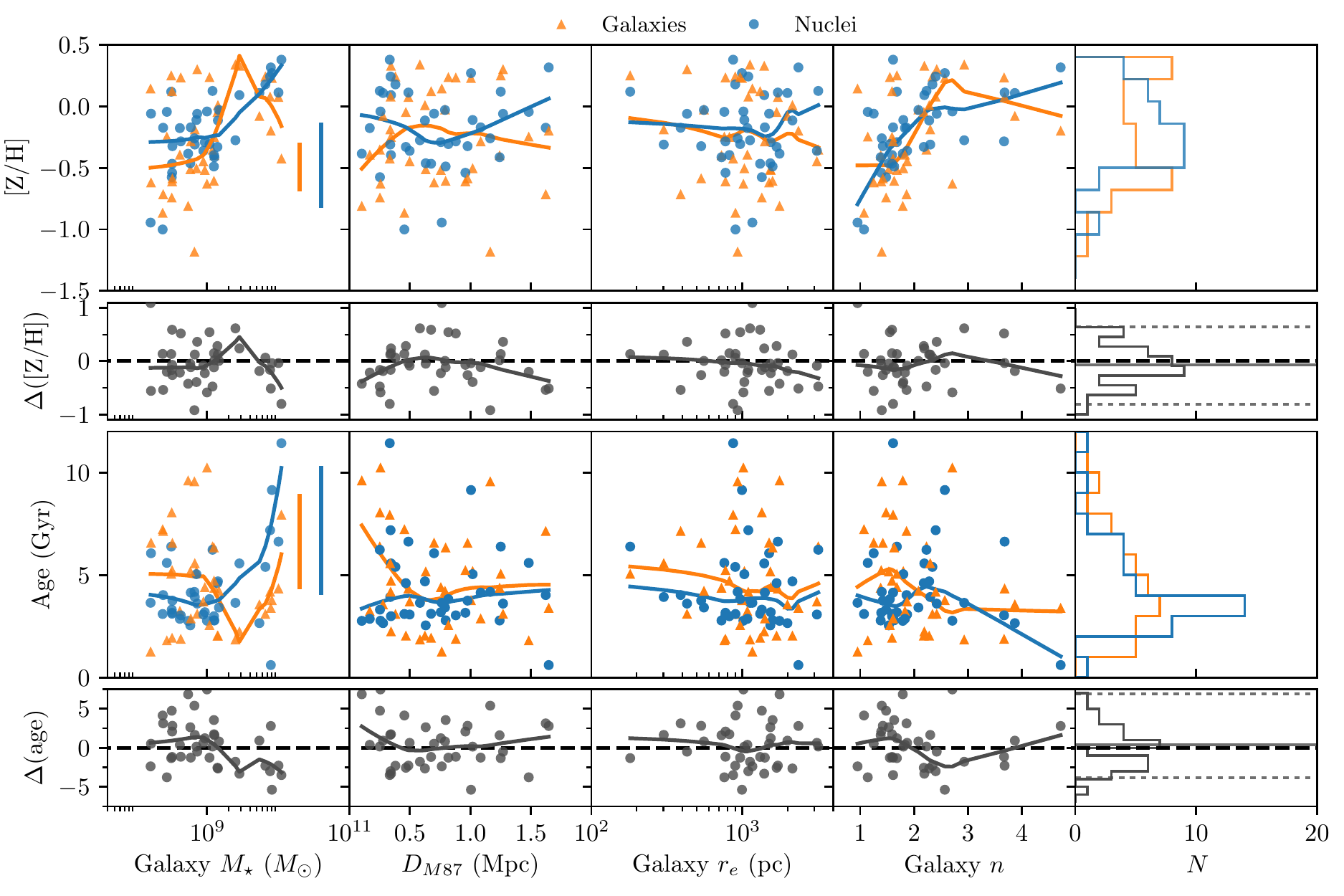} 
   \caption{Metallicities (first row), metallicity differences (galaxy [Z/H] - nucleus [Z/H]; second row), ages (third row), and age differences (galaxy age - nucleus age; fourth row) for galaxies (orange triangles) and nuclei (blue circles). These parameters are plotted, from left to right, against galaxy stellar mass, distance to M87 (as a proxy for environment density), galaxy effective radius, and galaxy S\'ersic index. LOWESS \citep{Cleveland:1979aa} fits are shown as solid lines in each panel. Typical error bars are shown in the first column in orange and blue for galaxies and nuclei, respectively. Distributions for each population are shown in the far right column, with the mean (solid line) and 2.5\ and 97.5\ percentiles (dotted lines) shown for the metallicity and age difference distributions. Metallicity and age values are derived using the BC03 models with a Chabrier IMF.}
   \label{fig:agez_bc03}
\end{figure*}

In this section, we examine the photometrically derived SSP ages and metallicities, focusing mainly on the estimates found using a Chabrier IMF and the BC03 SSP models (Table~\ref{tab:sedagez}). 

A comparison of the masses and metallicities for galaxies and nuclei is shown in Figure~\ref{fig:massz}. For comparison, we also show results for UCDs based on the sample of \citet{Liu:2015aa}. We used the published $u^*giz$ photometry (and, when available, $r$ and $K_s$ photometry) to estimate mass, metallicity and age using the SED fitting procedure described in \S\ref{subsec:sedfits}. This figure shows that a remarkably similar mass-metallicity relation holds for both the nuclei and galaxies. Fitting an equation of the form $\log_{10}Z/Z_\odot = \alpha\log_{10}M_\star + \beta$ yields $\alpha = 0.41 \pm 0.05$ and $\beta = -2.80 \pm 0.35$ for the nuclei. We compare the galaxies to the relation produced by the simulations of \citet{Ma:2016aa}. The slope of this relation fits the data well; however, a shift of +0.3\,dex in metallicity is required to match the data. This offset may be explained by morphological or environmental differences --- since we are looking exclusively at early-type galaxies in a cluster environment --- or by different model assumptions. A few outliers are found to have metallicities even higher than the shifted relation. These include the compact ellipticals VCC~1192, VCC~1199, and VCC~1440, as well as the bright, structurally complex galaxies VCC~698 and VCC~1283. 

Nuclei and galaxies are found to occupy a similar, broad range in metallicity. It is interesting to note that, at fixed mass, the UCDs are systematically {\it less} metal rich (by $0.56 \pm 0.12$\,dex) than the nuclei. UCDs are thought to be either the stripped remains of nucleated galaxies, or simply the high-mass tail of the globular cluster luminosity function \citep[e.g.,][]{Janz:2016aa, Pfeffer:2014aa, Norris:2014aa, Mieske:2013aa}. In the mass range $10^6 - 10^8 M_\odot$, \citet{Janz:2016aa} find that the UCDs span a broad metallicity range $-1.1 \leq \mbox{[Z/H]} \leq 0.2$, consistent with a sample containing both stripped objects and GCs. We do not detect any solar or super-solar metallicity UCDs, which may suggest that our sample is predominantly high-mass GCs. However, a subset of the UCDs overlap with the nucleus sample, suggesting that these UCDs and nuclei are drawn from the same population. Perhaps the UCD hosts were stripped or disrupted at early times, removing the surrounding supply of gas and halting their chemical enrichment. However, firm conclusions regarding UCD metallicities would be premature because other factors may be at play. For instance, the sample of galaxies and nuclei examined in this paper are scattered throughout the entire cluster (see Figure~\ref{fig:footprint}), whereas the UCDs of \citet{Liu:2015aa} are drawn from the central $\sim$ 4 deg$^2$ and, thus, may be among the oldest stellar systems in Virgo \citep[e.g.,][]{Lisker:2009aa}.

In the top two rows of Figure~\ref{fig:agez_bc03}, we plot the nucleus and galaxy metallicities, as well as the metallicity {\it differences}, as a function of galaxy stellar mass, number density, galaxy effective radius, and galaxy S\'ersic index. We see that both galaxies and nuclei trend toward higher metallicities as mass increases. On average, the nuclei in our sample have metallicities statistically indistinguishable from those of their host galaxies, with a mean metallicity $0.07 \pm 0.3$\,dex higher than that of their hosts. However, if we exclude the galaxies (and their corresponding nuclei) that deviate from the mass-metallicity relation in Figure~\ref{fig:massz}, then the nuclei are, on average, $0.20 \pm 0.28$\,dex more metal rich than their hosts. This suggests that most nuclei are not formed primarily via GC infall, as GC systems typically have lower metallicities than their host galaxies \citep{Jordan:2004ab, Puzia:2005aa}. It is also apparent that the metallicity distributions, shown in the top right panel of Figure~\ref{fig:agez_bc03}, are shaped differently, with nuclei having a broad, single-peaked distribution, compared to a bimodal distribution for the galaxies.

The third column of Figure~\ref{fig:agez_bc03} suggests that the smallest galaxies in our sample exclusively have metallicities [Z/H] $\geq -0.5$. Although this may seem counterintuitive, the relation is, in fact, driven by the small number of intrinsically rare, compact ellipticals in our sample: e.g., VCC~1192, VCC~1199, and VCC~1627. These galaxies, despite their low masses and compact sizes, have some of the oldest ages and highest metallicities in our sample. This would be consistent with the notion that they represent the tidally-stripped relics of initially much more massive galaxies \citep{Guerou:2015aa}.

The relationship between metallicity and S\'ersic index in the fourth column of Figure~\ref{fig:agez_bc03} is likely another manifestation of the mass-metallicity relation. Dwarf ellipticals tend to have lower indices than their more massive counterparts \citep[e.g.,][]{Caon:1993aa,Ferrarese:2006aa,Mahajan:2015aa}, so the S\'ersic index effectively traces mass.

\subsection{$\alpha$-Element Abundances}

\begin{figure*}[htbp]
   \centering
   \epsscale{1.1}
   \plotone{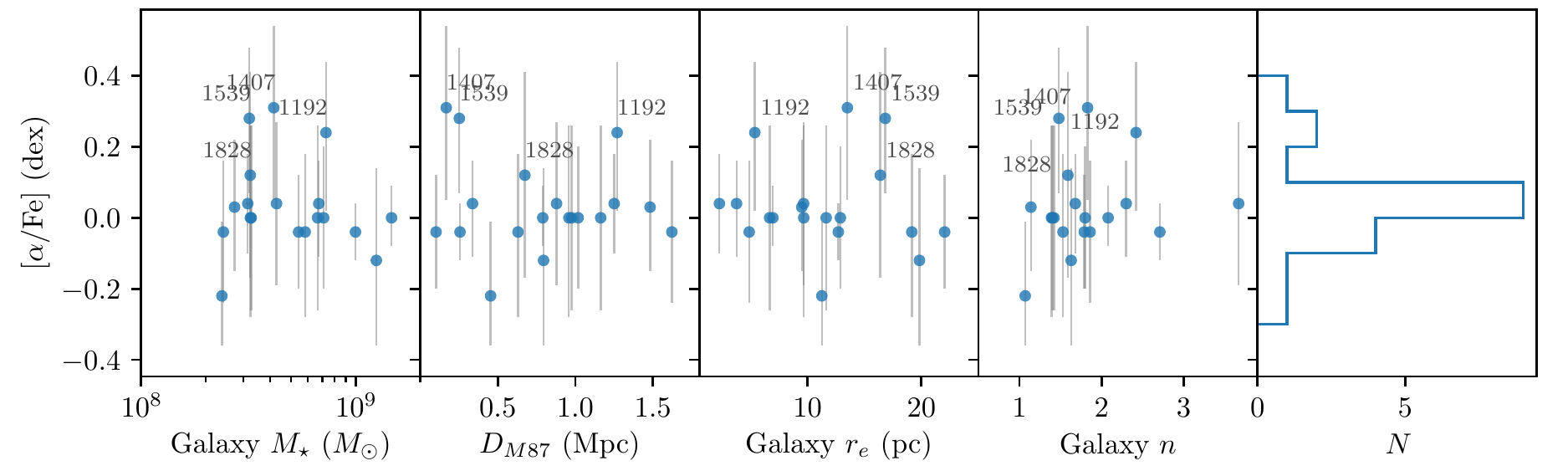} 
   \caption{Nucleus [$\alpha$/Fe] plotted as a function of, from left to right, galaxy stellar mass, distance to M87 (as a proxy for environmental density), galaxy effective radius, and galaxy S\'ersic index. The far right panel shows the [$\alpha$/Fe] histogram. Nuclei with [$\alpha\mbox{/Fe]} > 0.1$ have been labeled by VCC number.}
   \label{fig:all_alpha}
\end{figure*}

The $\alpha$-element abundance [$\alpha$/Fe] is known to trace star formation timescales, with short timescales corresponding to higher [$\alpha$/Fe] values. Type II supernovae from the most massive, rapidly-evolving stars eject relatively large amounts of $\alpha$ elements. The [$\alpha$/Fe] of the galaxy only begins to decrease after less $\alpha$-enhanced ejecta from Type Ia supernovae begin to appear. Previous results for low-mass galaxies show that their star formation timescales are regulated by the density of their environment, with galaxies in the densest regions having both super-solar [$\alpha$/Fe] and higher GC specific frequencies \citep{Liu:2016aa}. In addition, GCs in early type galaxies are known to have super-solar [$\alpha$/Fe] \citep{Puzia:2005aa}. Therefore the [$\alpha$/Fe] of nuclei may indicate the importance of environment and/or GC infall in nucleus formation.

While our spectroscopic analysis does not include the galaxies in our sample, we can still investigate any trends among the 19 nuclei for which we have measured [$\alpha$/Fe]. We show [$\alpha$/Fe] as a function of galaxy mass, distance to M87, galaxy $r_e$, and galaxy $n$ in Figure~\ref{fig:all_alpha}. There are no clear trends, with most nuclei having roughly solar [$\alpha$/Fe] regardless of environment or galaxy properties. However, it is interesting to note that the galaxies VCC~1407 and VCC~1539, which were found to have super-solar [$\alpha$/Fe] and high GC specific frequency by \citet{Liu:2016aa}, have similar [$\alpha$/Fe] in their nuclei. The VCC~1185 galaxy also has with a similarly high [$\alpha$/Fe], but its nucleus has [$\alpha$/Fe]$ = -0.04$ and it has a low GC specific frequency for its [$\alpha$/Fe]. In general, though, the lack of significant $\alpha$ enhancement suggests that nuclei have not formed through particularly brief star formation episodes. A more thorough investigation of the connections among nuclei, GC populations, and [$\alpha$/Fe] is beyond the scope of this paper.

\subsection{Ages} \label{subsec:ages}

As discussed previously, measuring accurate ages for SSP populations older than a few Gyr can be challenging, given the similarities among the model spectra within one model family, and the differences between predictions from various codes \citep{Powalka:2016aa}. Therefore, our age estimates are hampered by somewhat large uncertainties and any apparent trends should be considered with caution.

As can be seen in the left panel of the third row of Figure~\ref{fig:agez_bc03}, there appear to be two populations, clustered distinctly in age, among the galaxies and the nuclei. The most massive objects ($M_{\star,gal}\geqslant2\times10^9 M_{\odot}$) contain nuclei that are consistently older, with typical ages of $\sim$7 Gyr, than the lower-mass nuclei (which have ages scattered around 4 Gyr). The overall age distributions for galaxies and nuclei (shown in the far right panel of the third row) appear quite similar and there are no clear age offsets between the two types of object. However, the age differences in the bottom right panel do have a tail toward positive values in which the galaxy is older than the nucleus.

The density of the surrounding environment, as traced by distance from M87, does not appear to have a strong effect on the galaxies or the nuclei. There is perhaps some evidence that older objects tend to be found in higher density environments, as expected if the earliest objects to fall into the Virgo Cluster are now found close to the bottom of the cluster potential. Having been stripped of any gas upon infall, only old stellar populations would remain.

We reiterate that the derived ages are SSP-equivalent ages. The galaxies almost certainly host complex populations, and it is quite possible that nuclei consist of multiple stellar populations. As a result, fitting an SSP model introduces some bias and uncertainty. If the actual stellar content is primarily that of an old population, even a small contribution from a substantially younger population could skew the SSP-equivalent, luminosity-weighted age. Also, the results of this work do not consider the effect of dust extinction. Future analyses should consider stellar populations that are more complex than SSPs and explore the possibility of non-zero dust extinction.

\subsection{Structural Parameters of Nuclei} \label{subsec:structure}

\begin{figure*}[htbp]
   \centering
   \epsscale{1.1}
   \plotone{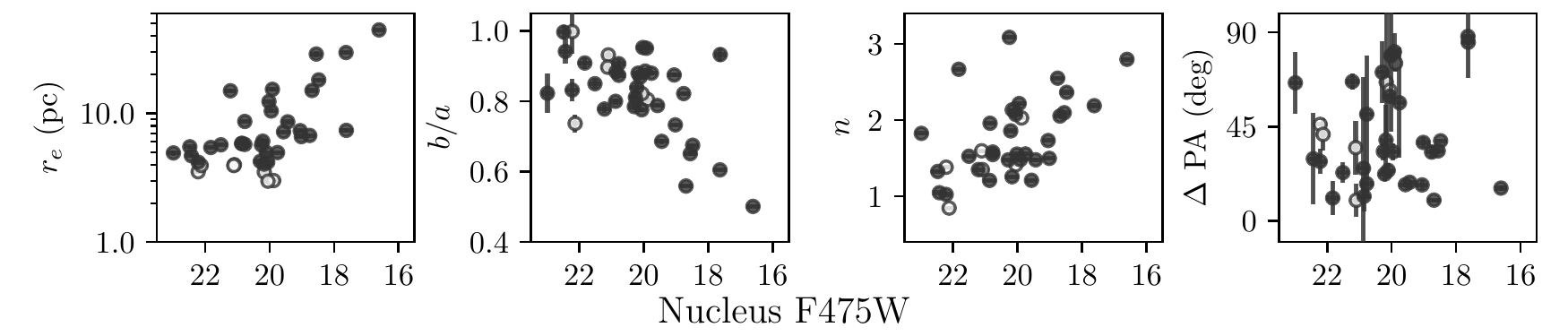} 
   \caption{Trends in nucleus effective radius, axis ratio, S\'ersic index and position angle offset between the nucleus and galaxy, plotted as a function of nucleus F475W magnitude. Open circles denote nuclei that we consider marginally resolved ($r_e < 4$\,pc or 0\farcs05) and as such have less robust structural measurements. All parameters have been measured using a single S\'ersic component fitted to the nucleus using \textsc{galfit} as described in \S\ref{subsec:galfit} with the exception of the galaxy position angle (see text for details).}
   \label{fig:nuc_struc}
\end{figure*}

By fitting each nucleus with a S\'ersic profile, we can measure not only its effective radius, $r_e$, but also its concentration index, $n$, axis ratio, $b/a$, and position angle, PA. However, for nuclei that are only marginally resolved, these structural parameters are clearly not very meaningful --- after convolving the model with the PSF, any intrinsic flattening would go unnoticed. Nevertheless, in this section, we look for possible trends in $r_e$, $b/a$, $n$, and the offset between nucleus and galaxy PA, $\Delta$PA. Because most of the galaxies have been modeled with multiple components in our 2D decomposition, there is no single PA that we can provide for the galaxy. Therefore, we estimate a mean PA from the \texttt{ELLIPSE} isophotes between 1\arcsec{} and 1 $r_e$, with $r_e$ determined through curve of growth analysis (Ferrarese et al. 2016).

We have searched for trends in these structural parameters as a function of magnitude, environment, metallicity, and color, finding that most parameters are tightly correlated with nucleus magnitude. In Figure~\ref{fig:nuc_struc}, we show nucleus $r_e$, $b/a$, $n$, and $\Delta$PA as a function of nucleus magnitude. \citet{Cote:2006aa} determined the resolution limit for ACSVCS images to be 2\,pc, so we consider nuclei with $r_e < 4$\,pc (0\farcs05) to be marginally resolved. Seven nuclei meet this criterion and have been indicated with open circles in the figure. Overlooking these nuclei, some trends emerge. Unsurprisingly, the brightest nuclei tend to be larger \citep{Cote:2006aa} and have larger S\'ersic indices compared to the fainter nuclei. Indeed, the trends in these nuclear parameters mimic what we observe for the galaxies. However, it is interesting to note that as nucleus luminosity increases, the nuclei become more flattened, and are weakly aligned with the semi-major axis of the host galaxy. If this flattening is indicative of rotation, then it is likely that these nuclei formed predominantly via dissipative processes when gas falls to the center of the galaxy and forms a rotating disk. On the other hand, recent dissipationless models have been able to produce a rotating, flattened nucleus as well \citep{Tsatsi:2017aa}. Understanding the significance of the observed trends may shed light on the various formation scenarios, although it is already clear that kinematic information with high spatial and spectral resolution would be extremely useful in discriminating between the competing models. For now, we simply note that these results suggest that different formation mechanisms may dominate in different regimes of nucleus and galaxy mass \citep{Turner:2012aa}.

\subsection{Co-existence with Supermassive Black Holes} \label{subsec:smbhs}

Many studies have established that both a nucleus and supermassive black hole (SMBH) exist in the Milky Way \citep[e.g.,][]{Ghez:2008aa, Schodel:2007aa, Becklin:1968aa}, a situation that is true of some other galaxies as well \citep{Seth:2008aa,Neumayer:2012aa}. In this section, we investigate whether any nuclei in our sample might contain a supermassive black hole as well --- a possibility that bears consideration following the recent discovery of a supermassive black hole in a Virgo UCD \citep{Seth:2014aa}. In addition, the various proposed modes of SMBH formation produce different occupation fractions in this regime of galaxy mass, so SMBH detections can be a valuable constraint in our understanding of SMBH formation and galaxy evolution \citep{Volonteri:2010aa, Greene:2012aa}.

To test this possible co-existence, we adopt the black hole mass to galaxy mass relation from \citet{McConnell:2013aa},
\begin{eqnarray} \label{eq:smbh}
   \log_{10}(M_\bullet) = (8.46 \pm 0.08) +\nonumber\\(1.05\pm0.11) \log_{10}(M_{\star,bulge}/10^{11} M_\odot),
   \end{eqnarray}
   where $M_\bullet$ is the mass of the SMBH and $M_{\star,bulge}$ is equal to galaxy stellar mass for early-type galaxies. For a simple, first-order approach, we assume that this relation applies for the total ``compact massive object" (CMO) mass \citep{Ferrarese:2006ab}, noting that previous work suggests nuclei and SMBHs may follow separate mass relations \citep{Balcells:2007aa,Scott:2013aa, Leigh:2012aa, Graham:2012aa}. We then highlight any nucleus that deviates from the expected relation at the $3\,\sigma$ level or more.
   
\begin{figure}[htbp]
   \centering
   \epsscale{1.1}
   \plotone{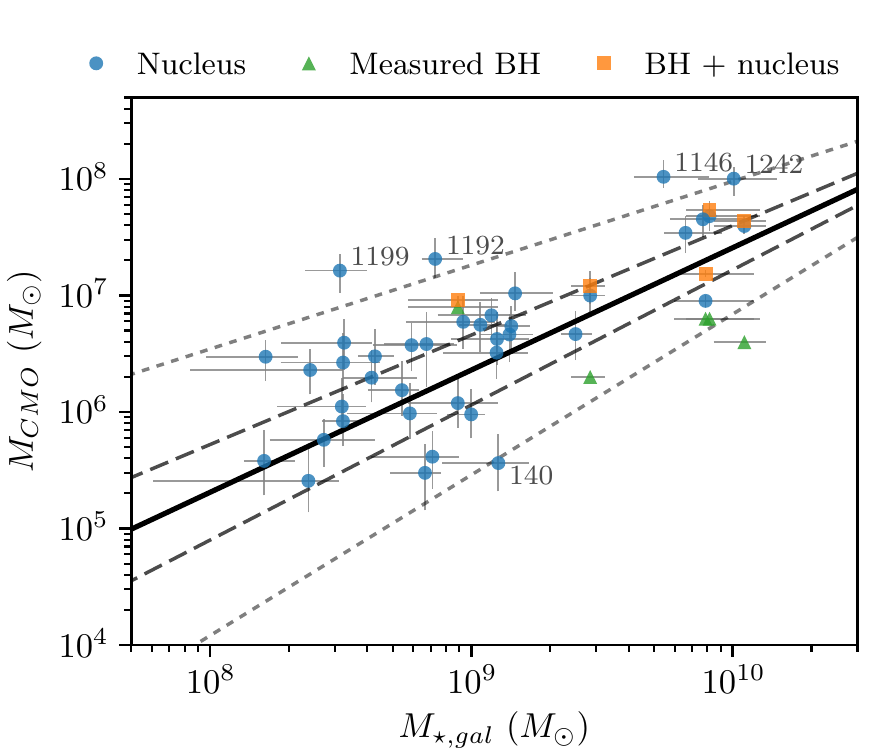} 
   \caption{Relationship between galaxy stellar mass and the mass of the central massive object (CMO), including confirmed black holes within nuclei. The solid line shows the relation from \citet{McConnell:2013aa} with 1\,$\sigma$ and 3\,$\sigma$ uncertainties indicated by the dashed and dotted lines, respectively. The stellar masses of our nuclei are shown as blue circles. Green triangles indicate black hole detections from AMUSE-Virgo \citep{Gallo:2010aa}, while orange squares show the total mass of these black holes and their corresponding nuclei. Objects that fall outside the 3\,$\sigma$ region are labeled with their VCC numbers.}
   \label{fig:bh_nuc}
\end{figure}

The results of this exercise are shown in Figure~\ref{fig:bh_nuc}, along with any confirmed SMBH masses based on X-ray detections in the AMUSE-Virgo survey \citep{Gallo:2010aa}. VCC~140 is an under-massive outlier nucleus; a SMBH of $\sim$$10^{6.4} M_\odot$ would be necessary to match the relation from Eq~\ref{eq:smbh}. We also consider galaxies in our sample that may have over-massive nuclei. Significant outliers in this region include the compact ellipticals, VCC~1192 and VCC~1199, and the bright galaxies, VCC~1146 and VCC~1242, which have structurally complex inner regions. One might expect to find compact ellipticals in this region if they are indeed the tidally-stripped relics of more massive galaxies that hosted a similarly massive nucleus. Meanwhile, the nuclei of galaxies such as VCC~1146 and VCC~1242 seem to be markedly different morphologically from other nuclei \citep{Turner:2012aa}, which may be evidence that these nuclei have experienced a different evolutionary path, perhaps leading to increased growth of the CMO.

\section{Summary} \label{sec:summary}

We have carried out a comprehensive analysis of the stellar populations and masses of 39 nuclei belonging to early-type galaxies in the Virgo cluster. The UV, optical and infrared datasets that form the basis of our analysis --- consisting of both imaging and spectroscopy --- are the most extensive ever used to characterize the stellar content of nuclei in nearby, early-type galaxies. Our photometric analysis rests on multi-band imaging from HST (ACS, WFPC2, and NICMOS) and CFHT (MegaCam and WIRCam) that was collected in the course of the {\it Virgo Redux} survey, {\it ACS Virgo Cluster Survey} \citep[ACSVCS;][]{Cote:2004aa}, and {\it Next Generation Virgo Cluster Survey} \citep[NGVS;][]{Ferrarese:2012aa}. For 19 of our program nuclei, we have also analyzed long-slit and/or IFU optical spectroscopy from the Keck~II (DEIMOS and ESI) and Gemini South (GMOS) telescopes.

Nucleus and galaxy magnitudes were extracted through two methods: (1) a two-component (nucleus and galaxy) surface brightness profile decomposition, using composite profiles created from the high resolution HST data combined with deep, wide-field CFHT data; and (2) a multi-component image decomposition, allowing for more complex galaxy structure, using \textsc{galfit}. After a careful comparison of the two methods, the two-dimensional approach was used to produce our final photometric measurements for the extracted nucleus and its host. Through Markov Chain Monte Carlo fitting of the spectral energy distributions to various sets of SSP models, we have determined robust mass and metallicity measurements, as well as broad age estimates. Parameters obtained from our photometric analysis have been compared to the spectroscopic results, derived homogeneously using the {\tt EZ-AGES} code \citep{Schiavon:2007aa, Graves:2008aa}.
 
The main results of this work can be summarized as follows:

\begin{itemize}
  \item Regardless of the choice of SSP model, there are no strong systematic trends in the derived properties. The \citet{Maraston:2005aa} models can produce a broader, younger range of ages for the sample, likely due to their treatment of the TP-AGB stellar evolutionary phase. Nuclei stellar population parameters derived from the Keck-DEIMOS, Keck-ESI and Gemini-GMOS spectra show good internal agreement, despite the different instrumental setups.
  \item A comparison of spectroscopic age and metallicity estimates in the literature for nuclei and galaxies in our sample shows a significant level of scatter among the measured parameters. Some variation may be due to differences in model assumptions (i.e., adopted isochrones, spectral libraries and stellar evolution treatments) or data analysis methods (i.e., the radius selected for analysis or decomposition techniques). Although homogeneous datasets should still provide reliable \emph{relative} age and metallicity estimates, this comparison suggests that conclusions on the stellar populations in the nuclei, and differences with respect to their host galaxies, should be viewed with caution.
  \item The photometric metallicities are in reasonable agreement with those derived from spectroscopy for the nuclei, with an RMS scatter of $\sim$0.3\,dex. Photometric ages are scattered around 4\,Gyr, although spectroscopic ages can be as old as 12\,Gyr.
This discrepancy may be caused by (1) loss of age sensitivity for old stellar populations in optical spectra; or (2) the possible presence of a small young stellar population that can enhance the blue-optical and UV fluxes in the nuclei. The limited age resolution at old and intermediate ages available from SED fitting is due, in part, to the modest SNR in the UV data, and the underlying assumption of pure SSP populations (which can display similar broadband features for most ages older than a few Gyr).
  \item Our computed stellar masses (measured from SED fitting to six to ten photometric bands) are accurate to typical precisions of 35\%. This is nearly a factor of two improvement over previous measurements that have usually been derived using just one or two photometric bands with assumed ages and/or metallicities \citep{Georgiev:2016aa, Leigh:2012aa, Ferrarese:2006aa}. Over the range of $10^{8.4}$ to $10^{10.3} M_\odot$ in galaxy stellar mass for our galaxies, the nuclei are found to contribute a fraction of $0.33^{+0.09}_{-0.07}$ percent of the total stellar mass, consistent with previous results for early-type galaxies based on less precise stellar masses for the nuclei and simpler nucleus-galaxy decompositions \citep{Turner:2012aa, Cote:2006aa, Ferrarese:2006aa}. The $M_{nuc}$ vs. $M_{gal}$ relation is also consistent with new results from S\'anchez-Janssen~(2017, in preparation) which extend the relation to much lower masses (after combing the observed $i$-band luminosities with an assumed $M/L=1$).
   \item The nuclei show evidence for a rather steep mass-metallicity relation of the form $\log_{10}Z/Z_\odot \propto \alpha\log_{10}M_\star$ with $\alpha = 0.41 \pm 0.05$. A similar trend is exhibited by UCDs (with masses $10^6 \leq M_\star \leq 10^{7.3} M_\odot$) in the Virgo core region, although the UCDs are more metal-poor, by $0.56 \pm 0.12$\,dex, at fixed mass. The galaxies follow the slope of the relation measured in \citet{Ma:2016aa}, but systematically shifted to higher metallicities by $\sim$0.3\,dex.
  \item Nuclei metallicities are statistically indistinguishable from those of their hosts, appearing $0.07 \pm 0.3$\,dex more metal-rich on average. However, excluding outlier galaxies (i.e., compact ellipticals and morphologically unusual galaxies) that do not follow the mass-metallicity, nuclei are $0.20\pm0.28$\,dex more metal-rich than their hosts, qualitatively consistent with conclusions from previous studies \citep{Paudel:2011aa, Chilingarian:2009aa, Koleva:2011aa}. There is no clear age difference, with nuclei ages showing a broad distribution between $\sim$3 and $\sim$12 Gyr.
  \item There is a clear trend for the brightest nuclei to be the most flattened; these bright nuclei may also be more closely aligned with the major axes of their hosts. Due to the barely-resolved sizes of the fainter nuclei, it is unclear how these trends manifest in the fainter regimes. However, this suggests that the largest nuclei --- which belong to galaxies with stellar masses greater than $\sim$10$^{9.5}M_{\odot}$ --- may be formed predominantly through dissipative processes that can induce flattening and rotation.
\end{itemize}

A number of questions regarding the stellar populations and the formation of nuclei remain unanswered. This work has adopted a fairly simple approach to the stellar populations of the nuclei. Future work should consider the effects of internal extinction as well as more complex stellar populations, which may provide a better understanding of the systematics of the age estimates. Given current predictions from formation scenarios, it is still difficult to determine how much each mechanism might contribute to the formation of a particular nucleus, or whether certain processes become more important in different regimes of mass, environment, or other properties. This could be addressed with model predictions for not only size, mass and velocity dispersion relations, but also ages and abundances, which would provide more points of comparison with observations. While some models have presented qualitative statements about the ages of nuclei relative to their hosts \citep{Bekki:2007aa}, precise predictions for relative or absolute ages could prove useful. Simulations that include multiple processes of nucleus formation, such as those in \citet{Antonini:2015aa}, would be ideal for investigating the relative contributions of dissipative and dissipationless processes, and any differences these may produce in abundance and age distributions.

The present study has focused on a somewhat limited sample of nuclei --- in the sense that we have explored a restricted morphological type and mass range for the host galaxies --- so we do not yet have a complete picture of the nucleus population. Fortunately, many hundreds of nucleated galaxies are available in the NGVS survey area. By applying our methods to the full sample in NGVS, albeit with a smaller number of photometric bands, it should be possible to examine the nucleation fraction, stellar population parameters, and scaling relations for a greatly expanded number of nuclei. This will include exploring a new and important regime in galaxy mass (see, e.g., S\'anchez-Janssen et al. 2017 for first results). The large numbers of UCDs and GCs detected in the NGVS imaging should also enable a full comparison of the properties of compact stellar systems within a single, homogeneous dataset \citep{Powalka:2016aa, Liu:2015aa}.


%% Included in this acknowledgments section are examples of the
%% AASTeX hypertext markup commands. Use \url without the optional [HREF]
%% argument when you want to print the url directly in the text. Otherwise,
%% use either \url or \anchor, with the HREF as the first argument and the
%% text to be printed in the second.

\acknowledgments

The authors would like to thank Joachim Janz and Thorsten Lisker for providing the data used in Figure~\ref{fig:maghist}. This research made use of: {\tt APLpy}, an open-source plotting package for Python hosted at {\tt http://aplpy.github.com}; {\tt Astropy}, a community-developed core Python package for Astronomy (Astropy Collaboration, 2013); and the NASA/IPAC Extragalactic Database (NED) which is operated by the Jet Propulsion Laboratory, California Institute of Technology, under contract with the National Aeronautics and Space Administration. This work was supported in part by the Canadian Advanced Network for Astronomical Research (CANFAR) which has been made possible by funding from CANARIE under the Network-Enabled Platforms program. This research also used the facilities of the Canadian Astronomy Data Centre operated by the National Research Council of Canada with the support of the Canadian Space Agency. The authors thank the directors and staff of the Canada-France-Hawaii Telescope for their outstanding efforts in support of the NGVS. E.W.P. acknowledges support from the National Natural Science Foundation of China through Grant No. 11573002.


%% copy editing.  Individual instruments can be provided in parentheses,
%% after the keyword, but they will not be verified.

\facility{CFHT (MegaCam, WIRCam)} \facility{HST (ACS, WFPC2, NICMOS)} \facility{Keck:II (DEIMOS, ESI)} \facility{Gemini:South (GMOS)}
\software{IRAF, Astropy, APLpy, SciPy, Matplotlib, emcee \citep{Foreman-Mackey:2013aa}, corner.py \citep{Foreman-Mackey:2016aa}}



\clearpage
\bibliographystyle{aasjournal}
\bibliography{references}

\end{document}